\pdfoutput=1
\documentclass[12pt]{article}
\usepackage{jheppub}
\usepackage{ifpdf}
\usepackage{graphicx}
\usepackage{microtype}
\usepackage{amsmath,amssymb}
\usepackage[utf8x]{inputenc}
\usepackage[T1]{fontenc}
\usepackage{ae}
\usepackage{aecompl}
\usepackage{titlesec}
\def\bfseries{\fontseries \bfdefault \selectfont \boldmath}

\usepackage{bm}
\usepackage{subfigure}

\usepackage[normalem]{ulem} % either use this (simple) for \sout or
\usepackage{xcolor,soul}
\definecolor{olive}{rgb}{0.3, 0.54, .1}

\DeclareMathOperator{\extdm}{d}
\newcommand{\extd}{\extdm \!}

\titleformat*{\section}{\large\bfseries}
\titleformat*{\subsection}{\bfseries}
\titleformat*{\subsubsection}{\bfseries}
\newcommand{\eps}{\epsilon}
\newcommand{\al}{\alpha}

\newcommand{\de}{\delta}

\newcommand{\la}{\lambda}

\newcommand{\De}{\Delta}

\newcommand{\La}{\Lambda}

\newcommand{\eq}[2]{\begin{equation}
#1\label{#2}
\end{equation}}

\newcommand{\bes}{\begin{eqnarray*}}
\newcommand{\ees}{\end{eqnarray*}}
\newcommand{\bel}[1]{\begin{eqnarray}\label{#1}}
\newcommand{\be}{\begin{eqnarray}}
\newcommand{\ee}{\end{eqnarray}}
\newcommand{\bea}{\begin{eqnarray}}
\newcommand{\eea}{\end{eqnarray}}

\newcommand{\nn}{\nonumber}

\newcommand{\GN}{G_3}
\newcommand{\cut}{\ell_{\textrm{\tiny{cut}}}}

\newcommand{\cen}{\textrm{\tiny{cen}}}

\newcommand{\source}{j}

\preprint{TUW--20--01}
\newcommand{\mytitle}{QNEC$_2$ in deformed holographic CFTs}

\title{\boldmath\mytitle}

\author[1,2,3]{C.~Ecker,}
\author[4]{D.~Grumiller,}
\author[5,6,7]{H.~Soltanpanahi}
\author[4]{and P.~Stanzer}

\affiliation[1]{Institute for Theoretical Physics and Center for Extreme Matter and Emergent Phenomena\\
Utrecht University, Leuvenlaan 4, 3584 CE Utrecht, The Netherlands}
\affiliation[2]{Institute Lorentz for Theoretical Physics, Leiden University\\
 P.O. Box 9506, Leiden 2300RA, The Netherlands}
\affiliation[3]{Institut f\"ur Theoretische Physik, Goethe Universit\"at,
 Max-von-Laue-Str. 1, 60438 Frankfurt am Main, Germany}
\affiliation[4]{Institute for Theoretical Physics, TU Wien, Wiedner Hauptstrasse 8--10, A-1040 Vienna, Austria}
 \affiliation[5]{Guangdong Provincial Key Laboratory of Nuclear Science, Institute of Quantum Matter, South China Normal University, Guangzhou 510006, China}
 \affiliation[6]{Guangdong-Hong Kong Joint Laboratory of Quantum Matter, Southern Nuclear Science Computing Center, South China Normal University, Guangzhou 510006, China}
 \affiliation[7]{Institute of Theoretical Physics, Jagiellonian University, S. Lojasiewicza 11, PL 30-348 Krakow, Poland}

\emailAdd{ecker@itp.uni-frankfurt.de}
\emailAdd{grumil@hep.itp.tuwien.ac.at}
\emailAdd{hesam@m.scnu.edu.cn}
\emailAdd{pstanzer@hep.itp.tuwien.ac.at}

\abstract{
We use the quantum null energy condition in strongly coupled two-dimensional field theories (QNEC$_2$) as diagnostic tool to study a variety of phase structures, including crossover, second and first order phase transitions. We find a universal QNEC$_2$ constraint for first order phase transitions with kinked entanglement entropy and discuss in general the relation between the QNEC$_2$-inequality and monotonicity of the Casini--Huerta $c$-function. We then focus on a specific example, the holographic dual of which is modelled by three-dimensional Einstein gravity plus a massive scalar field with one free parameter in the self-interaction potential. We study translation invariant stationary states dual to domain walls and black branes. Depending on the value of the free parameter we find crossover, second and first order phase transitions between such states, and the $c$-function either flows to zero or to a finite value in the infrared. We present evidence that evaluating QNEC$_2$ for ground state solutions allows to predict the existence of phase transitions at finite temperature.
}

\begin{document}

\maketitle

%%%%%%%%%%%%%%%%%%%%%%%%%%%%%%%%%%%%%%%%%%
\section{Introduction}
%%%%%%%%%%%%%%%%%%%%%%%%%%%%%%%%%%%%%%%%%%

While the main job description of holography is to teach us how quantum gravity works, its favorite pastime is to enlighten us about strongly coupled quantum field theories (QFTs). Starting with the seminal work by Ryu and Takayanagi (RT) \cite{Ryu:2006bv} on the holographic computation of entanglement entropy (EE), the last 1.5 decades have led to a cross-fertilization between quantum information and holography, yielding numerous new tools and insights, see e.g.~\cite{Harlow:2014yka,Hayden:2016cfa,VanRaamsdonk:2016exw,Almheiri:2020cfm} and refs.~therein. 

In the present work we focus on a particular such tool, namely the Quantum Null Energy Condition (QNEC) \cite{Bousso:2015mna} with the aim of using it to diagnose strongly coupled QFTs, their phase structure,  and their ultraviolet (UV) and infrared (IR) behavior. We always assume that the QFT under consideration is two-dimensional, has a conformal field theory (CFT$_2$) fixed point in the UV and a holographic description in terms of a three-dimensional gravity theory with asymptotically anti-de~Sitter (AdS$_3$) solutions.

QNEC locally constrains the expectation value of null projections of the energy-momentum tensor $\langle T_{kk}\rangle$ in terms of lightlike variations (denoted by prime) of EE $S$ 
\eq{
\mathrm{QNEC:} \quad  2\pi\, \langle T_{kk}\rangle\geq\frac{1}{\sqrt{h}}\, S''\qquad\qquad \langle T_{kk}\rangle:=\langle T_{\mu\nu}k^\mu k^\nu\rangle \qquad\forall k^2=0
}{eq:QNEC}
where $h$ is the determinant of the induced metric at the boundary of the entangling region (for details see \cite{Bousso:2015mna}).  Proofs of QNEC exist for free bosonic \cite{Bousso:2015wca} and fermionic \cite{Malik:2019dpg} theories, for CFTs with holographic duals \cite{Koeller:2015qmn} and for interacting QFTs in spacetime dimension $d>2$ \cite{Balakrishnan:2017bjg,Ceyhan:2018zfg}.  Recently the notion of QNEC has been extended also to non-relativistic theories \cite{PhysRevLett.123.121602}. 

In CFT$_2$ QNEC takes the stronger form 
\begin{equation}\label{eq:QNEC2}
\mathrm{QNEC_2:} \quad  2\pi\,\langle T_{kk}\rangle\geq S''+\frac{6}{c}\,\big(S'\big)^2
\end{equation}
where $c$ is the central charge of the CFT$_2$ and the additional positive contribution $\frac{6}{c}(S')^2$ follows from the conformal transformation properties of EE \cite{Wall:2011kb}.
The lightlike variations denoted by prime are defined as follows. At the spacetime point where the left hand side of the QNEC$_2$ inequality is evaluated one of the two endpoints of the entangling interval is anchored, while the second one can be chosen arbitrarily. The first endpoint is deformed into the null direction $k^\mu$, parametrized by an infinitesimal parameter $\lambda$. EE then depends on the deformation parameter $\lambda$ and prime means derivative with respect to it. We shall be more explicit about this construction and how to evaluate QNEC$_2$ in our review in section \ref{sec:1}.
For sake of brevity, when there is no chance of confusion, sometimes we refer to the expression on the right hand side of the inequality \eqref{eq:QNEC2} and sometimes to the inequality itself as QNEC$_2$. 

The first part of our discussion will be general, where we review some known properties of QNEC$_2$ and address also some novel ones, related to kinked EE, first order phase transitions and monotonicity of the Casini--Huerta $c$-function. Later on we focus on bulk matter in AdS$_3$ described by a massive scalar field $\phi$, which on the field theory side corresponds to a deformation of the CFT$_2$ by a scalar operator $\mathcal{O}_\phi$. For concrete examples we assume a family of scalar field potentials with a single free parameter, similar to those studied in \cite{Gubser:2008ny,Janik:2015iry,Attems:2016ugt,Janik:2016btb,Attems:2017ezz,Janik:2017ykj}, which lead to a non-trivial phase structure. The study of this phase structure and in particular of crossovers, second and first order phase transitions using QNEC$_2$ and EE is one of the main goals of our work.

The paper is structured as follows: 
in section \ref{sec:1} we summarize main aspects of QNEC$_2$, holographic EE and AdS$_3$/CFT$_2$, in particular a convexity constraint on kinked EE and the relation between QNEC$_2$ and monotonicity of the Casini--Huerta $c$-function;
in section \ref{sec:model} we review the holographic model, its formulation through a superpotential, its holographic renormalization, domain wall and black brane solutions;
in section \ref{sec:thermo} we present results for the thermodynamic quantities, including free energy, entropy density and the speed of sound for different choices of the potential leading to different kinds of phase transitions;
in section \ref{sec:entanglement} we present the results for the holographic EE and the Casini--Huerta $c$-function, perturbatively for small and large entangling intervals and numerically in between;
in section \ref{sec:QNEC} we present results for QNEC$_2$, in particular for the ground state, where we see evidence for first and second order phase transitions at finite temperature;
in section \ref{sec:Summary} we conclude with a summary and an outlook to generalizations. 

%%%%%%%%%%%%%%%%%%%%%%%%%%%%%%%%%%%%%%%%%%
\section{\texorpdfstring{QNEC$_2$ and AdS$_3$/CFT$_2$}{QNEC2 and AdS3/CFT2} }\label{sec:1}
%%%%%%%%%%%%%%%%%%%%%%%%%%%%%%%%%%%%%%%%%%

In this section we review salient features of QNEC$_2$, including holographic aspects, and present also novel features. In section \ref{sec:1.1} we give a lightning review of AdS$_3$/CFT$_2$. In section \ref{sec:1.2} we display a uniformized result for (holographic) EE valid for all states dual to vacuum solutions of the Einstein equations. In section \ref{sec:1.3} we summarize the proof that QNEC$_2$ saturates for all such states. In section \ref{sec:1.6} we recall relevant features when QNEC$_2$ does not saturate, in particular the half-saturation effect for quenches. In section \ref{sec:1.7} we present a shortcut to QNEC$_2$ for boost invariant states. In section \ref{sec:1.4} we show how to determine efficiently the QNEC$_2$ combination of EE variations for general states. In section \ref{sec:1.5} we demonstrate that QNEC$_2$ poses a convexity constraint on kinks in EE. Finally, in section \ref{sec:1.8} we show a relationship between QNEC$_2$ and the Casini--Huerta $c$-function.
 
%%%%%%%%%%%%%%%%%%%%%%%%%%%%%%%%%%%%%%%%%%
\subsection{\texorpdfstring{AdS$_3$/CFT$_2$}{AdS3/CFT2}}\label{sec:1.1}
%%%%%%%%%%%%%%%%%%%%%%%%%%%%%%%%%%%%%%%%%%

We work mostly on the gravity side and are not specific about the dual QFT, except that it must have a UV fixed point corresponding to a CFT$_2$ without gravitational anomaly and with central charge $c=\frac{3\ell_{\textrm{\tiny{AdS}}}}{2\GN}\gg 1$, where $\ell_{\textrm{\tiny{AdS}}}$ is the AdS radius (which we set to one) and $\GN$ is Newton's constant. The inequality is necessary for the validity of the (super-)gravity approximation. The CFT is put either on a torus, a cylinder or the plane. On the cylinder we use standard coordinates $\extd s^2=-\extd t^2+\extd\varphi^2$ with $\varphi\sim\varphi+2\pi$, and on the plane we use  $\extd s^2=-\extd t^2+\extd x^2$.

The gravity theory we consider is AdS$_3$ Einstein gravity with scalar matter, reviewed in detail in the next section. In the absence of matter the boundary conditions are the seminal ones by Brown and Henneaux \cite{Brown:1986nw} and the solutions to this theory are given by the Ba\~nados metrics \cite{Banados:1998gg}
\eq{
\extd s^2 = \frac{\extd z^2-\extd x^+ \extd x^-}{z^2}+{\cal L}^+(x^+)\big(\extd x^+\big)^2+{\cal L}^-(x^-)\big(\extd x^-\big)^2 - z^2{\cal L}^+(x^+){\cal L}^-(x^-)\,\extd x^+ \extd x^-
}{eq:i1}
where we used lightcone coordinates $x^\pm = t\pm\varphi$. The solution ${\cal L}^\pm = -\tfrac14$ (${\cal L}^\pm = 0$) describes global (Poincar\'e patch) AdS$_3$, while constant positive ${\cal L}^\pm$ yield non-extremal BTZ black holes with horizons located at $r_\pm = |\sqrt{\cal L^+}\pm\sqrt{\cal L^-}|$, temperature $T=(r_+^2-r_-^2)/(2\pi r_+)$ and angular velocity $\Omega=r_-/r_+$.

The holographic dictionary relates the Ba\~nados geometries \eqref{eq:i1} to CFT$_2$ states $|{\cal L}^+,{\cal L}^-\rangle$ (global AdS$_3$ corresponds to the vacuum state $|0\rangle$). Key relations for us are the expectation values of the flux components of the CFT$_2$ energy-momentum tensor expressed in terms of the metric functions ${\cal L}^\pm$,
\eq{
2\pi\big<{\cal L}^+,{\cal L}^-\big|T_{\pm\pm}(x^\pm)\big|{\cal L}^+,{\cal L}^-\big> = \frac c6\,{\cal L}^\pm(x^\pm)\,.
}{eq:i2}
The fact that all Ba\~nados geometries are locally diffeomorphic to each other leads to corresponding uniformization results on the CFT-side. We summarize below this uniformization for EE, which is a necessary ingredient for QNEC$_2$.

%%%%%%%%%%%%%%%%%%%%%%%%%%%%%%%%%%%%%%%%%%
\subsection{Uniformized result for holographic entanglement entropy}\label{sec:1.2}
%%%%%%%%%%%%%%%%%%%%%%%%%%%%%%%%%%%%%%%%%%

There is a simple and uniform result for EE in CFT$_2$ for all states dual to Ba\~nados geometries, namely
\eq{
S(x_1^\pm,\,x_2^\pm) = \frac{c}{6}\,\ln\big(\ell^+(x_1^+,\,x_2^+)\ell^-(x_1^-,\,x_2^-)/\cut^2\big)
}{eq:i3}
where $x_1^\pm$ and $x_2^\pm$ are the two endpoints of the entangling interval, $\cut$ is a UV cutoff (that tends to zero when the cutoff is removed) and the functions $\ell^\pm=\psi_1^\pm(x^\pm_1)\psi_2^\pm(x^\pm_2)-\psi_2^\pm(x^\pm_1)\psi_1^\pm(x^\pm_2)$ are bilinears in solutions to Hill's equation, $\psi^{\pm\,\prime\prime}-{\cal L}^\pm\,\psi^\pm=0$, subject to unit Wronskians, $\psi_1^\pm\psi_2^{\pm\,\prime}-\psi_2^\pm\psi_1^{\pm\,\prime}=\pm 1$. 

On the gravity side the geometric reason for this uniformization is that all solutions to the vacuum Einstein equations are locally AdS$_3$ and therefore there is a diffeomorphism that maps any such geometry locally to Poincar\'e patch AdS$_3$ with coordinates $x^\pm_{\textrm{\tiny{PP}}}$ and $z_{\textrm{\tiny{PP}}}$ \cite{Roberts:2012aq,Sheikh-Jabbari:2016znt}. The coordinate transformation involves the solutions to Hill's equation above, $x^\pm_{\textrm{\tiny{PP}}}=\int\extd x^\pm/\psi^{\pm\,2} - z^2\psi^{\mp\,\prime}/[\psi^{\pm\,2}\psi^{\mp}(1-z^2/z_h^2)]$ and $z_{\textrm{\tiny{PP}}}=z/[\psi^+\psi^-(1-z^2/z_h^2)]$, where $z_h=[\psi_a^+\psi_b^-/(\psi_c^{+\,\prime}\psi_d^{-\,\prime})]^{1/2}$ ($a,b=1,2$ and $c,d=1,2$ in some permutation) is the locus of one of the Killing horizons; since we want to map the outside causal patch to Poincar\'e patch AdS$_3$ for us $z_h$ is always the black hole event horizon, so we have to choose $a,b,c,d$ accordingly.

For Poincar\'e patch AdS (${\cal L}^\pm=0$) Hill's equation is solved by $\psi_2^+=1=\psi_1^-$, $\psi_1^+=x^+$ and $\psi_2^-=x^-$ leading to $\ell^\pm=\pm x_1^\pm\mp x_2^\pm$, thereby recovering the well-known result for EE of a constant-time interval $\ell=|x_2^+-x_1^+|=|x_1^--x_2^-|$ in a CFT$_2$ on the plane \cite{Holzhey:1994we}
\eq{
S_{\textrm{\tiny{PP}}} = \frac c3\,\ln\frac{\ell}{\cut}\,.
}{eq:i4}
By slight abuse of notation we refer to logarithmic behavior in the entangling interval as `area law'. Following the RT prescription \cite{Ryu:2006bv} the result \eqref{eq:i4} is obtained on the gravity side from the length of a geodesic anchored at the endpoints of the entangling interval.

For non-extremal BTZ black holes or black branes (${\cal L}^\pm = \textrm{const.} > 0$) Hill's equation is solved by $\psi_1^\pm=\exp{(\sqrt{{\cal L}^\pm})}/(4{\cal L}^\pm)^{1/4}$ and $\psi_2^\pm=\mp\exp{(-\sqrt{{\cal L}^\pm})}/(4{\cal L}^\pm)^{1/4}$ leading to $\ell^+=\sinh{[\sqrt{{\cal L}^+}(x_2^+-x_1^+)}]$ and $\ell^-=\sinh{[\sqrt{{\cal L}^-}(x_1^--x_2^-)}]$, thereby recovering as special case (non-rotating BTZ, ${\cal L}
^+={\cal L}^-=\pi/\beta$) the result for EE of thermal states in a CFT$_2$ \cite{Calabrese:2004eu}
\eq{
S_{\textrm{\tiny{thermal}}} = \frac c3\, \ln\Big(\frac{\beta}{\pi\cut}\, \sinh\frac{\pi\ell}{\beta}\Big)
}{eq:i5}
where $\beta$ is inverse temperature and $\ell=|x_2^+-x_1^+|=|x_1^--x_2^-|$ is again the constant-time interval defining the entangling region. In the large $\ell$ limit EE \eqref{eq:i5} obeys the volume law 
\eq{
S_{\textrm{\tiny{thermal}}}(\ell\gg 1) = \frac{c\pi\ell}{3\beta} + \textrm{subleading}\,.
}{eq:i6}

%%%%%%%%%%%%%%%%%%%%%%%%%%%%%%%%%%%%%%%%%%
\subsection{\texorpdfstring{QNEC$_2$}{QNEC2} saturation for vacuum-like states}\label{sec:1.3}
%%%%%%%%%%%%%%%%%%%%%%%%%%%%%%%%%%%%%%%%%%

By virtue of the uniformized result \eqref{eq:i3} for EE it is straightforward to prove that the QNEC$_2$ inequality \eqref{eq:QNEC2} saturates for all CFT$_2$ states dual to Ba\~nados geometries \eqref{eq:i1} \cite{Ecker:2019ocp}. Namely, defining a function resembling a vertex-operator, $V=\exp{(6S/c)}=V^+V^-$ with $V^\pm=\ell^\pm/\cut$, leads to the right hand side of QNEC$_2$, $c V^{\pm\,\prime\prime}/(6V^\pm)=S''+\tfrac6c\,(S')^2$, where prime denotes $\extd/\extd x_1^+$ for the upper sign and $\extd/\extd x_1^-$ for the lower sign. On the other hand, the explicit form of the vertex-functions $V^\pm$ shows that they obey Hill's equation, $V^{\pm\,\prime\prime}={\cal L}^\pm V^\pm$. Using the holographic dictionary \eqref{eq:i2} then establishes the equality
\eq{
2\pi\big<{\cal L}^+,{\cal L}^-\big|T_{\pm\pm}(x^\pm)\big|{\cal L}^+,{\cal L}^-\big> = S'' + \frac{6}{c}\,\big(S'\big)^2
}{eq:i6a}
which is the saturated version of the QNEC$_2$ inequality \eqref{eq:QNEC2}.

A related way to understand QNEC$_2$ saturation is to analyze the transformation properties of EE under bulk diffeomorphisms or boundary conformal transformations generated by some [anti-]holomorphic function $\xi(x^+)$ [$\xi(x^-)$]. As shown by Wall EE transforms like an anomalous scalar \cite{Wall:2011kb}.
\eq{
\delta_\xi S = \xi S' - \frac{c}{12}\,\xi'
}{eq:i19}
The first term is the usual Lie-derivative expression for a scalar field and the second, anomalous, term results from dilatation relative to the cutoff. The QNEC$_2$ combination then transforms with the same infinitesimal Schwarzian derivative
\eq{
\delta_\xi\Big(S'' + \frac{6}{c}\,\big(S'\big)^2\Big) = \xi\, \Big(S'' + \frac{6}{c}\,\big(S'\big)^2\Big)^\prime + 2\xi'\,\Big(S'' + \frac{6}{c}\,\big(S'\big)^2\Big) - \frac{c}{12}\,\xi'''
}{eq:i20}
as the boundary stress tensor. This means that whenever QNEC$_2$ saturates for one particular state/geometry it also saturates for all states/geometries related by conformal transformations/diffeomorphisms, which explains the idea of the proof above.

%%%%%%%%%%%%%%%%%%%%%%%%%%%%%%%%%%%%%%%%%%
\subsection{\texorpdfstring{QNEC$_2$}{QNEC2} non-saturation and half-saturation in presence of bulk matter}\label{sec:1.6}
%%%%%%%%%%%%%%%%%%%%%%%%%%%%%%%%%%%%%%%%%%

When bulk matter is present QNEC$_2$ does not saturate in general, and in particular it never saturates when the RT surface intersects regions with bulk matter \cite{Khandker:2018xls}. A sufficient condition for QNEC$_2$ to hold is that bulk matter obeys the null energy condition \cite{Koeller:2015qmn}, which will always be the case in the present work.

An intriguing aspect discovered (but not explained) in \cite{Ecker:2019ocp} is that there is universal half-saturation of QNEC$_2$ for quenches (modelled on the gravity side by Vaidya-type of metrics) in the sense that the ratio of left- and right-hand sides of the QNEC$_2$ inequality \eqref{eq:QNEC2} approaches $\tfrac12$ in the limit of large entangling intervals. Evidence for half-saturation was extracted from numerical and perturbative calculations. This phenomenon was explained more recently in a work by Mezei and Virrueta \cite{Mezei:2019sla} who studied quantum quenches and QNEC$_2$ constraints imposed on them. Their beautifully simple explanation of the ratio $\tfrac12$ is that it corresponds to $\tfrac 1d$ in CFT$_d$, i.e., one over the spacetime dimension of the CFT.\footnote{
Their results not only explain the half-saturation observed in \cite{Ecker:2019ocp} for CFT$_2$, but also the `curious ratio $0.25$' mentioned in an AdS$_5$/CFT$_4$ context in the first numerical study of QNEC, see the caption of Figure~3 in \cite{Ecker:2017jdw}.
} To be more explicit we recall their key statement. Shortly after the quench at time $t=0$ EE has the expansion\footnote{
The state is assumed to be time reflection symmetric at $t=0$.
}
\eq{
S(t,\,\ell) = s_0(\ell) + s_2(\ell)\,t^2 + {\cal O}(t^4)
}{eq:i7}
where $\ell$ is the entangling interval. The quantity $s_0$ drops out in the QNEC$_2$ evaluation since before the quench the Vaidya metric is a special case of a Ba\~nados geometry, so only $s_2$ enters there. The QNEC$_2$ inequality \eqref{eq:QNEC2} then leads to the bound $s_2 \leq 2\pi (e+p)/2$, where $e$ is the energy density and $p$ the pressure of the state under consideration (their sum $(e+p)/2$ corresponds to $\langle T_{kk}\rangle$). Mezei and Virrueta were able to holographically prove a stronger bound for states dual to Vaidya metrics in CFT$_d$, viz.~$s_2 \leq 2\pi (e+p)/(2d)$, which explains the half-saturation for CFT$_2$. (An intrinsic CFT$_2$ proof of this statement has not been found so far, but probably exists.)

When bulk matter is present EE and the right hand side of QNEC$_2$ in general can only be determined numerically, except for special states and in large- or small-interval limits. 

%%%%%%%%%%%%%%%%%%%%%%%%%%%%%%%%%%%%%%%%%%
\subsection{\texorpdfstring{QNEC$_2$}{QNEC2} for boost invariant states}\label{sec:1.7}
%%%%%%%%%%%%%%%%%%%%%%%%%%%%%%%%%%%%%%%%%%

In typical QFTs there is at least one boost invariant state, namely the Poincar\'e invariant vacuum. Whenever we have such a boost invariant state there is a simple way to obtain QNEC$_2$ by calculating EE as function of the interval length $\ell$ and taking a suitable combination of derivatives thereof. We show now how this works.

The null deformation of the interval requires the evaluation of EE for slices where time is not constant. However, if the state under consideration is boost invariant we can always boost to the rest frame and determine EE on a constant time slice in that frame.
\eq{
S(\lambda,\,\ell\pm\lambda) = S(0,\,\sqrt{(\ell\pm\lambda)^2-\lambda^2}) =: S_0(\sqrt{(\ell\pm\lambda)^2-\lambda^2})
}{eq:i21}
The expression on the left hand side denotes EE as a function of the temporal and spatial extent of the null deformed entangling interval in the original frame, while the right hand side is EE in the rest frame, which we denote as $S_0$. 

Expanding $S_0$ as function of proper length in powers of $\lambda$ up to second order establishes the desired relation between QNEC$_2$ and derivatives of EE with respect to the interval length.
\eq{
\frac{\extd^2 S_0}{\extd\lambda^2}\bigg|_{\lambda=0} + \frac{6}{c}\,\bigg(\frac{\extd S_0}{\extd\lambda}\bigg)^2\bigg|_{\lambda=0} = \frac{\extd^2 S_0}{\extd\ell^2} - \frac{1}{\ell}\,\frac{\extd S_0}{\extd\ell} + \frac{6}{c}\,\bigg(\frac{\extd S_0}{\extd\ell}\bigg)^2
}{eq:i22}
Thus, if one knows EE as function of the interval length $\ell$ for a boost invariant state then the right hand side of \eqref{eq:i22} yields the right hand side of the QNEC$_2$ inequality \eqref{eq:QNEC2} for this state.

The simplest holographic example is Poincar\'e patch AdS$_3$. The result \eqref{eq:i4} for EE yields 
\eq{
\frac{\extd^2 S_{\textrm{\tiny{PP}}}}{\extd\ell^2} - \frac{1}{\ell}\,\frac{\extd S_{\textrm{\tiny{PP}}}}{\extd\ell} + \frac{6}{c}\,\bigg(\frac{\extd S_{\textrm{\tiny{PP}}}}{\extd\ell}\bigg)^2 = -\frac{c}{3\ell^2} - \frac{c}{3\ell^2} + \frac{6}{c}\,\bigg(\frac{c}{3\ell}\bigg)^2 = 0\,.
}{eq:i23}
This result provides an alternative proof of QNEC$_2$ saturation for the state dual to Poincar\'e patch AdS$_3$, since also the boundary stress tensor vanishes for this state.

A more interesting example is a situation where the QFT flows from a CFT$_2$ in the UV with central charge $c=c_\textrm{\tiny{UV}}$ to a different CFT$_2$ in the IR with central charge $c_\textrm{\tiny{IR}}$ so that for large values of the interval $\ell$ we have EE obeying an area law
\eq{
\lim_{\ell\gg 1} S_0 = \frac{c_\textrm{\tiny{IR}}}{3}\,\ln\ell + {\cal O}(1)
}{eq:i27}
but with the IR value of the central charge. We then obtain from \eqref{eq:i22} the QNEC$_2$ expression
\eq{
\frac{\extd^2 S_0}{\extd\lambda^2}\bigg|_{\lambda=0,\,\ell\gg 1} + \frac{6}{c}\,\bigg(\frac{\extd S_0}{\extd\lambda}\bigg)^2\bigg|_{\lambda=0,\,\ell\gg 1} = -\frac{2c_{\textrm{\tiny{IR}}}}{3\ell^2}\,\Big(1-\frac{c_{\textrm{\tiny{IR}}}}{c_{\textrm{\tiny{UV}}}}\Big) + \dots
}{eq:i28}
where the ellipsis denotes terms that vanish more quickly than the terms displayed in the limit of large intervals. This means that the well-known inequality $c_{\textrm{\tiny{UV}}} \geq c_{\textrm{\tiny{IR}}}$ can be considered as a consequence of the QNEC$_2$ inequality \eqref{eq:QNEC2}.

For small $\ell$ we expect EE to be close to the Poincar\'e patch AdS$_3$ result \eqref{eq:i4},
\eq{
S_0 = \frac c3\,\ln\ell + \sum_{n=0}^\infty\,s_n\ell^n
}{eq:i33}
with some coefficients $s_n$ that depend on the boost invariant state. According to \eqref{eq:i22} QNEC$_2$ contains a piece that diverges at small $\ell$.
\eq{
\frac{\extd^2 S_0}{\extd\lambda^2}\bigg|_{\lambda=0} + \frac{6}{c}\,\bigg(\frac{\extd S_0}{\extd\lambda}\bigg)^2\bigg|_{\lambda=0} = \frac{3s_1}{\ell} + {\cal O}(1)
}{eq:i32}
So despite of being close to AdS$_3$ in general there is a large correction to QNEC$_2$ at small $\ell$. The QNEC$_2$ inequality \eqref{eq:QNEC2} with $T_{kk}=0$ requires non-positivity of the coefficient $s_1$.
\eq{
s_1 \leq 0
}{eq:i31}

%%%%%%%%%%%%%%%%%%%%%%%%%%%%%%%%%%%%%%%%%%
\subsection{\texorpdfstring{QNEC$_2$}{QNEC2} with paper and pencil}\label{sec:1.4}
%%%%%%%%%%%%%%%%%%%%%%%%%%%%%%%%%%%%%%%%%%

Having discussed the definition and main properties of QNEC$_2$ we elaborate on how to calculate the right hand side of the QNEC$_2$ inequality \eqref{eq:QNEC2} holographically. For concreteness and because this is the only case considered in our work we focus on states dual to geometries with two commuting Killing vectors (either stationary axi-symmetric or stationary homogeneous geometries). The presence of two commuting Killing vectors means that there is always an adapted set of coordinate systems where these Killing vectors read $\partial_t$ and $\partial_x$. We use such a coordinate system from now on. The metric $\extd s^2 = g_{\mu\nu}(z)\,\extd x^\mu \extd x^\nu$ then depends only on the holographic (`radial') coordinate $z$.

Thanks to RT we just need to calculate the lengths of a one-parameter family of geodesics, where the family parameter $\lambda$ corresponds to the null deformation of the entangling region required to generate the QNEC$_2$ expression. At each step we can work perturbatively in $\lambda$ to second order, since in the end we take at most two derivatives with respect to $\lambda$ and set it to zero afterwards. Using the spatial coordinate $x$ as affine parameter the geodesic Lagrangian (dot means derivative with respect to $x$)
\eq{
{\cal L}(\dot t,\,\dot z,\,z) = \sqrt{g_{tt}(z)\dot t^2+g_{zz}(z)\dot z^2+g_{xx}(z)+g_{tz}(z)\dot t\dot z+g_{tx}(z)\dot t+g_{zx}(z)\dot z}
}{eq:i8}
yields the area
\eq{
{\cal A}(\lambda,\,\ell,\,z_{\textrm{\tiny{cut}}}) = 2 \int\limits_0^{(\ell+\lambda)/2-\omega}\extd x\,{\cal L}(\dot t,\,\dot z,\,z)
}{eq:i9}
where $\ell$ is the length of the entangling interval, $\lambda$ is the aforementioned deformation parameter, $\omega$ is a specific function of the cutoff $z_{\textrm{\tiny{cut}}}$ in the radial coordinate and the overall factor $2$ comes about because we integrate the geodesic from its turning point in the bulk to the anchor point at the cutoff surface and use the fact that the geodesic is mirror symmetric around $x=0$.

Since the Lagrangian ${\cal L}$ is $x$-independent we have the usual Noether-charge associated with $x$-translation invariance.
\eq{
Q_1 = \dot z\,\frac{\partial\cal L}{\partial\dot z} +\dot t\,\frac{\partial\cal L}{\partial\dot t} - \cal L 
}{eq:i10} 
It is convenient to evaluate the Noether charge $Q_1$ at the turning point $z=z_\ast$. The Lagrangian ${\cal L}$ is also $t$-independent, which yields a second Noether charge.
\eq{
Q_2 = \frac{\partial\cal L}{\partial\dot t}
}{eq:i11} 
The two Noether charges allow to express the velocity of the time coordinate and the velocity of the radial coordinate as functions of radial coordinate and turning point. However, it is more convenient to relabel their dependence on the Noether charges as dependence on the turning point $z_\ast$ and a specific combination $\Lambda$ of the two Noether charges that vanishes when the deformation parameter goes to zero, $\lambda\to 0$.
\eq{
\dot t = \Lambda\,h(z,\,z_\ast,\,\Lambda) \qquad\qquad \dot z=f(z,\,z_\ast,\,\Lambda)
}{eq:i18}
For the special case of diagonal metrics $\Lambda$ is given by the ratio of the Noether charges, $\Lambda=Q_2/Q_1$.

The temporal part of the deformed entangling interval is obtained by integrating $\extd t$.
\eq{
\frac{\lambda}{2} = \int\limits_0^{\lambda/2}\extd t = \int\limits_{z_\ast}^0\extd z\,\frac{\dot t}{\dot z} = \Lambda\,\int\limits_{z_\ast}^0\extd z\,\frac{h(z,\,z_\ast,\,\Lambda)}{f(z,\,z_\ast,\,\Lambda)} 
}{eq:i13}
The spatial part of the deformed entangling interval is obtained by integrating $\extd x$.
\eq{
\frac{\ell+\lambda}{2} = \int\limits_0^{(L+\lambda)/2}\extd x = \int\limits_{z_\ast}^0\frac{\extd z}{\dot z} = \int\limits_{z_\ast}^0\frac{\extd z}{f(z,\,z_\ast,\,\Lambda)} 
}{eq:i12}
Finally, the area integral \eqref{eq:i9} can be recast as
\eq{
{\cal A} = 2 \int\limits_{z_\ast}^{z_{\textrm{\tiny{cut}}}}\extd z\,\frac{{\cal L}( \Lambda\,h(z,\,z_\ast,\,\Lambda),\,f(z,\,z_\ast,\,\Lambda),\,z)}{f(z,\,z_\ast,\,\Lambda)}
}{eq:i14}
where $z_{\textrm{\tiny{cut}}}$ denotes the cutoff on the radial coordinate; for concreteness we assume that the limit $z_{\textrm{\tiny{cut}}}\to 0^+$ corresponds to removing the cutoff. Evaluating the temporal interval integral \eqref{eq:i13} yields $\Lambda$ as function of $\lambda$ and $z_\ast$. Since we can drop terms of order ${\cal O}(\lambda^3)$ and to leading order $\Lambda$ is already linear in $\lambda$, we can expand the functions $h$ and $f$ before integrating, which usually simplifies these integrals considerably. The evaluation of the spatial interval integral \eqref{eq:i12} allows to express the turning point $z_\ast$ in terms of the interval length $\ell$ and the deformation parameter $\lambda$. Again one can expand in $\Lambda$ and keep only the first few terms, dropping everything of order ${\cal O}(\lambda^3)$. These results allow to express the area integral \eqref{eq:i14} entirely in terms of $\ell$ and $\lambda$, so performing this integral then yields deformed EE as function of the interval length $\ell$ and the deformation parameter $\lambda$.

For most practical purposes these three integrals cannot be performed by hand since the functions $h$ and $f$ can be quite complicated, even when the metric functions are known in closed form. However, in the limit of small $\ell$ or large $\ell$ drastic simplifications occur that can allow to perform the first two integrals. The third integral diverges when the cutoff is removed, so it is practical to use instead a renormalized area
\begin{multline}
{\cal A}_{\textrm{\tiny{ren}}}(\lambda,\,\ell) =  2 \int\limits_{z_\ast}^0\extd z\,\bigg(\frac{{\cal L}( \Lambda\,h(z,\,z_\ast,\,\Lambda),\,f(z,\,z_\ast,\,\Lambda),\,z)}{f(z,\,z_\ast,\,\Lambda)} - \frac{\extd A_{\textrm{\tiny{ct}}}(z)}{\extd z}\bigg) - 2 A_{\textrm{\tiny{ct}}}(z_\ast) \\
= {\cal A}_0(\ell) + \lambda\, {\cal A}_1(\ell) + \frac{\lambda^2}{2}\,{\cal A}_2(\ell) + {\cal O}(\lambda^3)
\label{eq:i15}
\end{multline}
where $A_{\textrm{\tiny{ct}}}(z)$ is a counter-term added in such a way that the additional term in the renormalized area is independent from the interval length $\ell$ and the deformation parameter $\lambda$, i.e., it only depends on the cutoff $z_{\textrm{\tiny{cut}}}$. The integral in \eqref{eq:i15} is now finite and has $0$ as one of its boundaries, which considerably simplifies its evaluation. Inserting the solutions for $\Lambda$ and $z_\ast$ in terms of $\ell$ and $\lambda$ yields the functions ${\cal A}_i(\ell)$ in the second line in \eqref{eq:i15}.

The RT-formula 
\eq{
S(\lambda,\,\ell)=\frac{{\cal A}(\lambda,\,\ell)}{4\GN}=\frac{{\cal A}_{\textrm{\tiny{ren}}}(\lambda,\,\ell)}{4\GN} + \lambda\textrm{-independent terms} 
}{eq:i16}
establishes the final expression appearing on the right hand side of QNEC$_2$ 
\eq{
\frac{\extd^2 S}{\extd\lambda^2}\bigg|_{\lambda=0} + \frac{6}{c}\, \bigg(\frac{\extd S}{\extd\lambda}\bigg)^2\bigg|_{\lambda=0} = \frac{c}{6}\,\big({\cal A}_2(\ell) + {\cal A}_1(\ell)^2\big)
}{eq:i17}
where we replaced Newton's constant by the central charge, $\tfrac{1}{4\GN}=\tfrac{c}{6}$. In later sections we shall provide some examples where ${\cal A}_2(\ell)$ and ${\cal A}_1(\ell)$ are calculated in the limits of small and/or large entangling interval $\ell$.

For boost invariant states, like domain wall solutions, we can instead determine EE as function of the interval length $\ell$ and apply \eqref{eq:i22}. In the algorithm above this means that $\lambda$ and $\Lambda$ can be set to zero and the time integral \eqref{eq:i13} need not be calculated, which makes the calculation a bit shorter.

\subsection{\texorpdfstring{QNEC$_2$}{QNEC2} constraint on kinked entanglement}\label{sec:1.5}
%%%%%%%%%%%%%%%%%%%%%%%%%%%%%%%%%%%%%%%%%%

Sometimes holographic EE leads to two or more branches of geodesics \cite{Myers:2012ed,Liu:2012eea}, so there can be a critical interval value where one jumps from one of these branches to another. If this happens then EE as a function of the interval $\ell$ has a kink, and there is a first order phase transition (the converse is not necessarily true: there can be first order phase transitions without kinks in EE). We show now that QNEC$_2$ imposes a constraint on the behavior of EE near such a kink.

Let us assume EE as function of the entangling region has the following form
\eq{
S(\lambda,\,\ell+\lambda) = f_L(\lambda,\,\ell+\lambda)\,\theta\big(f_L(\lambda,\,\ell+\lambda)\big) + f_R(\lambda,\,\ell+\lambda)\,\theta\big(f_R(\lambda,\,\ell+\lambda)\big) 
}{eq:kink1}
where $f_L$ and $f_R$ are sufficiently smooth functions of the spatial length of the entangling interval $\ell$ and the null deformation parameter $\lambda$ used in QNEC$_2$. The subscripts $L,R$ refer to `left' and `right' of the kink at $\ell+\lambda=\ell_0$. Taylor expanding these functions without loss of generality yields ($\alpha_L\neq\alpha_R$)
\begin{subequations}
\label{eq:kink2}
\begin{align}
f_L &= \alpha_L(\lambda) (\ell_0 - \ell - \lambda) + {\cal O}(\lambda^2) + {\cal O}(\ell+\lambda-L_0)^2\\
f_R &= \alpha_R(\lambda) (\ell + \lambda - \ell_0) + {\cal O}(\lambda^2) + {\cal O}(\ell+\lambda-L_0)^2\,.
\end{align}
\end{subequations}
The expression $\ell+\lambda$ is the proper length of the entangling interval, up to irrelevant higher order terms, since $\sqrt{(\ell+\lambda)^2-\lambda^2}= \ell + \lambda + {\cal O}(\lambda^2)$. 

With the assumptions above EE \eqref{eq:kink1} is smooth except at $\ell+\lambda=\ell_0$. In the following we investigate the QNEC$_2$ combination of first and second derivatives of EE with respect to the null deformation parameter $\lambda$.

Consider the first derivative of EE. If the derivative acts on the step function we obtain an expression of the form $x\,\delta(x)$ which vanishes. 
\eq{
\frac{\partial S}{\partial\lambda} = \frac{\partial f_L}{\partial\lambda}\,\theta(f_L) + \frac{\partial f_R}{\partial\lambda}\,\theta(f_R)
}{eq:kink3}
This means that the first derivative of EE is piecewise continuous.

Consider now the second derivative of EE evaluated at vanishing $\lambda$. 
\eq{
\frac{\partial^2 S}{\partial\lambda^2}\bigg|_{\lambda=0} = \bigg(\frac{\partial f_L}{\partial\lambda}\bigg)^2\,\delta(f_L) + \bigg(\frac{\partial f_R}{\partial\lambda}\bigg)^2\,\delta(f_R) + \textrm{piecewise\;continuous}
}{eq:kink4}

Since we are interested in the behavior near the kink we use now the Taylor expansions \eqref{eq:kink2} as well as the identity $\delta(\alpha(x-x_0))=\delta(x-x_0)/|\alpha|$ and obtain the QNEC$_2$ combination
\eq{
\frac{\partial^2 S}{\partial\lambda^2}\bigg|_{\lambda=0} + \frac{6}{c}\,\bigg(\frac{\partial S}{\partial\lambda}\bigg)^2\bigg|_{\lambda=0} = (\alpha_R^0 - \alpha_L^0)\,\delta(\ell-\ell_0) + \textrm{piecewise\;continuous}
}{eq:kink5}
where $\alpha_{L,R}^0=|\alpha_{L,R}(\lambda=0)|$. At the kink $\ell=\ell_0$ the QNEC$_2$ combination has an infinite peak due to the $\delta$-function. The sign of this peak must be negative since otherwise the QNEC$_2$ inequality \eqref{eq:QNEC2} would be violated. Therefore, consistency with QNEC$_2$ imposes the convexity condition
\eq{
\alpha_L^0 > \alpha_R^0\,.
}{eq:kink6}
For ground states the inequality \eqref{eq:kink6} is a consequence of RG-flow monotonicity. In the next subsection we make the relation between QNEC$_2$ and
RG-flows more precise.

%%%%%%%%%%%%%%%%%%%%%%%%%%%%%%%%%%%%%%%%%%
\subsection{Relation between \texorpdfstring{QNEC$_2$}{QNEC2} and Casini--Huerta \texorpdfstring{$c$}{c}-function}\label{sec:1.8}
%%%%%%%%%%%%%%%%%%%%%%%%%%%%%%%%%%%%%%%%%%

If one views EE as function of the interval length $\ell$ mechanically as a trajectory then QNEC$_2$ is the natural acceleration associated with this trajectory. It is then suggestive to ponder whether the velocity associated with this trajectory also has a physical meaning. At least for boost invariant states, which we shall refer to as ground states, the answer is affirmative --- the velocity is the Casini--Huerta $c$-function \cite{Casini:2004bw,Casini:2006es,Myers:2012ed}. We show now how this relationship works, denoting EE in a general frame by $S$ and in the rest frame by $S_0$.

The Casini--Huerta $c$-function for a deformed CFT$_2$ 
\eq{
c(\ell) = 3\ell\,\frac{\extd S_0(\ell)}{\extd\ell}
}{eq:i24}
is proportional to the first derivative with respect to the null deformation parameter (to reduce sign clutter we assume here an outgoing null deformation)
\eq{
\frac{\extd S(\lambda,\,\ell+\lambda)}{\extd\lambda}\bigg|_{\lambda=0} = \frac{\extd S_0(\sqrt{(\ell+\lambda)^2-\lambda^2})}{\extd\lambda}\bigg|_{\lambda=0} =  \frac{\extd S_0(\ell)}{\extd\ell} = \frac{1}{3\ell}\,c(\ell)\,.
}{eq:i25}
So up to an overall factor of $1/(3\ell)$ the Casini--Huerta $c$-function \eqref{eq:i24} is indeed the `velocity' of EE in essentially the same sense that QNEC$_2$ is its `acceleration'. It obeys the monotonicity relation
\eq{
\frac{\extd c(\ell)}{\extd\ell} \leq 0\,.
}{eq:i42}
Integrating this monotonicity inequality for EE using the definition \eqref{eq:i24} yields a bound on ground state EE
\eq{
S_0(\ell) \leq \frac{\hat c}{3}\,\ln\frac{\ell}{\cut}\qquad\qquad \hat c,\,\cut\in\mathbb{R}^+
}{eq:i41}
which means in particular that at large $\ell$ ground state entanglement cannot grow faster than the area law \eqref{eq:i4}.

Consistently, the first derivative of the Casini--Huerta $c$-function (up to a factor $3\ell$)
\eq{
\frac{1}{3\ell}\,\frac{\extd c(\ell)}{\extd\ell} %= \frac{\extd^2 S_0(\ell)}{\extd\ell^2} + \frac{1}{\ell}\,\frac{\extd S_0(\ell)}{\extd\ell} 
=  \frac{\extd^2 S_0(\ell)}{\extd\ell^2} - \frac{1}{\ell}\,\frac{\extd S_0(\ell)}{\extd\ell} + \frac{6}{c(\ell)}\,\bigg(\frac{\extd S_0(\ell)}{\extd\ell}\bigg)^2
}{eq:i26}
yields essentially the QNEC$_2$ combination, as evident from our result \eqref{eq:i22} for boost invariant states. 

Thus, for boost invariant states the right hand side of the QNEC$_2$ inequality can be re-interpreted as the rate by which the Casini--Huerta $c$-function changes in the UV. The monotonicity property of this $c$-function \eqref{eq:i42} is then equivalent to the QNEC$_2$ inequality \eqref{eq:QNEC2} with $\langle T_{kk}\rangle=0$ for Poincar\'e invariant states, except that $c$ is replaced by $c(\ell)$. (By the constant $c$ we always mean the UV-value $\lim_{\ell\to 0}c(\ell)$ of the Casini--Huerta $c$-function, which in our context is the Brown--Henneaux central charge.) Monotonicity of the Casini--Huerta $c$-function \eqref{eq:i42} is stronger than QNEC$_2$ since $c\geq c(\ell)$.

For general states that are not boost invariant the function \eqref{eq:i24} need not be monotonically decreasing. For instance, thermal states dual to BTZ black branes have EE \eqref{eq:i5} leading to a monotonically increasing function \eqref{eq:i24}, $c(\ell)=\frac{c\pi\ell}{\beta}\,\coth\frac{\pi\ell}{\beta}$. Nevertheless, the QNEC$_2$ inequality \eqref{eq:QNEC2} holds in full generality even for states that are not boost invariant.

%%%%%%%%%%%%%%%%%%%%%%%%
\section{Holographic model}\label{sec:model}
%%%%%%%%%%%%%%%%%%%%%%%%

In our work we consider deformed holographic CFTs that are dual to three-dimensional Einstein gravity with a minimally coupled massive self-interacting scalar field. In this section we summarize the gravity side of the holographic model that we study in the rest of this work. 

In section \ref{sec:3.1} we present the Einstein--Klein--Gordon action, including its boundary terms and the explicit form of the scalar potential in terms of a superpotential. In section \ref{sec:3.2} we derive solutions corresponding to the ground state of the dual field theory. In section \ref{sec:3.3} we discuss solutions corresponding to thermal states in the dual field theory.

%%%%%%%%%%%%%%%%%%%%%%%%%%%%%%%%%%%%%%%%%%
\subsection{Action}\label{sec:3.1}
%%%%%%%%%%%%%%%%%%%%%%%%%%%%%%%%%%%%%%%%%%

The action of the gravity system 
\begin{equation}\label{eq:action}
\Gamma=\frac{1}{16\pi \GN}\int_{\mathcal{M}}\!\!\extd^{3}x\sqrt{-g}\, \Big(R-\frac{1}{2}(\partial \phi)^2-V(\phi)\Big) + \frac{1}{8\pi \GN}\int_{\partial\mathcal{M}}\!\!\!\!\extd^2x\sqrt{-\gamma}\,K+I_{\textrm{ct}}
\end{equation}
entails the Ricci scalar $R$ of the bulk geometry on a manifold $\cal M$ with boundary $\partial\cal M$ and bulk metric $g_{\mu\nu}$, the trace of the extrinsic curvature $K$ of the boundary geometry with induced metric $\gamma_{ij}$ and the counter-term  $I_{\textrm{ct}}$ that renders the variational principle well-defined (and also the on-shell action finite).

For asymptotically AdS$_3$ spacetimes the potential needs to have the small $\phi$ expansion
\begin{equation}\label{eq:potenital}
V(\phi)=-2+\frac{1}{2}m^2\phi^2+v_4\phi^4+\ldots
\end{equation}
where we assumed the $\mathbb{Z}_2$ symmetry $V(\phi)=V(-\phi)$  for simplicity. By the usual AdS/CFT dictionary the conformal weight $\Delta$ of the dual operator is related to the mass of the scalar field as $m^2 =\Delta(\Delta-2)$. We restrict to potentials that globally can be written in terms of a superpotential $W$.
\be
V(\phi)=-\frac{1}{2}W(\phi)^2+\frac{1}{2}W'(\phi)^2\label{V2W}
\ee 
Technical advantages of this choice are that one can find the ground state (domain wall) solution more easily, the counter-terms $I_{\textrm{ct}}$ are known explicitly in terms of the superpotential (see appendix \ref{appendix-HRG}) and neither in Fefferman--Graham- nor in Gubser-gauge [defined in \eqref{eq:GubserGauge}] the near boundary solution has logarithmic terms. The latter is related to the absence of the trace anomaly (again, see appendix \ref{appendix-HRG}). This makes the numerical analysis of geodesics and QNEC$_2$ on such backgrounds more stable.

It turns out that simple superpotentials characterised by a single parameter
\be\label{eq:superpotential}
W(\phi)=-2-\frac{1}{4}\,\phi^2-\frac{\alpha}{8}\,\phi^4
\ee
reveal already rich physical features. 
The associated potential \eqref{V2W} reads
\eq{
V(\phi)=-2-\frac{3}{8}\,\phi^2-\frac{1}{32}\,\phi^4-\frac{\alpha \left(1-4 \alpha\right)}{32}\,  \phi^6-\frac{\alpha^2}{128}\, \phi^8\,.
}{eq:angelinajolie} 

\begin{figure}[hbt]
    \centering
    \includegraphics[width=0.5\textwidth]{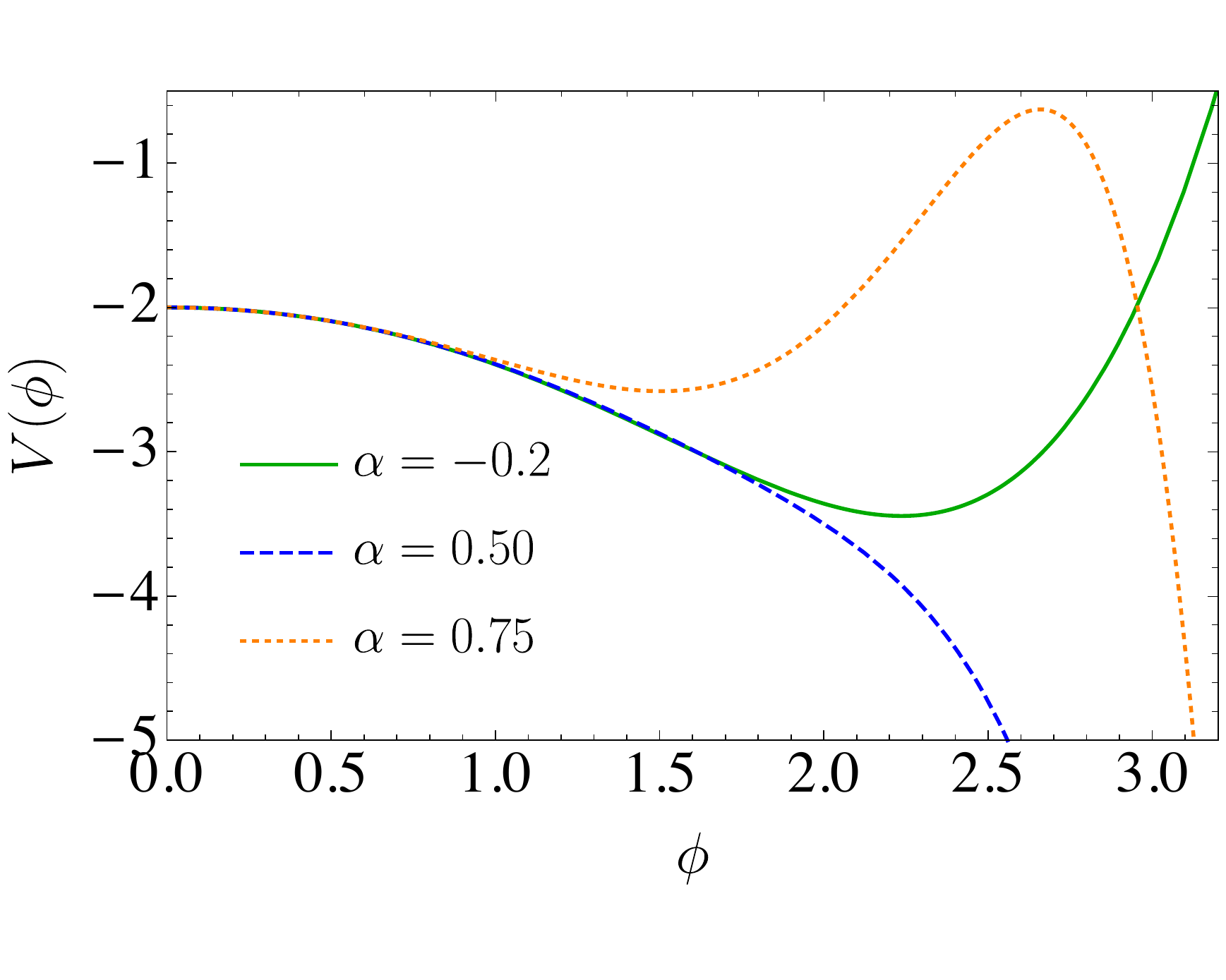}
    \caption{Potential \eqref{eq:angelinajolie} for three different holographic RG flows.}
    \label{fig:potentials}
\end{figure}

See Figure~\ref{fig:potentials} for three examples. The conformal weight of the dual operator is then given by $\Delta=\tfrac12$ or $\Delta=\tfrac32$, since $m^2=-\tfrac34$, and the value of the quartic self-interaction constant is fixed to $v_4=-\frac{1}{32}$. In the present work we always make the choice $\Delta=\tfrac32$ and consider exclusively the potential \eqref{eq:angelinajolie} for our examples. We comment on generalizations to arbitrary $m^2$ (respecting the Breitenlohner--Freedman bound \cite{Breitenlohner:1982bm,Breitenlohner:1982jf}) and potentials of other form than \eqref{eq:superpotential} in sections \ref{sec:predict} and \ref{sec:Summary}.

Extrema of the potential are obtained when either the superpotential has an extremum, $W'(\phi)=0$, leading to the real solution $\phi=\phi_0=0$ (for negative $\al$ there is an additional extremum with real scalar field $\phi=\phi_n=1/\sqrt{-\alpha}$), or when the superpotential obeys $W''(\phi_\pm)=W(\phi_\pm)$, leading to the (potentially) real solutions
\eq{
\phi_\pm^2 = 6 - \frac{1}{\al} \pm \frac{\sqrt{1 - 24\al + 36\al^2}}{\al} \,.
}{eq:extrema}
The extremum corresponding to $\phi_0$ is always a maximum. The extrema corresponding to $\phi_+$ are maxima and exist only for $\al>\tfrac16\,(2+\sqrt{3})$. The extrema corresponding to $\phi_-$ are maxima for $\al < 0$ and minima for $\tfrac16\,(2+\sqrt{3})<\al<1$. The extrema corresponding to $\phi_n$ are always minima. In the range $0 < \al < \tfrac16\,(2+\sqrt{3})$ the only extremum is at $\phi_0$. We are going to exploit this information when investigating holographic RG flows between UV and IR conformal fixed points.

%%%%%%%%%%%%%%%%%%%%%%%%%%%%%%%%%%%%%%%%%%
\subsection{Ground states}\label{sec:3.2}
%%%%%%%%%%%%%%%%%%%%%%%%%%%%%%%%%%%%%%%%%%

Poincar\'e$_2$-invariant domain wall solutions to the Einstein--Klein--Gordon equations 
\begin{subequations}
\label{eq:EOM}
\begin{align}
R_{\mu\nu}-\frac{1}{2}g_{\mu\nu}R&=\frac{1}{2}\partial_\mu\phi\,\partial_\nu\phi-\frac{1}{4}(\partial \phi)^2g_{\mu\nu}-\frac{1}{2}Vg_{\mu\nu}\\
\nabla^2\phi&=\frac{\partial V}{\partial \phi}
\end{align}
\end{subequations}
corresponding to the ground state of the dual deformed CFT can be found easily. The domain wall parametrization of the metric 
\be\label{eq:metricDomainWall}
\extd s^2=\extd\rho^2+e^{2A(\rho)}\left(-\extd t^2+ \extd x^2\right)
\ee
reduces the equations of motion \eqref{eq:EOM} in terms of the superpotential to first-order equations.
\bea
\frac{\extd A(\rho)}{\extd\rho}=-\frac{1}{2}W(\phi(\rho))\qquad \qquad \frac{\extd\phi(\rho)}{\extd\rho}=\frac{\extd W(\phi(\rho))}{\extd\phi(\rho)}\label{1stO-eqs}
\eea
The asymptotic region in this parametrization is at $\rho\rightarrow\infty$. 
Integrating the equations of motion \eqref{1stO-eqs} with the superpotential \eqref{eq:superpotential} yields 
\begin{subequations}
\label{eq:DomainWallSol}
\begin{align}
\phi(\rho)&=\frac{ {\source} e^{ - \rho/2}}{\sqrt{1- \alpha\source^2  e^{-\rho}}} %\qquad %\text{For real solutions:} \qquad  \alpha\source^2<1
\\
A(\rho)&=\Big(1-\frac{1}{16 \alpha}\Big) \rho-\frac{\source ^2}{16 \left(e^\rho-\alpha \source ^2\right)}+\frac{\log \left(e^\rho-\alpha \source ^2\right)}{16 \alpha}\, \label{eq:A}
\end{align}
\end{subequations}
where the integration constant $\source$ can be identified with the source of the dual operator. For real field configurations the inequality $\alpha\source^2<1$ holds. 

Using the near boundary expansion of this solution one can show that the expectation value of the boundary stress tensor and the operator dual to the scalar field \eqref{vevPhi-1}-\eqref{EMT-xx} vanish.
\bea
\langle T_{ij} \rangle=0=\langle {O_{\phi}} \rangle 
\eea
This shows that this solution has vanishing free energy, $F=-\langle T_{xx} \rangle$, and thus corresponds to the ground state of the dual field theory. We shall see that even this elementary background exhibits remarkable features in EE and QNEC$_2$ studies.

%%%%%%%%%%%%%%%%%%%%%%%%%%%%%%%%%%%%%%%%%%
\subsection{Thermal states}\label{sec:3.3}
%%%%%%%%%%%%%%%%%%%%%%%%%%%%%%%%%%%%%%%%%%

We are also interested in more general (thermal) states and their entanglement and QNEC$_2$ properties. To describe their gravity duals we make the ansatz
\bea\label{eq:metric}
\extd s^2=e^{2A}\left(-H \extd t^2+\extd x^2\right)+e^{2B}\,\frac{\extd r^2}{H} 
\eea
where $A,B$ and $H$ are functions of the radial coordinate $r$ only. The ansatz \eqref{eq:metric} encodes solutions that are invariant under spacetime translations. If $H$ has a simple zero at some $r=r_h$, then the geometry is a black brane with regular event- and Killing horizon at $r=r_h$, which gives rise to finite temperature and entropy density of the dual field theory state. 

The ansatz \eqref{eq:metric} has a residual gauge freedom, namely reparametrizations of the radial coordinate. We fix this freedom by using Gubser gauge \cite{Gubser:2008ny}, where the radial coordinate is identified with the corresponding value of the scalar field.
\begin{equation}\label{eq:GubserGauge}
r := \phi(r)
\end{equation}
The equations of motion \eqref{eq:EOM},
\begin{align}
 H\left(B'-2 A'\right)-H'+e^{2B}V' &= 0\label{eq:EOM1}\\
 2\left(A'B'-A''\right)-1 &= 0\label{eq:EOM2}\\
 H''+(2A'-B')H' &=0 \label{eq:EOM3}\\
 2 A' H'+H \left(4 A'^2-1\right)+2 e^{2 B} V &=0 \label{eq:EOM4}
\end{align}
can be rephrased as a single master equation
\begin{equation}\label{eq:master}
2G \left(G V'+V\right)G''=\left(6 G V'+2V\right)G'^2+\left(4 G^3 V'+2 G^2 V''+4 G^2 V+3 G V'+V\right)G'
\end{equation}
for the master field $G(\phi):= A'(\phi)$, where prime denotes derivatives with respect to $\phi$.

For a given potential $V$ a solution of \eqref{eq:master} allows to express $A,B$ and $H$ in terms of integrals of simple functions of $G$ and $G'$ only. The solution for $A$ can be obtained by integrating the definition of $G$
\begin{equation}
A(\phi)=A(\phi_h)+\int\limits_{\phi_h}^\phi \mathrm{d}\phi' G(\phi')\,.
\end{equation}
The solution for $B$ follows from integrating (\ref{eq:EOM2})
\begin{equation}
B(\phi)=B(\phi_h)+\log\frac{G(\phi)}{G(\phi_h)}+\int\limits_{\phi_h}^\phi \frac{\mathrm{d}\phi'}{2G(\phi')}\,.
\end{equation}
Knowing $B$ allows to express $H$ algebraically by combining \eqref{eq:EOM1}, \eqref{eq:EOM2} and \eqref{eq:EOM4}
\begin{equation}
H(\phi)=-\frac{e^{2B}(V+GV')}{G'}\,.
\end{equation}

For certain simple choices of $V$ it is possible to solve the second order ordinary differential equation \eqref{eq:master} in closed form \cite{Gubser:2008ny}, but in general the master equation \eqref{eq:master} needs to be solved numerically. In that case it is useful to extract the  divergent asymptotic behavior of the master field $G$ inherited from the asymptotic behavior of $A$
\begin{equation}
A(\phi)=\frac{\log(\phi)}{\Delta-2}+\ldots\quad\implies\quad G(\phi)=\frac{1}{(\Delta -2)\phi}+\tilde{G}(\phi)
\end{equation}
where $\tilde{G}(\phi)$ remains finite at the boundary $\phi\to 0$. As discussed in  \cite{Gubser:2008ny} such a near boundary behavior of the fields corresponds to a relevant deformation of the CFT$_2$, namely
\be
\mathcal{L}=\mathcal{L}_{\textrm{CFT}_2}+\source^{2-\Delta}\mathcal{O}_\phi\ .
\ee
To find the equation of state, we set the source $\source$ of the dual operator to one in units of AdS radius, $\source=1$. This leaves the horizon value of the scalar field $\phi_h$ (which is equal to the horizon radius in Gubser gauge) as the only free parameter. 

%%%%%%%%%%%%%%%%%%%%%%%%
\section{Thermodynamics}\label{sec:thermo}
%%%%%%%%%%%%%%%%%%%%%%%%
In this section we analyze the thermodynamics of the system for different choices of the scalar field (super-)~potential resulting in different types of phase transitions. We do this for a deformation by an operator with fixed conformal weight $\Delta=\tfrac{3}{2}$, and modify the quartic term of the superpotential \eqref{eq:superpotential} by choosing different values for $\alpha$.
The entropy density and the temperature of the system can be expressed in terms of horizon data
\begin{equation} \label{eq:thermodynamics}
s=\frac{1}{4 \GN}\,e^{A(\phi_h)} \qquad \qquad T=\frac{1}{4\pi}\,e^{A(\phi_h)+B(\phi_h)}\left|V'(\phi_h)\right|\,.
\end{equation}

\begin{figure}[htb]
    \centering
    \includegraphics[width=0.32 \textwidth]{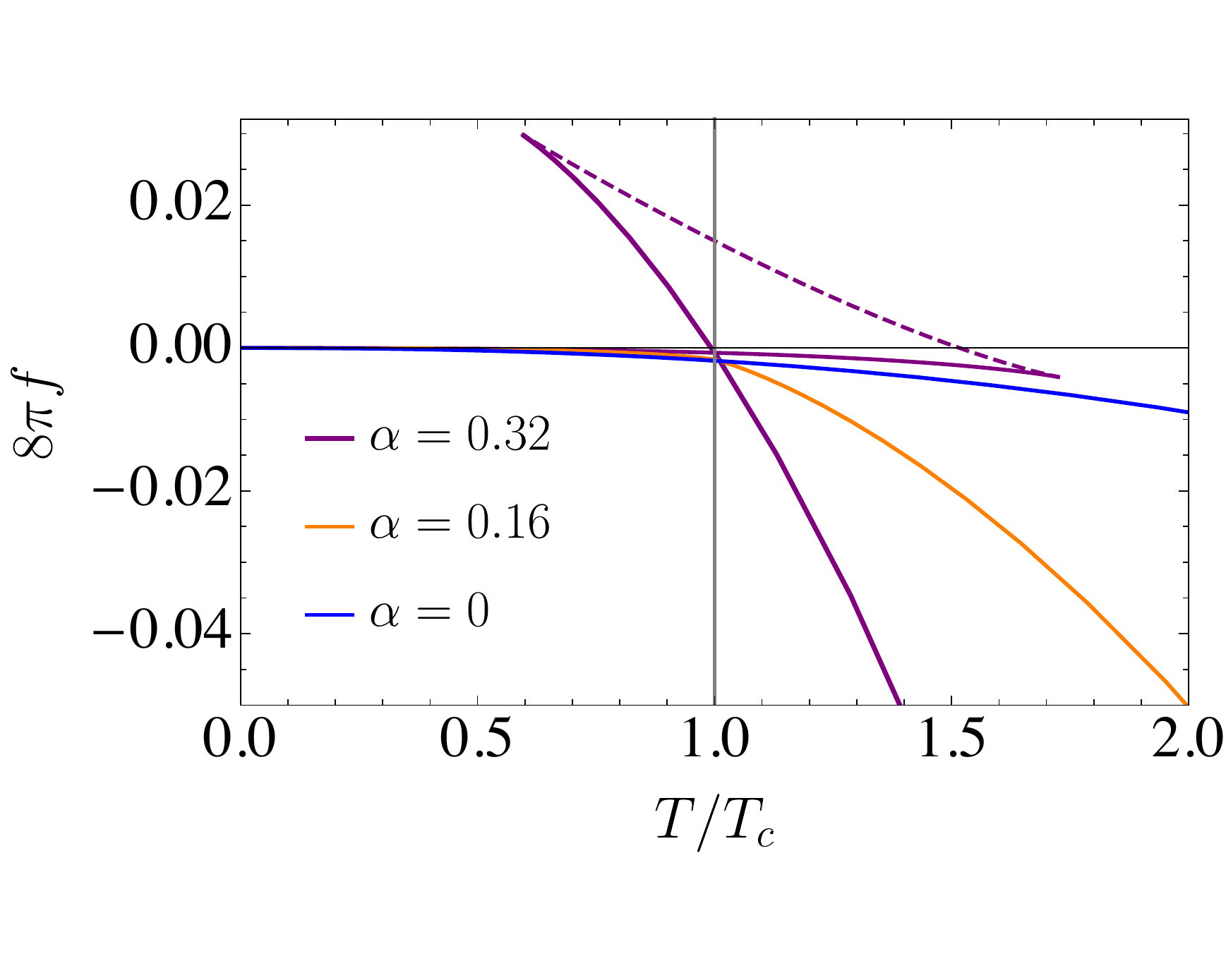}\quad\includegraphics[width=0.31\textwidth]{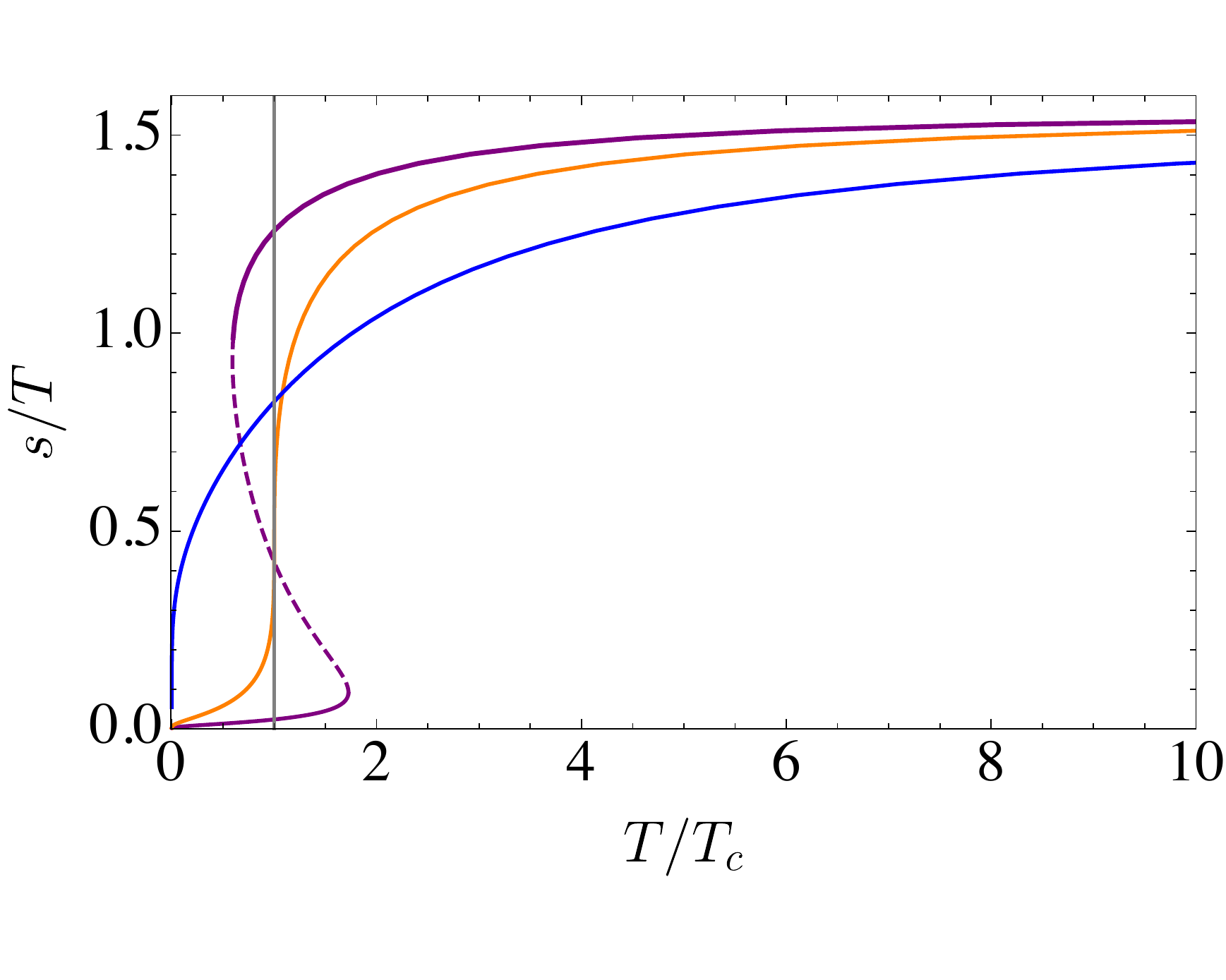}\quad\includegraphics[width=0.31 \textwidth]{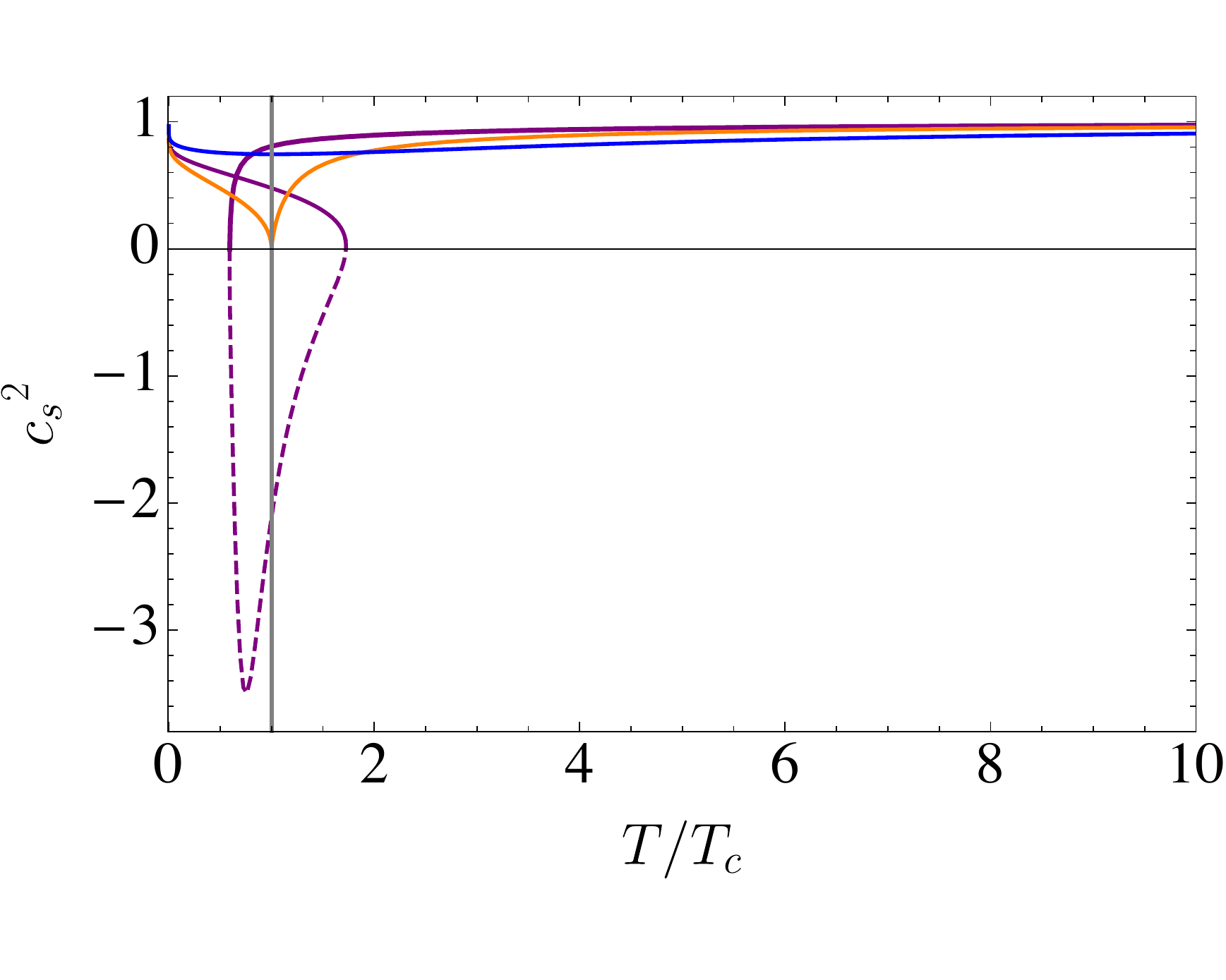}
    \caption{Free energy density (left), entropy density over temperature (center), speed of sound squared (right). Blue:  crossover ($\alpha=0$). Orange: second order phase transition ($\alpha=0.16$). Magenta: first order phase transition ($\alpha=0.32$). Vertical gray: locus of phase transition. Dashed: thermodynamically unstable solutions with imaginary speed of sound.}
    \label{fig:fsc}
\end{figure}

Figure~\ref{fig:fsc} displays free energy density $f$, entropy density over temperature $s/T$ and the speed of
sound squared $c_s^2$ for three different potentials (we set Newton's constant $\GN = 1$ from now on in all
numerical results). By `density' we mean that we divide by the trivial but infinite volume along the black brane. The free energy of the boundary theory is determined holo\-gra\-phi\-cally
from the on-shell action $\Gamma$ and given by the purely spatial component of the boundary
stress-tensor, which in turn is the pressure
\be
f= -\langle T_{xx}\rangle = -p
\ee
while the speed of sound squared of the boundary theory can be computed from horizon- or boundary-data as
\be
c_s^2=\frac{\extd\  \ln{T}}{\extd\ \ln{s}}=\frac{\extd p}{\extd e}
\ee
where 
\eq{
e=\langle T_{tt}\rangle
}{eq:e}
is the energy density of the boundary theory.

At large temperatures the entropy density becomes proportional to the temperature. This indicates that in this limit the corresponding states in all cases are close to the CFT$_2$. 
For $\alpha<0.1605$ the system undergoes a smooth crossover as the temperature increases.\footnote{Hereafter we refer to this specific value as $\alpha\approx0.16$ for brevity.} 
For $\alpha=0$ the speed of sound has a global minimum at $T\approx0.01$ and we call this the critical temperature of the crossover.
For this choice, the free energy, the entropy density and the speed of sound are smooth functions of temperature. 
For $\alpha\approx0.16$ we find a second order phase transition. The speed of sound vanishes at the critical temperature $T_c\approx0.024$ and the entropy density shows critical behavior close to $T_c$
\newcommand{\crit}{\gamma}
\begin{equation}\label{eq:fit}
s(T)=s_0+s_1\left(\frac{T-T_c}{T_c}\right)^{1-\crit}
\end{equation}
where we estimate $\crit\approx 0.66$ for the critical exponent (see Figure~\ref{fig:fit}) and $s_1\approx 0.02$.
\begin{figure}
    \centering
    \includegraphics[width=0.48\textwidth]{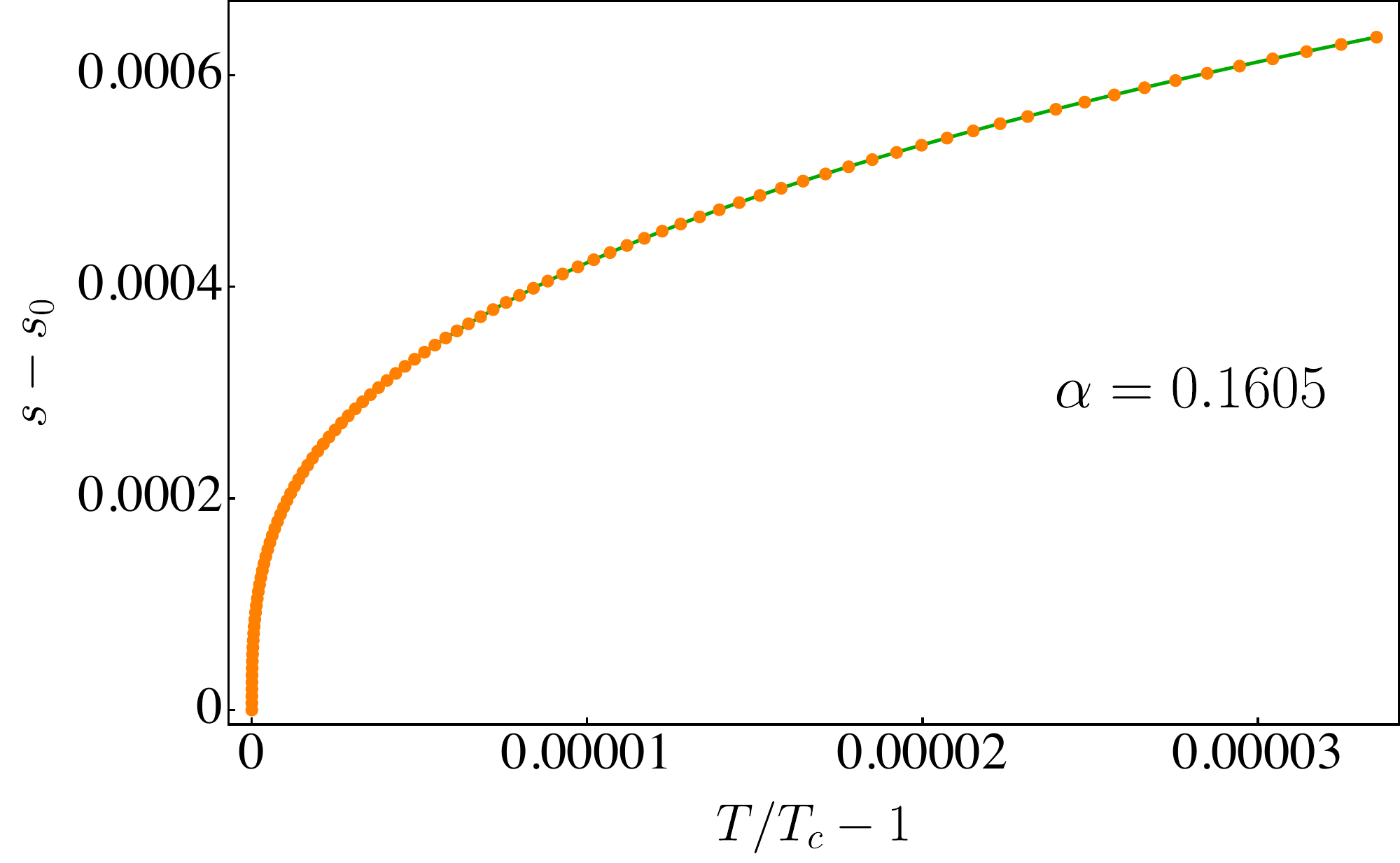}
    \caption{Entropy density close to second order phase transition. Orange dots: numerical results. Green line: fit \eqref{eq:fit} with $\crit=0.66$ and $s_1=0.02$.
    }
    \label{fig:fit}
\end{figure}

For $\alpha>0.16$ the system has a first order phase transition. On the gravity side it follows from the existence of three different black brane solutions around the critical point. They have the same Hawking temperature but different free energies. In this family we choose $\alpha=0.32$ as example, which leads to a first order phase transition at $T_c\approx0.04$ between a large and small black brane geometry. The imaginary speed of sound in the middle branch (dashed line in Figure~\ref{fig:fsc}) corresponds to unstable quasinormal modes in the sound channel which is known as spinodal instability \cite{Chomaz:2003dz}. The dynamics of the first order phase transition was studied in detail in 4D \cite{Janik:2017ykj} and 5D \cite{Attems:2017ezz} and a static black brane geometry with inhomogenous horizon and Hawking temperature $T=T_c$ \cite{Janik:2017ykj,Bea:2020ees} identified as the thermodynamically stable classical solution. This solution is a mixed state dual to an inhomogeneous arrangement of large and small black branes of equal temperature $T_c$.

We checked that different choices for $\Delta$ lead to qualitatively equivalent thermodynamic properties. In particular, the critical exponent $\gamma$ turns out to be remarkably universal. By computing $\gamma$ for various values of the conformal scaling dimension, for different forms of the potential and for different numbers of spacetime dimensions we always find values consistent with $\gamma=2/3$ discovered in \cite{Gubser:2008ny} for $d=4$. In Table \ref{tab:0} we summarize the different cases we have checked. In each row we show the boundary dimension $d$ in the first column and the (super)potential in the second column. For each row we sampled (equi-distantly) nine different values of conformal weights of the dual scalar operator in the interval given in the third column. The last column shows the mean and standard deviation of the critical exponent $\gamma$ defined in \eqref{eq:fit} averaged over the sample of nine different conformal weights. 

\begin{table}[htb]
    \centering
    \begin{tabular}{c|c|c|c}
        $d$ & (super)potential & $\Delta$ & $\gamma$   \\ 
        \hline
        2  & $W(\phi)=-2+\frac{\Delta-2}{2}\,\phi^2+{\widetilde{\alpha}}\,\phi^4$ & $[1,2]$ & %$[0.66x,0.66y]$ or
        0.6681 $\pm$ 0.0041 \\
        \hline
        2  & $V(\phi)=-2 \cosh(\frac{\phi}{\sqrt{3}})+\frac{3\Delta^2-6\Delta+2}{6}\,\phi^2+{\widetilde{\alpha}} \, \phi^4$ & $[1,2]$ & 0.6658 $\pm$ 0.0062\\
         \hline
        3  & $W(\phi)=-4+\frac{\Delta-3}{2}\,\phi^2+{\widetilde{\alpha}}\,\phi^4$ & $[\frac{3}{2},3]$ & 0.6670 $\pm$ 0.0045   \\
        \hline
        3  & $V(\phi)=-6 \cosh(\frac{\phi}{\sqrt{3}})+\frac{\Delta^2-3\Delta+2}{2} \, \phi^2+{\widetilde{\alpha}} \, \phi^4$ & $[\frac{3}{2},3]$ & 0.6635 $\pm$ 0.0055\\
        \hline
        4  & $W(\phi)=-6+\frac{\Delta-4}{2}\,\phi^2+{\widetilde{\alpha}}\,\phi^4$ & $[2,4]$ & 0.6667 $\pm$ 0.0078   \\
        \hline
        4  & $V(\phi)=-12 \cosh(\frac{\phi}{\sqrt{2}})+\frac{\Delta^2-4\Delta+6}{2}\phi^2+{\widetilde{\alpha}} \phi^4$ & $[2,4]$ & 0.6691 $\pm$ 0.0080\\
    \end{tabular}
    \caption{Means and standard deviations of critical exponents averaged over conformal weights}
    \label{tab:0}
\end{table}

From Table \ref{tab:0} we see that our hypothesis that the critical exponent $\gamma$ is independent from the dimension, the choice of potential and the conformal weight is supported by the data. Using the whole data set of 54 data points yields the critical exponent averaged over dimensions, potentials and conformal weights.
\eq{
\bar\gamma = 0.6667 \pm 0.0062
}{eq:gammamean}
The averaged critical exponent \eqref{eq:gammamean} is in excellent agreement with our hypothesis that the true critical exponent always is given by $\gamma=2/3$.

%%%%%%%%%%%%%%%%%%%%%%%%
\section{Entanglement entropy}\label{sec:entanglement}
%%%%%%%%%%%%%%%%%%%%%%%%

In this section we calculate EE for our theory holographically using the RT formula. In some cases this means we need to numerically determine geodesics. 

In section \ref{sec:5.1} we summarize briefly the main formulas for the special class of geometries discussed in sections \ref{sec:3.2} and \ref{sec:3.3}. In section \ref{sec:5.2} we focus on ground state entanglement, which already exhibits non-trivial features such as the appearance of different branches that we discuss in detail. In section \ref{sec:5.3} we calculate EE holographically for thermal states, finding a rich phase structure.

%%%%%%%%%%%%%%%%%%%%%%%%%%%%%%%%%%%%%%%%%%
\subsection{Entanglement entropy from geodesics}\label{sec:5.1}
%%%%%%%%%%%%%%%%%%%%%%%%%%%%%%%%%%%%%%%%%%

We adopt the algorithm described in section \ref{sec:1.4} for QNEC$_2$ to calculate EE holographically, basically by setting the deformation parameter $\lambda$ in that section to zero, specializing to static metrics
\be
\extd s^2=g_{tt}(z)\,\extd t^2+g_{zz}(z)\,\extd z^2+g_{xx}(z)\,\extd x^2\,.
\ee
EE for an interval with a separation $\ell$ along the spatial direction $x$ holographically is given by the geodesic length [times a factor $1/(4\GN)$] \cite{Ryu:2006bv}
\eq{
S_\textrm{EE}(\ell)=\frac{1}{2\GN} \int\limits_{z_\ast}^{z_\textrm{\tiny{cut}}} \extd z\,\frac{\sqrt{g_{zz}(z)}}{\sqrt{1-\frac{g_{xx}(z_\ast)}{g_{xx}(z)}}}
}{EE-general}
where we assumed that the asymptotic AdS boundary is at $z=0$, the cutoff surface that regulates UV divergences is at $z=z_\textrm{\tiny{cut}}$, and the geodesic hanging from the boundary into the bulk has a turning point at $z=z_\ast$. The quantity ${\cal A}_0(\ell)/(4\GN)$ in \eqref{eq:i15} is a renormalized version of EE \eqref{EE-general} where the cutoff term is subtracted. Using renormalized EE has the technical advantage that the integral extends from $0$ to $z_\ast$ and thus is a bit simpler to perform and also more stable to evaluate numerically, if required. Whenever we evaluate EE \eqref{EE-general} numerically we set Newton's constant to unity, $\GN=1$.

The Noether charge \eqref{eq:i10} establishes $\dot z =\sqrt{g_{xx}(z)/g_{zz}(z)}\,\sqrt{g_{xx}(z)/g_{xx}(z_\ast)-1}$, which allows to evaluate the spatial interval integral \eqref{eq:i12}. The ensuing relation between the turning point $z_\ast$ and the spatial separation $\ell$ 
\bea\label{l-as-rs}
\ell=2\int\limits_{0}^{z_\ast} \frac{\extd z}{g_{xx}(z)}\,\frac{\sqrt{g_{zz}(z) g_{xx}(z_\ast)}}{\sqrt{1-\frac{g_{xx}(z_\ast)}{g_{xx}(z)}}}  
\eea
may not be one to one and thus can lead to several branches, as we shall see explicitly in some of the examples below.

We are going to evaluate the integrals above numerically for arbitrary interval lengths $\ell$ and perturbatively in the limits of small and large $\ell$ (the former works since we need to know only the asymptotic expansion of the metric, while the latter works only without numerics if we have closed form expressions for the metric). For the numerical evaluation one can use simple methods, such as Mathematica's {\tt NIntegrate}.

The perturbative treatment at small $\ell$ schematically works as follows. The turning point has to be close to the AdS boundary, so we parametrize it using a parameter $\delta$ that tends to zero when the AdS boundary is approached (e.g.~in the coordinates used above we could use $z_\ast=\delta$) and make expansions in $\delta$ for all quantities. This simplifies the integrals so that often they can be performed in closed form. The integral \eqref{l-as-rs} then establishes a series in $\delta$ for the turning point $z_\ast$ as function of the interval length $\ell$. Plugging this series into the renormalized version \eqref{eq:i15} of the area integral \eqref{EE-general} by virtue of the RT formula then leads to a generalized power series for EE
\eq{
S(\ell\ll 1) = \frac c3\,\ln\ell + \sum_{n=0}^\infty s_n \ell^n + S_{\textrm{\tiny{cut}}}
}{eq:EEsmall}
where the universal leading order term comes from the asymptotic AdS behavior and $S_{\textrm{\tiny{cut}}}$ is an $\ell$-independent term depending on the UV cutoff that drops out in renormalized EE. The Taylor--Maclaurin coefficients $s_n$ are determined from relatively simple integrals, as described above. We are going to display the first few coefficients for the examples studied in subsection \ref{sec:5.2} below.

At large $\ell$ the situation is similar, but technical details regarding the evaluation of integrals are quite different, essentially because one can get arbitrarily close to branch cuts in integrands. There are basically three different scenarios at large $\ell$. 

In the first one the radial coordinate, and thus the turning point, is unbounded, by which we mean that there is neither a center nor a horizon that would provide an IR cutoff on the radial coordinate. If this happens the situation is similar to the small $\ell$ expansion, except that the integrals now extend into the deep IR. As a technical consequence the final expression for EE is not a generalized power series \eqref{eq:EEsmall} but can have a more complicated functional behavior on the (large) interval $\ell$. For example, as we shall see in the next subsection the leading order term can start with a double log, $S(\ell\gg 1)\propto \ln\ln\ell + {\cal O}(1)$. 

In the second scenario there is an IR cutoff at $z=z_\cen$ due to a center (which may be a regular center or feature a curvature singularity). The turning point is then close to this IR cutoff, so we introduce a small parameter $\delta$, e.g.~$z_\ast=z_\cen\,(1-\delta)$. While the integrals now remain bounded in the IR, the relation between the expansion parameter $\delta$ and the (large) interval length $\ell$ again leads to an expression for EE that is not a generalized power series of the form \eqref{eq:EEsmall}. In the example studied in the next subsection the leading behavior is going to be $S(\ell\gg 1)\propto S_0 + 1/\ln\ell + {\cal O}(\ln\ln\ell/\ln^2\ell)$ with some constant $S_0$. The physical reason behind these drastic changes of EE is the IR behavior of the Casini--Huerta $c$-function. 

Finally, in the third scenario there is an IR cutoff at $z=z_h$ due to a black brane horizon with inverse temperature $\beta$. Again the turning point is close to this IR cutoff, so we use e.g.~$z_\ast=z_h\,(1-\delta)$. In this scenario the small parameter $\delta$ turns out to be suppressed exponentially in the (large) interval length, $\delta\sim\exp{(-2\pi\ell/\beta)}$. If one is interested only in the leading order term in EE then no complicated integrals need to be evaluated and the final result is the volume law \eqref{eq:i6} plus exponentially suppressed corrections. If EE is calculated for QNEC$_2$ purposes the story is conceptually the same, but technically slightly more complicated since the structure of the integrals is more delicate. We discuss this rather explicitly in appendix \ref{app:int} for the domain wall and black brane examples studied in section \ref{sec:QNEC} below.

%%%%%%%%%%%%%%%%%%%%%%%%%%%%%%%%%%%%
\subsection{Ground states}\label{sec:5.2}
%%%%%%%%%%%%%%%%%%%%%%%%%%%%%%%%%%%%

We calculate now EE for ground states dual to the domain wall solutions \eqref{eq:DomainWallSol}.  These states are Poincar\'e invariant and have vanishing temperature, so a useful cross-check on the correctness of the results is positivity and monotonicity of the Casini--Huerta $c$-function, see section \ref{sec:1.8}. There are three qualitatively different possibilities for the parameter $\alpha$ in the potential \eqref{eq:angelinajolie}: Case 0, where $\alpha=0$, Case I, where $0<\alpha<1$ and Case II, where $\alpha<0$.

Geometrically, the difference between these three cases is as follows. Case 0 develops a curvature singularity at large negative values of the radial coordinate $\rho$ in the domain wall metric \eqref{eq:metricDomainWall}. Case I has a curvature singularity at the finite value $\rho=\ln\alpha$. Case II has a second locally asymptotically AdS$_3$ region at large negative values of $\rho$, with AdS-radius $\ell^\textrm{\tiny{IR}}_\textrm{\tiny{AdS}}$ given by 
\eq{
\frac{\ell^\textrm{\tiny{IR}}_\textrm{\tiny{AdS}}}{\ell^\textrm{\tiny{UV}}_\textrm{\tiny{AdS}}} = \frac{1}{1-1/(16\alpha)} 
}{eq:AdSradii}
which is smaller than its UV pendant. From these geometric considerations we expect the Casini--Huerta $c$-function to flow to the trivial value $c_{\textrm{\tiny{IR}}}\to 0$ for Cases 0 and I, while for Case II it should flow to the non-trivial value $c_{\textrm{\tiny{IR}}}=c_{\textrm{\tiny{UV}}}/(1-1/(16\alpha))$ where $c_{\textrm{\tiny{UV}}}=c$ is the Brown--Henneaux central charge. Below we show, among other things, that these expectations turn out to be correct.

We consider first Case 0. The metric function \eqref{eq:A} simplifies to $A=\rho-\exp(-\rho)/8$. The integrals in section \ref{sec:1.4}~\footnote{In order to compare with sections \ref{sec:1.4} and \ref{sec:5.1} we need to redefine the radial coordinate as $z=\exp{(-\rho)}$. The respective turning points $z_\ast=\exp{(-\rho_\ast)}$ and UV cutoffs $z_{\textrm{\tiny{cut}}}=\exp{(-\rho_{\textrm{\tiny{cut}}})}$ are related correspondingly.} 
\begin{align}
\ell e^{A_\ast} &= 2\int\limits_{\rho_\ast}^\infty\extd\rho\,\Big(\frac{1}{\sqrt{1-y}}-\sqrt{1-y}\Big) & A_\ast &:= A(\rho_\ast)\\
{\cal A}_{\textrm{\tiny{ren}}}(\ell) &= \ell e^{A_\ast} - 2\rho_\ast + 2\int\limits_{\rho_\ast}^\infty\extd\rho\,\big(\sqrt{1-y}-1\big) & y &:= \exp{\big(2A_\ast-2A(\rho)\big)} \label{eq:y}
\end{align}
can be evaluated perturbatively for small $\ell$ (using $\rho_\ast=-\ln\delta$ with $\delta\ll 1$, see appendix \ref{app:int})
\eq{
S_0(\ell\ll 1) = \frac{c}{3}\,\ln\ell - \frac{c\pi}{192}\,\ell + \frac{c(64-3\pi^2)}{73728}\,\ell^2 -\frac{c\pi(\pi^2-5)}{1179648}\,\ell^3 + {\cal O}(\ell^4) + S_{\textrm{\tiny{cut}}}
}{eq:S0small}
and for large $\ell$ (using $\rho_\ast=\ln\delta$ with $\delta\ll 1$)
\eq{
S_0(\ell\gg 1) = \frac{c}{3}\,\ln\ln\ell + c\,\ln 2 - \frac{c\,\ln 2}{\ln\ell} + {\cal O}(1/\ln^2\ell) + S_{\textrm{\tiny{cut}}}\,.
}{eq:S0large}

The Casini--Huerta $c$-function \eqref{eq:i24} in these limits
\begin{subequations}
\label{eq:CHc0}
\begin{align}
c(\ell\ll 1) &= c\,\Big(1-\frac{\pi}{64}\,\ell + \frac{64-3\pi^2}{12288}\,\ell^2 + {\cal O}(\ell^3)\Big) \\ 
c(\ell\gg 1) &= \frac{c}{\ln\ell} + \frac{3c\ln2}{\ln^2\ell} + {\cal O}(1/\ln^3\ell)
\end{align}
\end{subequations}
is indeed a strictly monotonically decreasing positive function of $\ell$. The large $\ell$ behavior in \eqref{eq:CHc0} shows that the central charge flows to zero in the IR in such a way that for an RG scale given by the (large) interval $\ell$ we have $c_{\textrm{\tiny{UV}}}=c_{\textrm{\tiny{IR}}}\,\ln\ell + \dots$. For Case I the small $\ell$ computation is very similar to Case 0 and yields
\eq{
S_I(\ell\ll 1) = S_0(\ell\ll 1) - \frac{c\alpha\,\ell^2}{192} + \Big(\frac{11\alpha}{98304} - \frac{\alpha^2}{1536}\Big)\,\pi c\,\ell^3 + {\cal O}(\ell^4)
}{eq:SIsmall}
which has a smooth limit to \eqref{eq:S0small} for vanishing $\alpha$. At large $\ell$ there is a qualitative change: there is a minimal value of the radial coordinate $\rho_{\textrm{\tiny{min}}}=\ln\alpha$, as opposed to Case 0 where $\rho$ could take arbitrarily negative values. This implies a different large $\ell$ behavior of EE
\eq{
S_I(\ell\gg 1) = -\frac{c}{3}\,\ln\alpha - \frac{c}{48 \alpha\, \ln\ell}  +\frac{c\,(32\alpha-1)}{768\alpha^2}\,\frac{\ln\ln\ell}{\ln^2\ell}
+ {\cal O}(1/\ln^2\ell) + S_{\textrm{\tiny{cut}}}
}{eq:SIlarge}
that has no smooth limit to \eqref{eq:S0large} for vanishing $\alpha$. The fact that EE does not grow at large $\ell$ can be understood geometrically from the presence of a center at $\rho=\rho_{\textrm{\tiny{min}}}$: since most of the geodesic is close to this center at large values of $\ell$ and the center is geometrically a point, there is practically no contribution to the length of the geodesic, which means that changing $\ell$ from a large to an even larger value almost does not affect the geodesic length. It would be interesting to find a field theoretic explanation of this plateau behavior of EE at large $\ell$.

Also for Case I the Casini--Huerta $c$-function \eqref{eq:i24} is a positive strictly monotonically decreasing function of $\ell$ in both limits. The relation between UV and IR values now reads $c_{\textrm{\tiny{UV}}}=c_{\textrm{\tiny{IR}}}\,16 \alpha\,\ln^2\ell + \dots \,
\,$. 

\begin{figure}[htb]
	\begin{center}
	\includegraphics[width=0.48\textwidth]{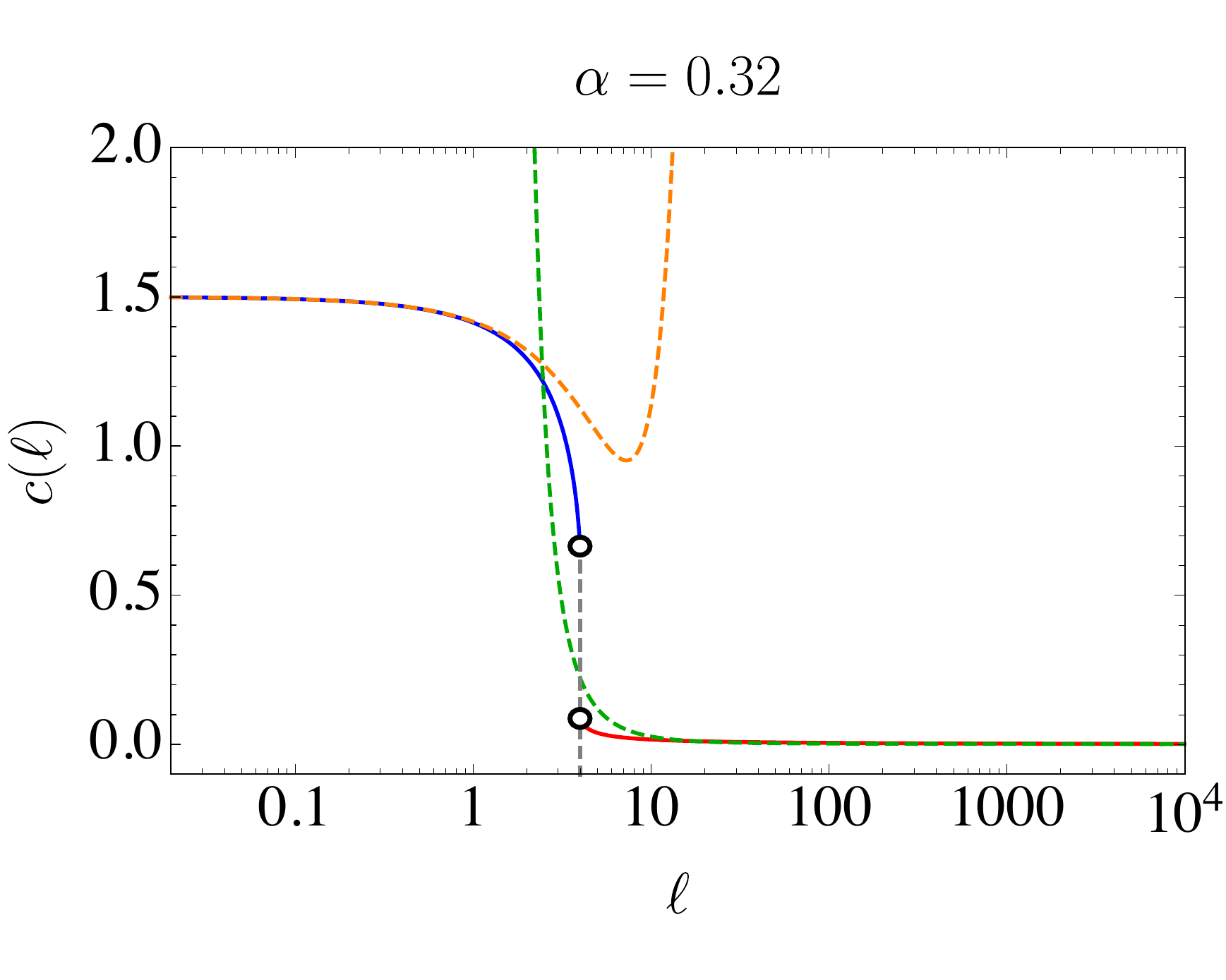}\includegraphics[width=0.48\textwidth]{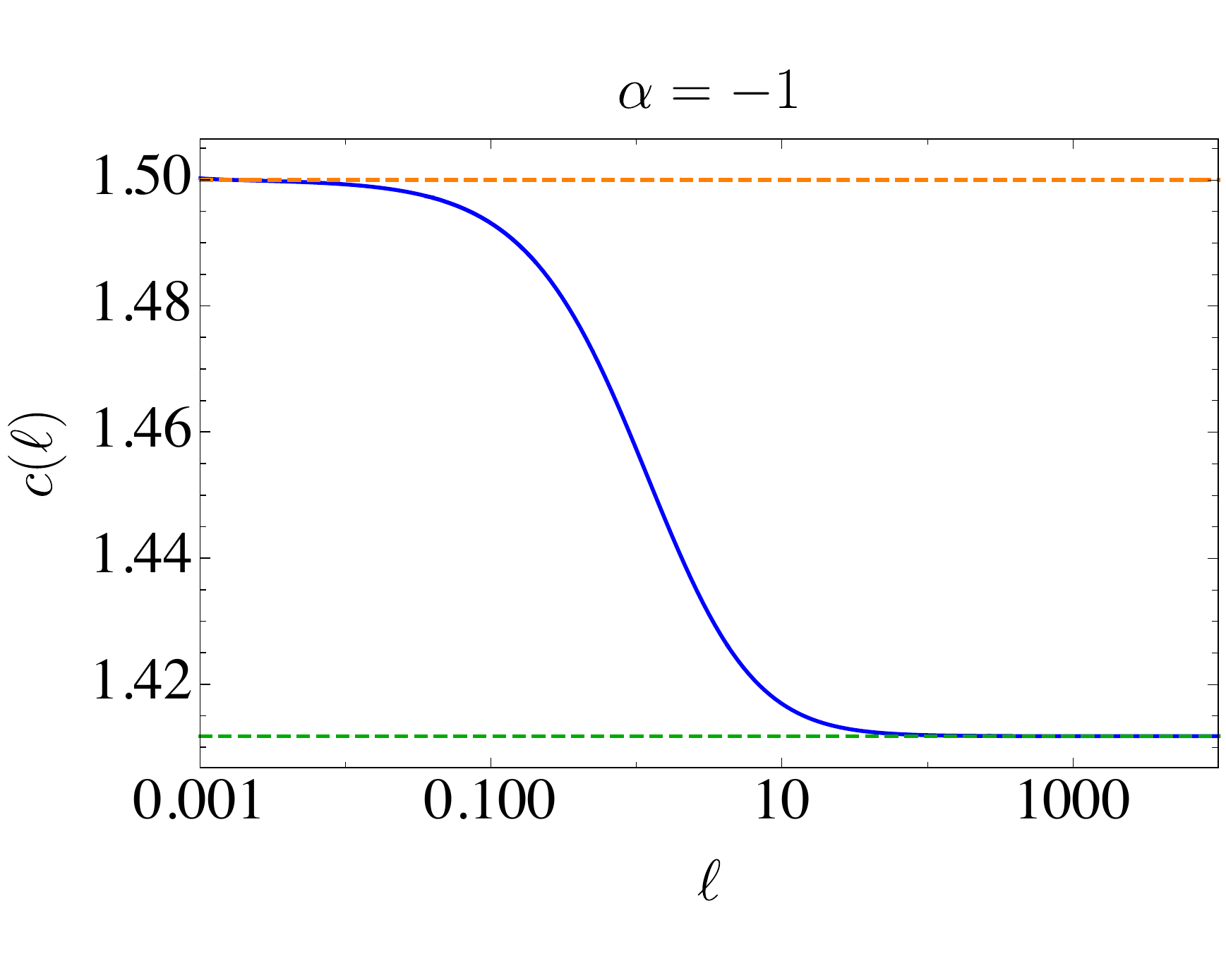}
	\caption{
		Left plot: Casini--Huerta $c$-function for $\alpha=0.32$. Blue and red solid lines: short and long RT surfaces on left and right hand sides of kink at $\ell^\ast$. Orange and green dashed lines: perturbative results for small \eqref{eq:SIsmall} and large \eqref{eq:SIlarge}.
		Right plot: Casini--Huerta $c$-function for $\alpha=-1$. Blue solid: numerical result.
		Orange dashed line: UV result, $c_\textrm{\tiny{UV}}=3/2$. Green dashed line: IR result \eqref{eq:cIR}, $c_\textrm{\tiny{IR}}=24/17\approx 1.41$.
	}
	\label{fig:c-function}
	\end{center}
\end{figure}

Case II, $\al<0$, is the only situation where we flow to another CFT fixed point in the IR. It is possible to evaluate all geodesic integrals exactly in the limit of large negative $\alpha$, yielding a result for EE
\eq{
S_{II}(\ell\gg 1) = \frac{c_{\textrm{\tiny{IR}}}}{3}\,\ln\ell + \dots
}{eq:SIIlarge}
that obeys the area law \eqref{eq:i4}, but with a central charge that is smaller than the UV result 
\eq{
c_{\textrm{\tiny{IR}}} = \frac{c_{\textrm{\tiny{UV}}}}{1-1/(16\alpha)}\,.
}{eq:cIR}
The ellipsis in \eqref{eq:SIIlarge} denotes terms that are $\ell$-independent or vanish at large $\ell$, as well as terms that vanish like $\alpha^{-3}$. Despite the assumption of large absolute values of $\alpha$, the IR value of the central charge \eqref{eq:cIR} is correct at arbitrary negative values of $\alpha$. Consistently, the ratio of IR and UV central charges coincides with the ratio of IR and UV AdS radii \eqref{eq:AdSradii}. The right plot in Figure~\ref{fig:c-function} shows an example of this case where $\alpha=-1$. The red solid line is the numerical results and the green dashed line is the IR expression given in \eqref{eq:cIR}.  By contrast, the left plot in Figure~\ref{fig:c-function} displays the Casini--Huerta $c$-function for the Case I example $\alpha=0.32$.

\begin{figure}[htb]
	\begin{center}
	\includegraphics[height=0.28\textheight]{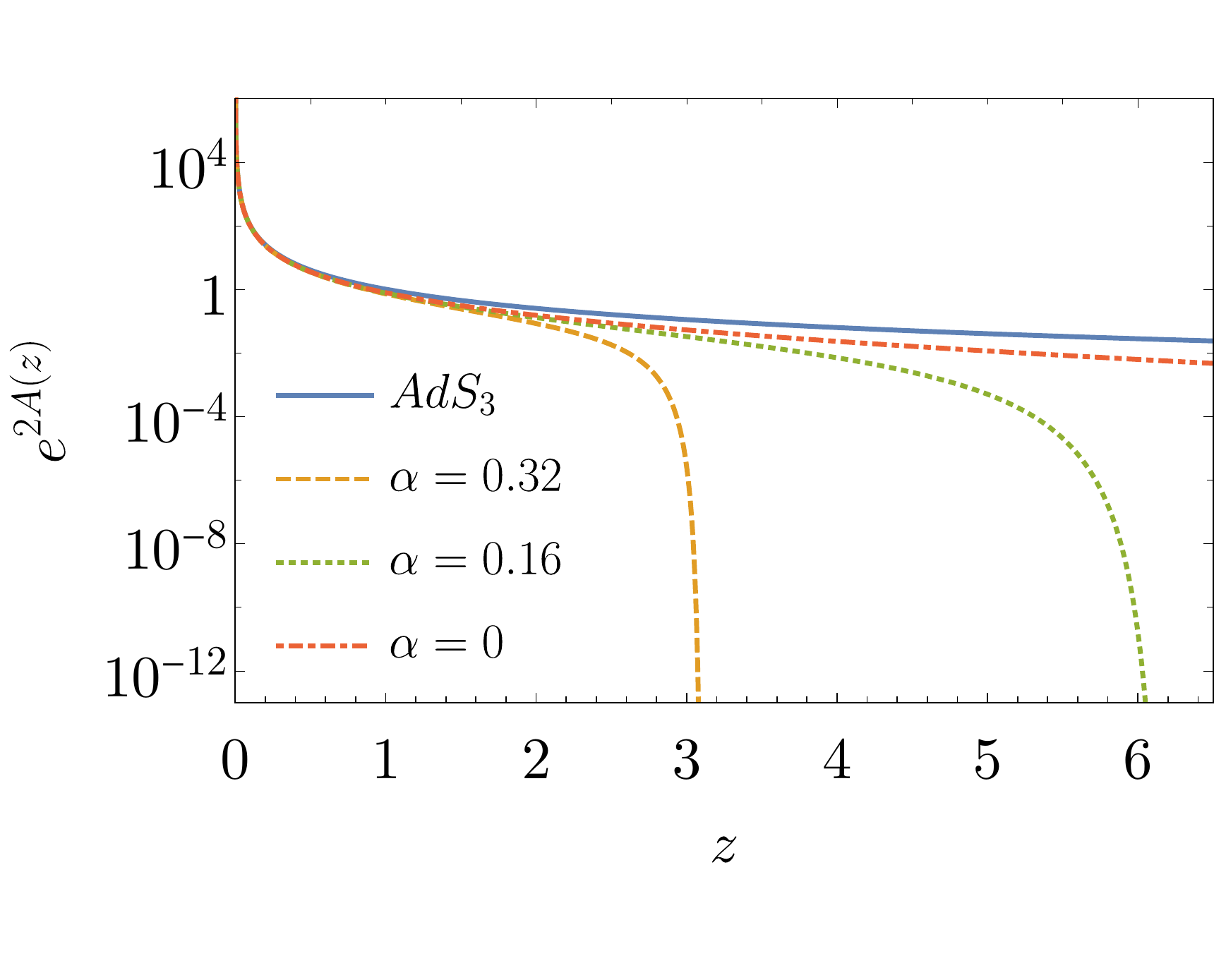}\quad \includegraphics[height=0.28\textheight]{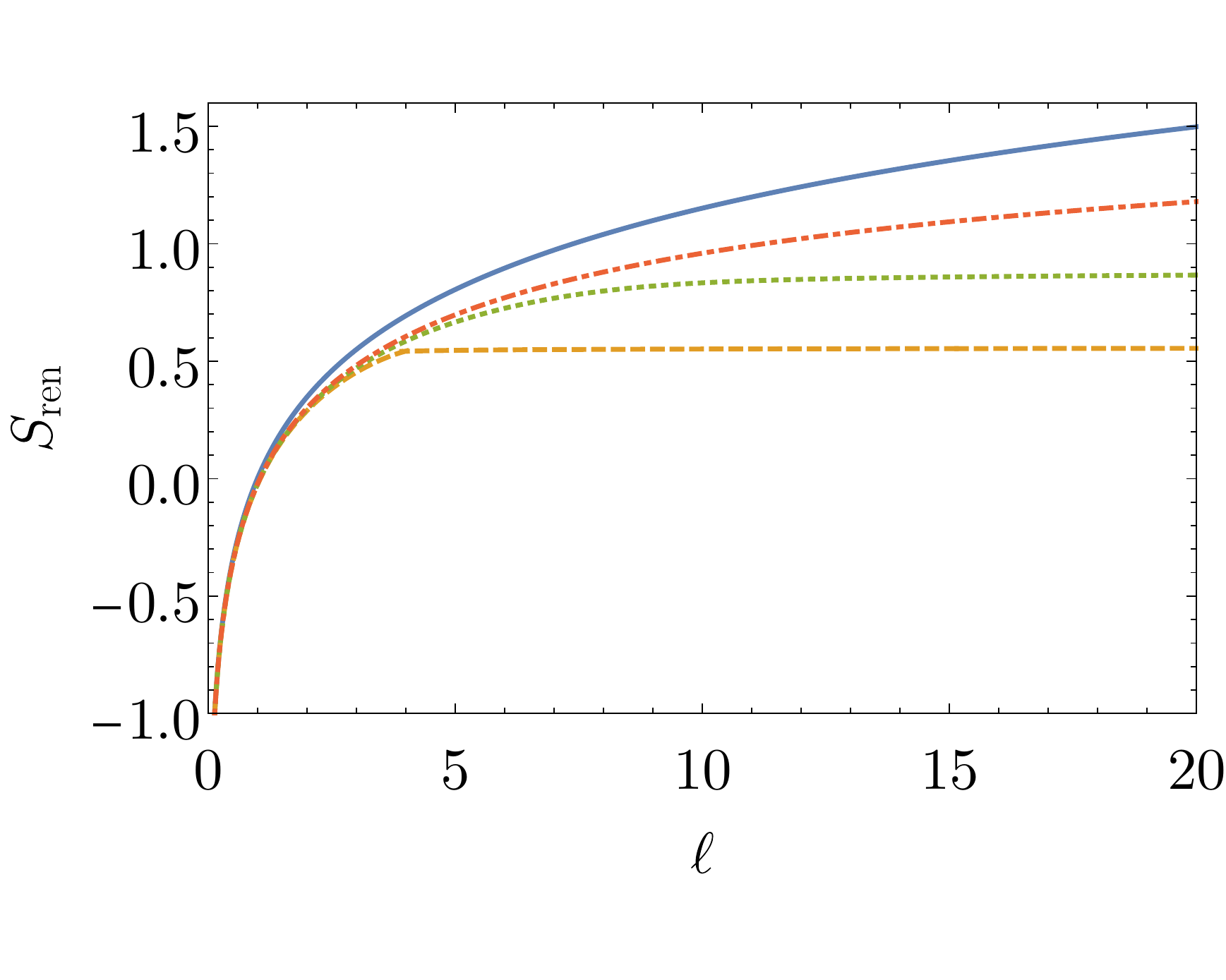}
		\caption{Left: Scale factor in domain wall geometry \eqref{eq:metricDomainWall} for different potentials. Right: Renormalized EE for four cases in left plot.
		}
	\label{fig:DomainWallGeom}
	\end{center}
\end{figure}

In order to get EE beyond the small or large interval expansions we resort to numerics and study three different values $\alpha=0.32,\,0.16,\,0$ of the free parameter in the potential \eqref{eq:angelinajolie}, i.e.~two examples of Case I and Case 0. Since Case II does not lead to interesting phase transitions we do not continue studying it numerically. The central value of the radial coordinate can be seen in the left panel of Figure~\ref{fig:DomainWallGeom} where we plot $e^{2A(z)}$ for the three cases together with the corresponding metric function $1/z^2$ of pure AdS$_3$. For small values of the radial coordinate $z$ all geometries converge to AdS$_3$. For $\alpha>0$ the geometry degenerates to a central point at $z=z_\cen=\tfrac{1}{\alpha}$ where $e^{2A(z_\cen)}=0$.

In the right panel of Figure~\ref{fig:DomainWallGeom} we show renormalized EE as function of $\ell$. As expected from the discussion in section \ref{sec:1.8} we find that EE is a monotonic function of $\ell$.

Asymptotic AdS$_3$ boundary conditions in the UV (=~small $z$) yield the area law  \eqref{eq:i4} at small $\ell$. Geometrically this follows from the fact that RT surfaces with small $\ell$ reside predominantly in the asymptotically AdS$_3$ part of the geometry. This can be seen in the right plot of Figure~\ref{fig:DomainWallGeom}, where at small $\ell$ all curves collapse to the universal scaling of the vacuum state dual to empty AdS$_3$ (blue curve), compatible with the perturbative results \eqref{eq:S0small} and \eqref{eq:SIsmall}.

The IR (=~large $z$) properties of the geometry determine the large $\ell$ behavior of EE. For cases where the geometry ends at a finite value of $z$ EE develops a plateau in the sense that increase of entanglement with $\ell$ is suppressed, as evident from the perturbative result \eqref{eq:SIlarge}.

The UV and IR regimes are well-described by the perturbative small and large $\ell$ results above. We studied numerically the region at intermediate $\ell$ and found that its details depend crucially on the value of the parameter $\alpha$. Namely, while EE is a monotonic and continuous function of $\ell$ in all cases, there can be kinks, meaning that left and right derivatives can be different at some values of $\ell$. The geometric reason for such kinks is the existence of several saddle points in the area functional, which give rise to several extremal surfaces with the same $\ell$ but different areas. For $\alpha<\alpha^\ast\approx0.277$ the extremalization has a unique solution, but for each $\alpha>\alpha^\ast$ there exists a finite range of $\ell$ where the area functional has three different saddle points. EE is associated only with the dominant saddle point, defined as the one with the smallest surface area. At a critical interval $\ell^\ast$ two of the saddle points exchange dominance. As a consequence the first derivative of EE is discontinuous, which is an example for kinked entanglement discussed in section \ref{sec:1.5}.

\begin{figure}[htb]
	\begin{center}
	\includegraphics[height=0.28\textheight]{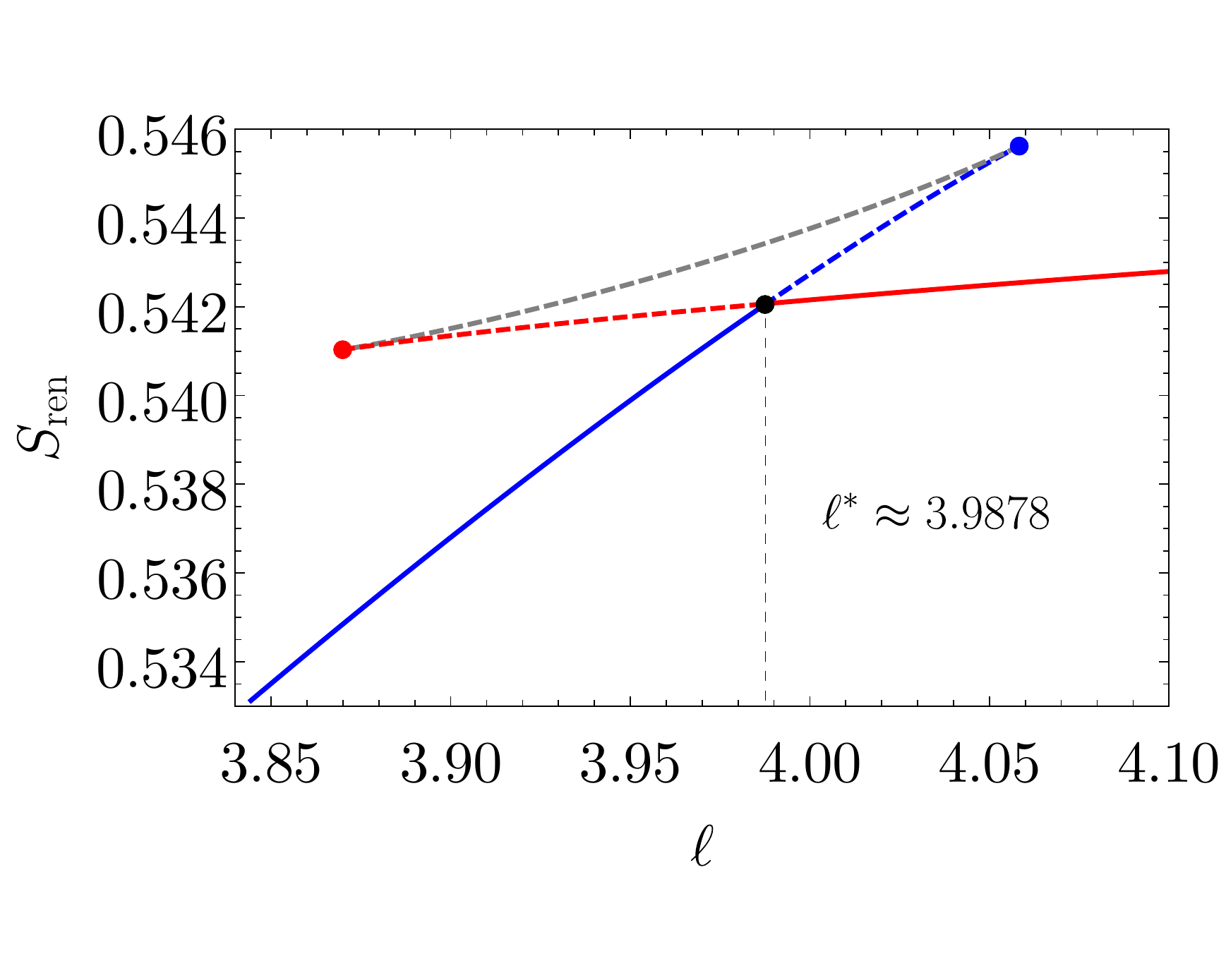}\quad\includegraphics[height=0.28\textheight]{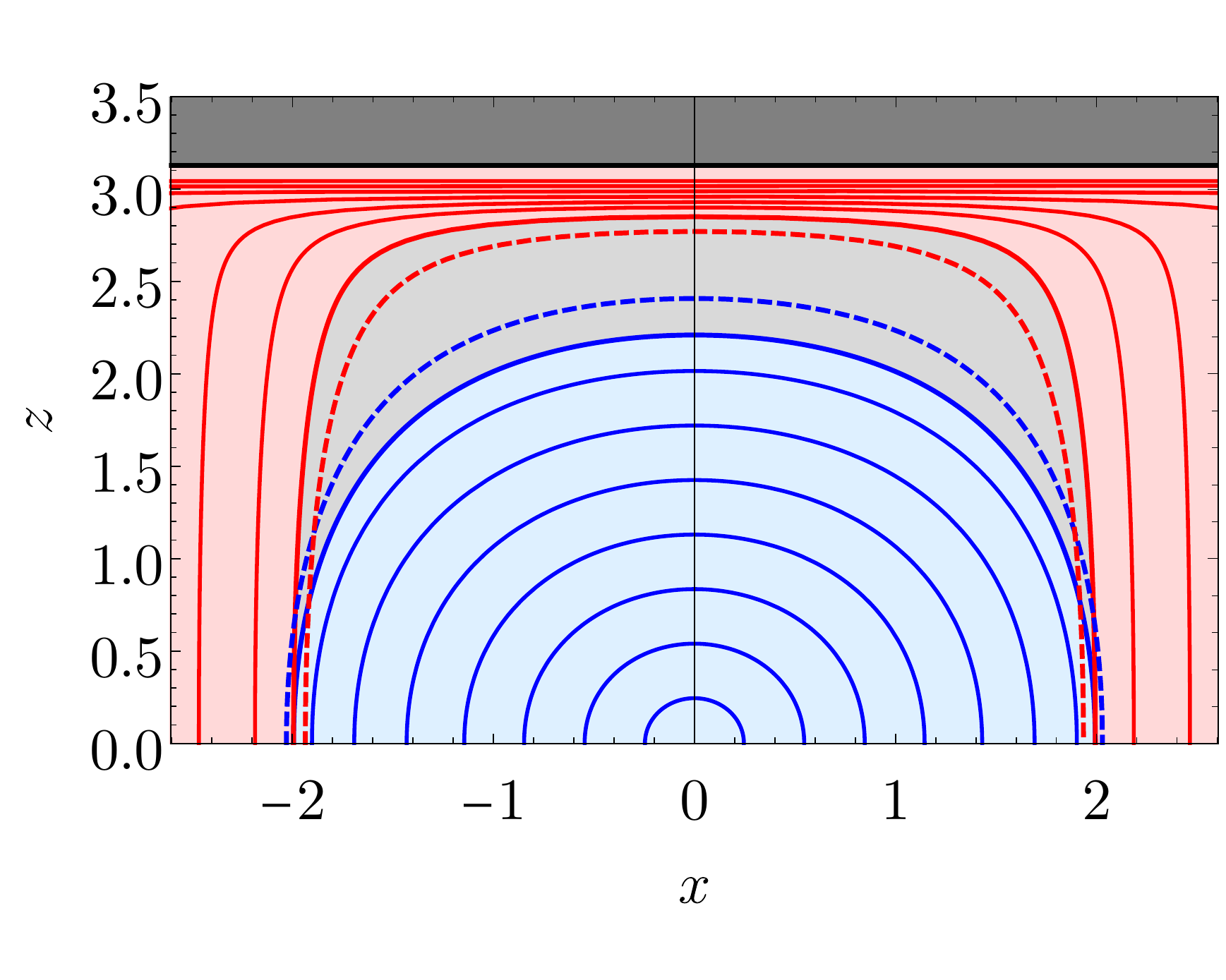}
	\caption{
		Left: EE in intermediate region. Right: Minimal surfaces in domain wall geometry.
	}
	\label{fig:DomainWallSurfaces}
	\end{center}
\end{figure}

We illustrate this situation in Figure~\ref{fig:DomainWallSurfaces} for the domain wall with $\alpha=0.32$. In the left plot we show the renormalized EE as a function of $\ell$ in the interesting region around the kink at $\ell^\ast$. In the range $3.870<\ell<4.058$ the area functional has three saddle points resulting in three different branches for the surface area shown in blue, red and gray. Dominant branches are shown as solid lines and sub-dominant ones with dashed lines. The red branch ceases to exist for $\ell<3.870$ and the blue one for $\ell>4.058$. At $\ell^\ast\approx 3.988$ the saddle points for the blue and red curve exchange dominance, while the third saddle point represented by the gray dashed line remains always sub-dominant. The plot on the right is a bulk picture of the corresponding RT surfaces. The black line indicates $z_\cen$ where the geometry degenerates to a point and the dark gray region at $z>z_\cen$ is not part of the geometry. Solid red and blue curves are representatives of minimal surfaces in the respective dominant branches. They can exist in the red and blue colored regions only. The dashed red and blue curves are subdominant surfaces with smallest and larges separation in the two branches marked by red and blue dots in the left plot. 

Curiously, the gray area remains untouched by dominant surfaces and is only traced out by sub-dominant ones. Thus, modulo $x$-translations of the interval these gray areas are entanglement shadows \cite{Balasubramanian:2014sra}, i.e.~bulk regions which no RT-surface can probe. However, permitting $x$-translations of the interval casts light on these shadows. This means that a field theory observer trying to reconstruct the bulk geometry will not be able to do so just by extending their entangling intervals to all possible values of interval length $\ell$; instead, they also have to shift the central position of the interval in order to reconstruct the gray areas in the bulk.

%%%%%%%%%%%%%%%%%%%%%%%%%%%%%%%%%%%%%%%%
\subsection{Thermal states}\label{sec:5.3}
%%%%%%%%%%%%%%%%%%%%%%%%%%%%%%%%%%%%%%%%%%

We consider now EE of thermal states that are dual to black brane geometries discussed in section \ref{sec:3.3}, first perturbatively and then numerically.

The perturbative analysis is most interesting in the large $\ell$ limit, since for small $\ell$ the existence of a horizon at finite temperature has almost no effect on EE. We consider additionally the limit of large temperature, which means that geometry is dominantly a BTZ black brane. This implies that the scalar field provides only a small perturbation of the thermal background, so one can solve the Klein--Gordon equation on a BTZ black brane background and then consider leading order backreactions on that background. We shall do this analysis in detail in section \ref{sec:QNECfiniteT} below when analyzing QNEC$_2$. The result \eqref{eq:area} for renormalized area yields the anticipated volume law \eqref{eq:i6} plus corrections that are suppressed exponentially like $e^{-2\pi\ell/\beta}$, as expected from \cite{Fischler:2012ca}.

In the remainder of this section we discuss numerical results for EE in Gubser gauge \eqref{eq:metric}-\eqref{eq:GubserGauge}, but using as radial coordinate $z=\phi^2/\phi_h^2$ where $\phi_h$ is the value of the scalar field at the horizon. The metric then expands near the boundary $z=0$ as
\be
\extd s^2=\frac{\phi_h^4\,\big(-\extd t^2+\extd x^2\big)+\extd z^2}{z^2}+\textrm{subleading}\,.
\ee
Renormalized EE and interval length are given by
\bea
&&S_\textrm{\tiny{ren}}=\frac{1}{2\GN}\int\limits^{z_\ast}_{z_\textrm{\tiny{cut}}} \left(\frac{e^{2B}}{H\left( 1-e^{2(A_\ast-A)} \right)} \right)^{1/2} \extd z
+\frac{1}{2\GN}\frac{\log(z_\textrm{\tiny{cut}})}{2-\Delta}\label{EE-hairyBH}\\
&&\ell=2\int\limits_{0}^{z_\ast} \frac{1}{e^{2A}}\left(\frac{e^{2(B+A_\ast)}}{H\left(1-e^{2(A_\ast-A)}\right)}\right)^{1/2}\extd z\,.
\eea
We evaluated the metric functions $A$, $B$ and $H$ and the two integrals above numerically.

\begin{figure}[hbt]
	\begin{center}
	\includegraphics[width=0.6\textwidth]{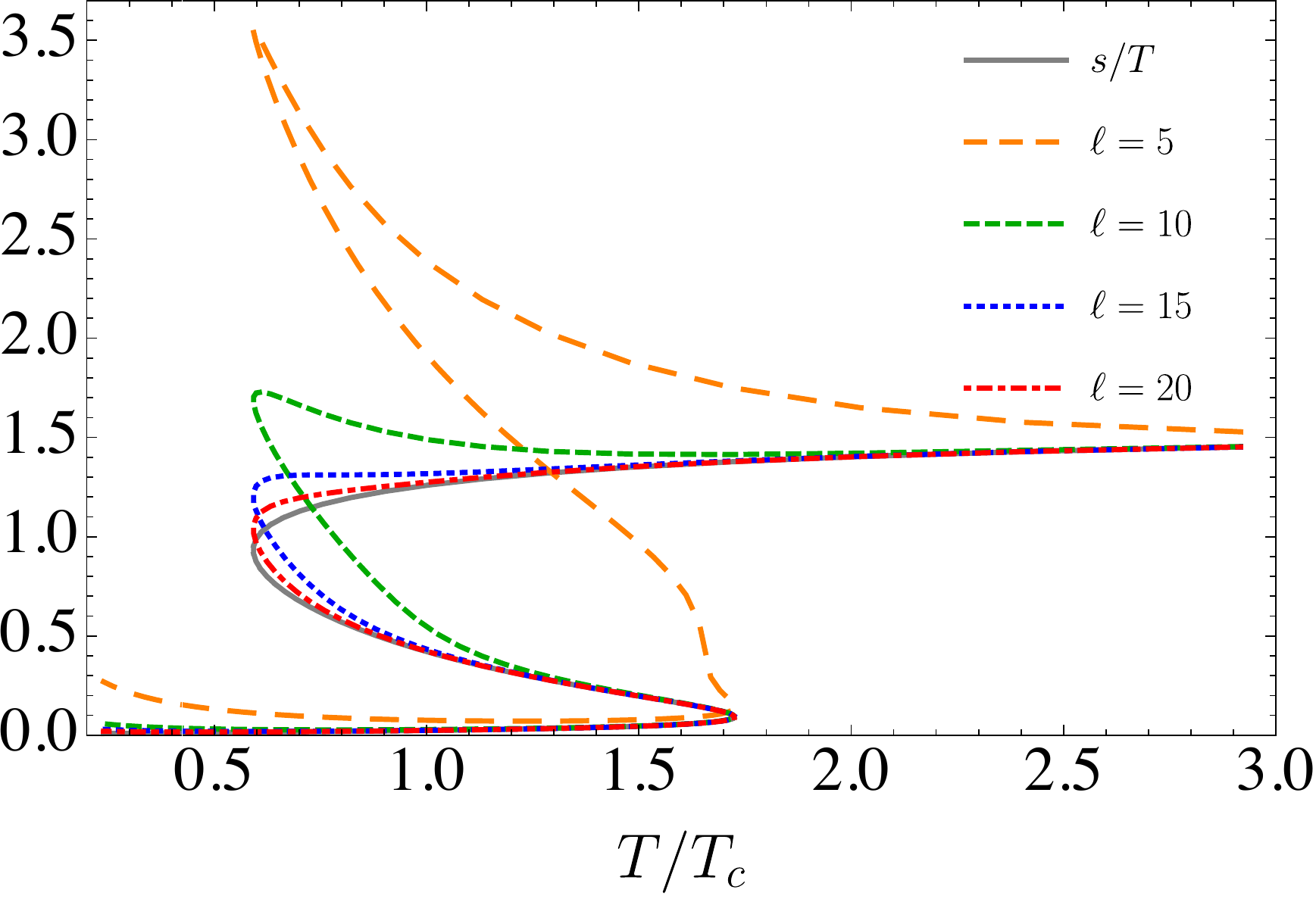}
	\caption{Comparing entropy density $s/T$ (Gray solid) with EE density $\sigma/T$ (Colored dashed) defined in \eqref{large-separation} for intervals $\ell=5,\, 10,\, 15,\, 20$.}
	\label{large-sep-fig}
	\end{center}
\end{figure}

Consider first again the large separation behaviour of EE. As stated above, EE obeys the volume law \eqref{eq:i6} with exponentially small corrections suppressed by $e^{-2\pi\ell/\beta}$, which provides a simple cross-check on the correctness of the numerics. For purposes of displaying results it is useful to plot EE density
\be
\sigma(\ell, T):=\frac{\extd S_\textrm{\tiny{ren}}}{\extd\ell}\label{large-separation}
\ee
since it is approximately constant at large values of $\ell$. In Figure~\ref{large-sep-fig} we show the temperature dependence of $\sigma(\ell, T)$ for several intervals.
We use the potential \eqref{eq:angelinajolie} with $\alpha=0.32$, which leads to a first order phase transition between large and small black branes at $T=T_c$.
As expected from the volume law, for large separations $\sigma(\ell, T)$ approaches the entropy density for all black brane geometries. The region around the critical temperature is more involved and shows that at the point where large black brane branch and spinodal unstable branch join together the agreement between EE and thermal entropy can be seen only for very large intervals. 

\begin{figure}[hbt]
	\begin{center}
	\includegraphics[width=0.48\textwidth]{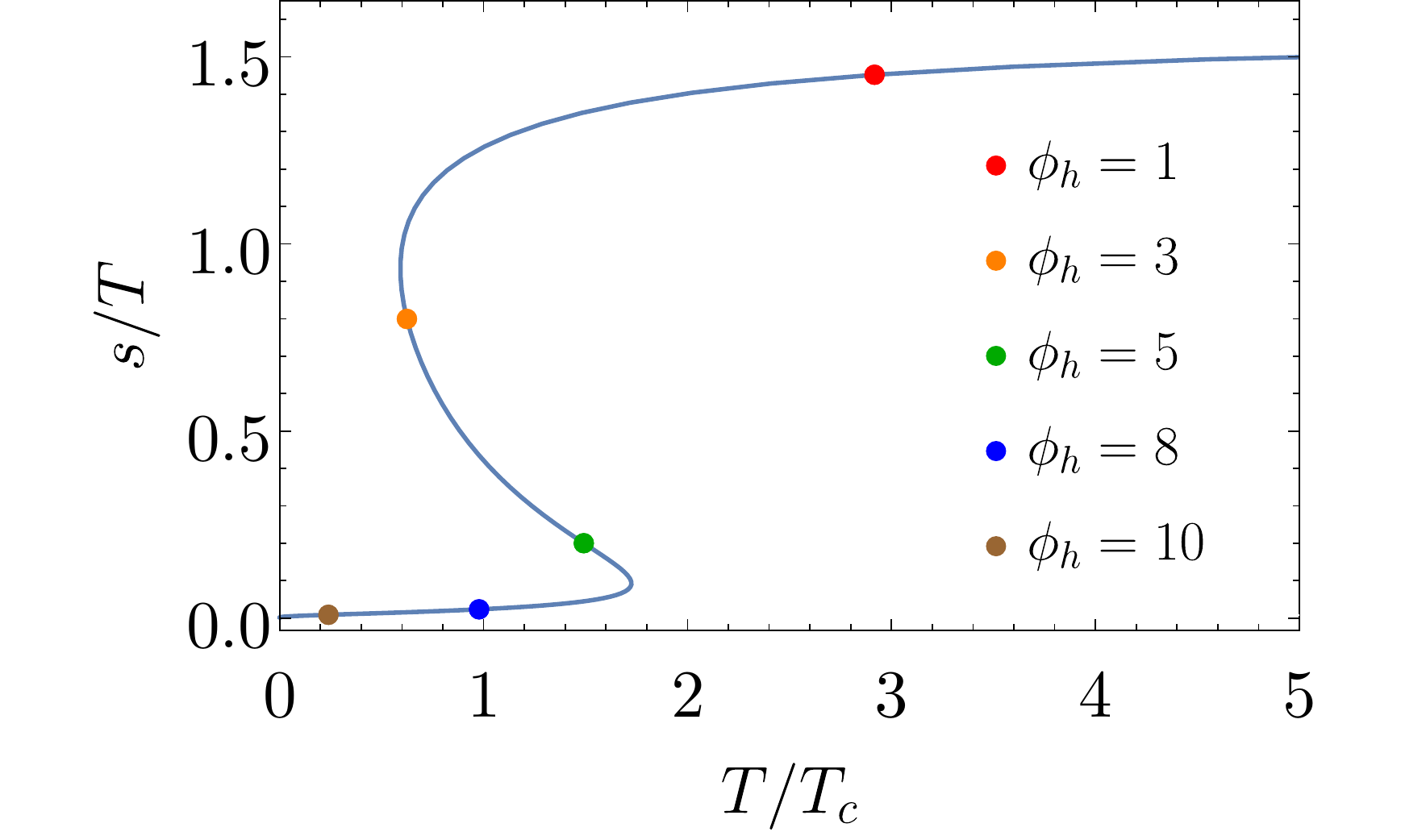}\includegraphics[width=0.48\textwidth]{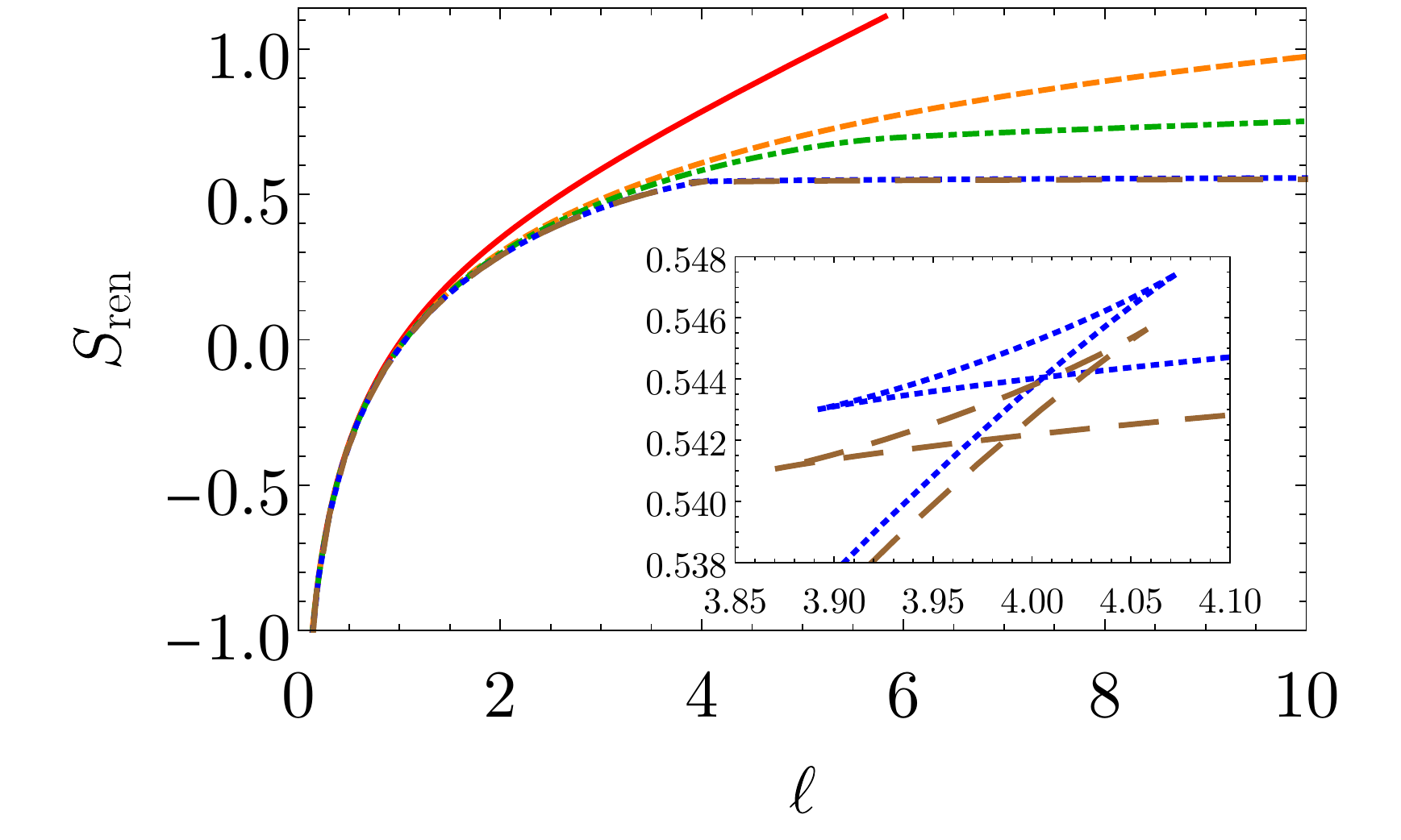}
	\caption{Left panel: Entropy density over temperature for first order phase transition ($\alpha=0.32$). Dots: horizon value of scalar field, $\phi_h=1, 3, 5, 8, 10$. 
	Right panel: EE computed using \eqref{EE-hairyBH} for different black brane solutions as function of interval $\ell$. Colors correspond to solutions marked with points in left panel.}
	\label{EE-1st-PT}
	\end{center}
\end{figure}

In Figure~\ref{EE-1st-PT} we show the renormalized EE as a function of the interval $\ell$ for different thermal states in the theory with $\alpha=0.32$ which has a first order phase transition.
A selection of black brane solutions is shown as colored points in the left panel.
The labels for each of these points correspond to the horizon value of the scalar field. We use the same colors for the renormalized EE in the right panel.

In the following we summarize the main features of EE in theories with first order phase transitions. For small intervals the turning point of the geodesic is close to the boundary where the metric is essentially locally AdS$_3$. Therefore, for all thermal states EE approaches the same value in this regime. For large intervals EE obeys the volume law. This is what one expects since most of the geodesic is close and parallel to the horizon as discussed around Figure~\ref{large-sep-fig}. In the insertion of the right panel in Figure~\ref{EE-1st-PT} we show a novel feature of EE for thermal states at low temperatures dual to small black branes. The RT surface is a multivalued function for a range of intervals, which is similar to the case we studied extensively in previous subsection for the ground state of the same theory. We found this phenomenon for $\phi_h> 5.6$, which is on the spinodal branch but close to the joining point to the small black brane branch. The smaller the value of $\phi_h$, the larger the critical value of the interval at which the transition between the RT surfaces occurs and the smaller the range of intervals with multivalued surfaces. We did not find this phenomenon in theories with crossover or second order phase transitions.

\begin{figure}[htb]
	\begin{center}
	\includegraphics[width=0.48\textwidth]{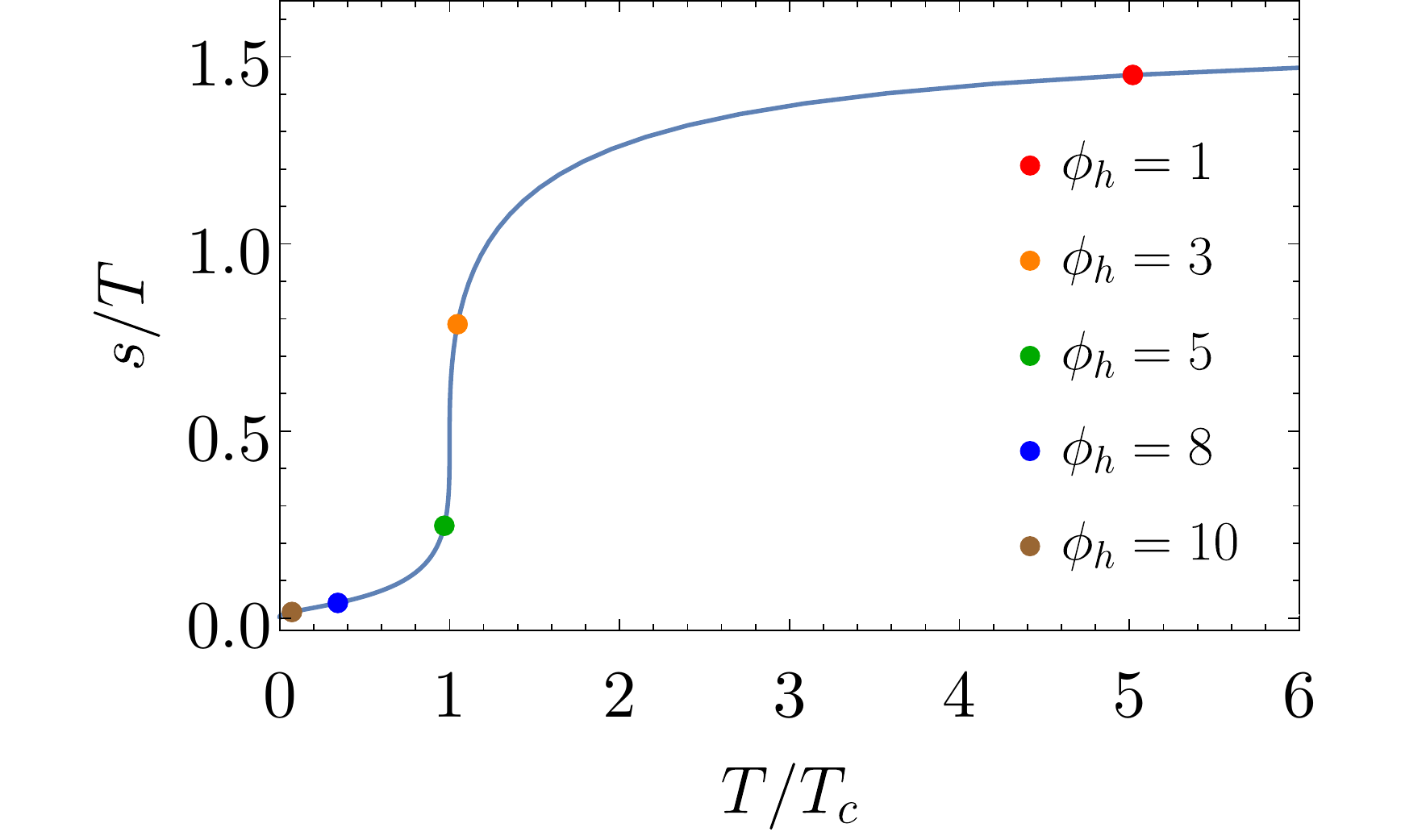}\includegraphics[width=0.5\textwidth]{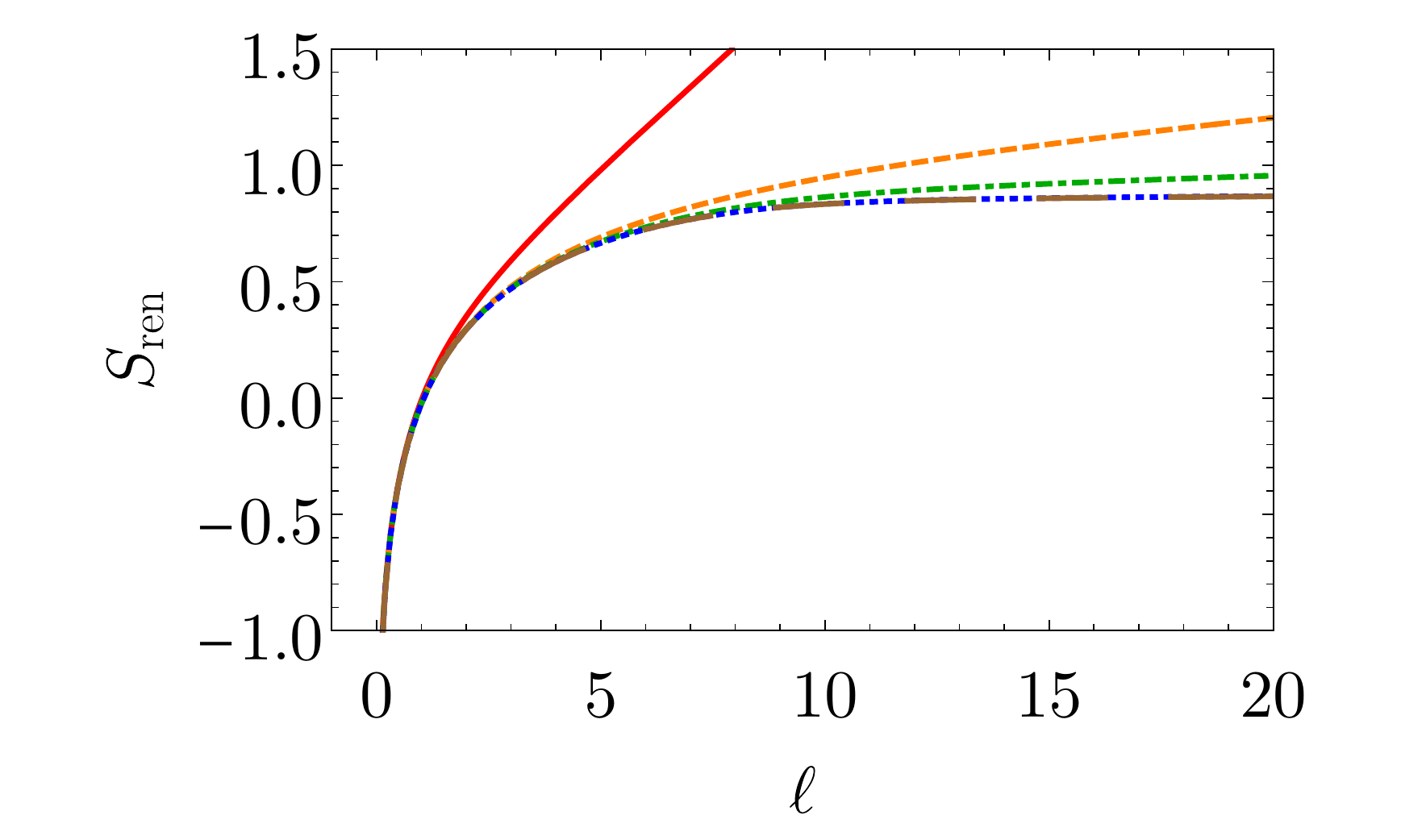}
	\caption{Left panel: Entropy density over temperature for second order phase transition ($\alpha=0.16$). Dots: horizon value of scalar field, $\phi_h=1, 3, 5, 8, 10$. 
	Right panel: EE computed using \eqref{EE-hairyBH} for different black brane solutions as function interval $\ell$. Colors correspond to solutions marked with points in left panel.}
	\label{EE-2nd-PT}
	\end{center}
\end{figure}

In Figure~\ref{EE-2nd-PT} we show the renormalized EE as a function of $\ell$ for various thermal states in a theory with second order phase transition ($\alpha=0.16$).
In contrast to the previously discussed case with first order phase transition, in a theory with second order phase transition EE is never kinked or discontinuous, but always a monotonic and smooth function of $\ell$ and $T$.
From the result for $\phi_h=8$ (dotted blue line) for example we can see that for large $\ell$ EE has only a very mild, plateau-like, volume law scaling at low $T$.
The reason for this is the small value of the entropy density \eqref{eq:thermodynamics} which gives the leading contribution to the slope of EE \eqref{large-separation} at large $\ell$. 
On the other hand EE shows a very rapid volume law scaling at large $T$, because the corresponding value of the thermal entropy is large. An example for this is the red curve for $\phi_h=1$ in the right plot of Figure~\ref{EE-2nd-PT}.

\begin{figure}[htb]
	\begin{center}
	\includegraphics[width=0.48\textwidth]{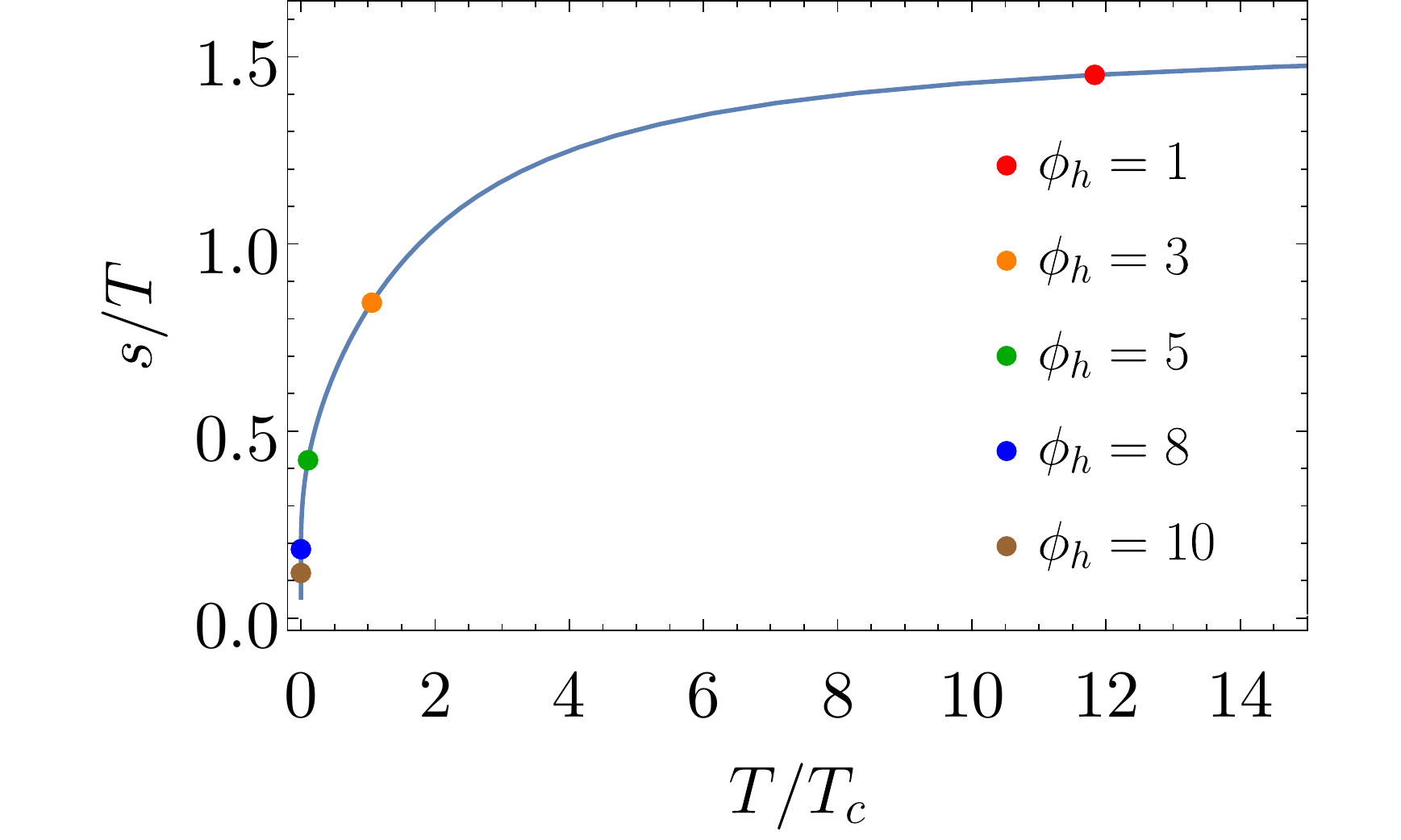}\includegraphics[width=0.5\textwidth]{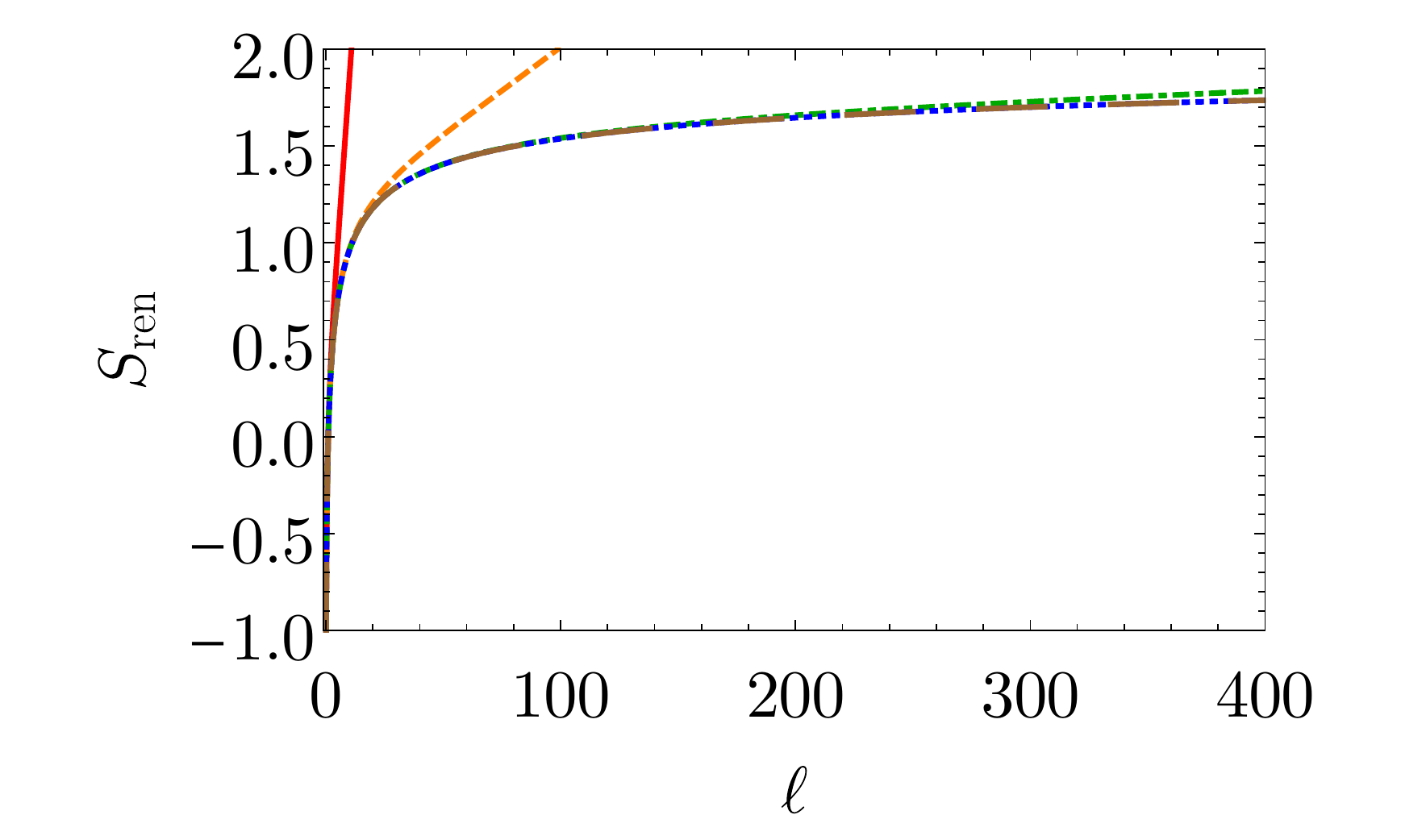}
	\caption{Left panel: Entropy density over temperature for crossover ($\alpha=0$). The dots show the horizon value of the scalar field at each point, $\phi_h=1, 3, 5, 8, 10$. 
	Right panel: EE computed by using the \eqref{EE-hairyBH} for different black hole solutions as function of separation. 
	Colors correspond to solutions marked with points in left panel.}
	\label{EE-CO-PT}
	\end{center}
\end{figure}

Finally, in Figure~\ref{EE-CO-PT} we show the renormalized EE as a function of $\ell$ in a theory with crossover ($\alpha=0$).
Also in this case EE is monotonic and smooth.
As discussed in subsection \ref{sec:5.2} $\alpha=0$ corresponds to Case 0 in which the ground state EE is not suppressed at large $\ell$, but grows like $\frac{c}{3}\ln\ln\ell$ instead, cf.~\eqref{eq:S0large}.
As a consequence at low temperatures the volume law scaling is approached extremely slow. This can be seen for example from the result for $\phi_h=8$ (dotted blue line) shown in the right Figure~\ref{EE-CO-PT}. The scaling of EE at large $T$ (red line in right Figure~\ref{EE-CO-PT}) is qualitatively similar to the other two cases discussed above. This universal (theory independent) behavior at large $T$ follows from \eqref{eq:area} obtained perturbatively in subsection \ref{sec:QNECfiniteT}.

We conclude this section with a brief summary table showing renormalized EE for various states dual to geometries that we studied in the limits of small and large intervals. QFT refers specifically to the deformed CFT that we study; the subscript refers to the sign of the parameter $\alpha<1$ in the scalar potential \eqref{eq:angelinajolie}.
\begin{table}[htb]
    \centering
    \begin{tabular}{l|l||l|l}
        State & Geometry & Small interval $\ell\ll 1$ & Large interval  $\ell\gg 1$ \\ \hline
         CFT ground & PP AdS & $\frac c3\,\ln\ell$ & $\frac c3\,\ln\ell$ \\
         QFT$_0$ ground & Dom.~wall & $\frac c3\,\ln\ell-\frac{c\pi}{192}\,\ell + {\cal O}(\ell^2)$ & $\frac{c}{3}\,\ln\big(8\,\ln\frac{\ell}{8}\big) + {\cal O}(\ln^{-2}\ell)$ \\
         QFT$_+$ ground & Dom.~wall & $\frac c3\,\ln\ell-\frac{c\pi}{192}\,\ell +{\cal O}(\ell^2)$ & $-\frac c3\,\ln(\alpha+\frac{1}{16\ln\ell})+ {\cal O}(\ln^{-2}\ell)$ \\
         QFT$_-$ ground &  Dom.~wall & $\frac c3\,\ln\ell-\frac{c\pi}{192}\,\ell +{\cal O}(\ell^2)$ & $\frac{c}{3(1-1/(16\alpha))}\,\ln\ell+{\cal O}(1)$ \\
         CFT thermal & BTZ & $\frac c3\,\ln\ell + {\cal O}(\ell^2)$ & $\frac{c\pi\ell}{3\beta}  + {\cal O}(e^{-2\pi\ell/\beta})$
    \end{tabular}
    \caption{Renormalized EE for various systems at small and large intervals.}
    \label{tab:1}
\end{table}

%%%%%%%%%%%%%%%%%%%%%%%%%%%%%%%%%%%%%%%%%%
\section{Quantum Null Energy Condition}\label{sec:QNEC}
%%%%%%%%%%%%%%%%%%%%%%%%%%%%%%%%%%%%%%%%%%
In this section we present our results for QNEC$_2$, starting with the analysis of vacuum states in section \ref{sec:QNECzeroT}, followed by remarks on how to predict phase transitions using QNEC$_2$ in section \ref{sec:predict}, and concluded by the analysis of finite temperature states in section \ref{sec:QNECfiniteT}. 

%%%%%%%%%%%%%%%%%%%%%%%%%%%%%%%%%%%%%%%%%%
\subsection{Ground states}\label{sec:QNECzeroT}
%%%%%%%%%%%%%%%%%%%%%%%%%%%%%%%%%%%%%%%%%%

For boost-invariant ground states small- and large-$\ell$ expansions of QNEC$_2$ follow directly from the corresponding expansions of EE in section \ref{sec:5.2}. Applying \eqref{eq:i22} to the small $\ell$-expansion \eqref{eq:SIsmall} yields
\begin{multline}
\lim_{\ell\ll 1}\Big(S^{\prime\prime} + \frac{6}{c}\,\big(S^\prime\big)^2\Big) = -\frac{c\pi}{64\ell} + \frac{c(128-3\pi^2)}{18432} - \frac{53\pi+9\pi^3}{1179648}\,c\,\ell\\
- \frac{c\alpha}{24} + \frac{229\pi}{98304}\,c\alpha\,\ell  - \frac{5\pi}{512}\,c\alpha^2\ell+ {\cal O}(\ell^2)\,.
\label{eq:QNECsmall}
\end{multline}
The result \eqref{eq:QNECsmall} is valid both for Case 0, I and II (in the former case only the first line contributes). The first term in \eqref{eq:QNECsmall} is negative and thrice the linear $\ell$-term in EE \eqref{eq:S0small}, compatible with the general results \eqref{eq:i32} and \eqref{eq:i31}. For the large $\ell$ expansion we need to discriminate between Case 0 ($\alpha=0$), where we apply \eqref{eq:i22} to \eqref{eq:S0large}
\eq{
\lim_{\ell\gg 1}\Big(S^{\prime\prime} + \frac{6}{c}\,\big(S^\prime\big)^2\Big) = -\frac{2c}{3\ell^2\,\ln\ell} - \frac{c(6\ln2-1)}{3\ell^2\,\ln^2\ell} + {\cal O}(1/(\ell^2\,\ln^3\ell))
}{eq:QNEClarge0}
Case I ($0<\alpha<1$), where we apply \eqref{eq:i22} to \eqref{eq:SIlarge}
\eq{
\lim_{\ell\gg 1}\Big(S^{\prime\prime} + \frac{6}{c}\,\big(S^\prime\big)^2\Big) = -\frac{c}{24\alpha\,\ell^2\,\ln^2\ell} + {\cal O}(1/(\ell^2\,\ln^3\ell))
}{eq:QNEClargeI}
and Case II ($\alpha<0$), where we apply \eqref{eq:i28} to \eqref{eq:cIR}
\eq{
\lim_{\ell\gg 1}\Big(S^{\prime\prime} + \frac{6}{c}\,\big(S^\prime\big)^2\Big) = -\frac{2c_{\textrm{\tiny{IR}}}}{3\ell^2(1-16\al)} + \dots
}{eq:QNEClargeII}

At intermediate values of $\ell$ we computed QNEC$_2$ by numerically integrating \eqref{EE-general} for a sufficiently large number of values for $z^\ast$ which together with \eqref{l-as-rs} allows to express EE as function of $\ell$. In the following we do this for several different examples and compare the results to the perturbative expressions.

\begin{figure}[htb]
	\begin{center}
	\includegraphics[height=0.37\textheight]{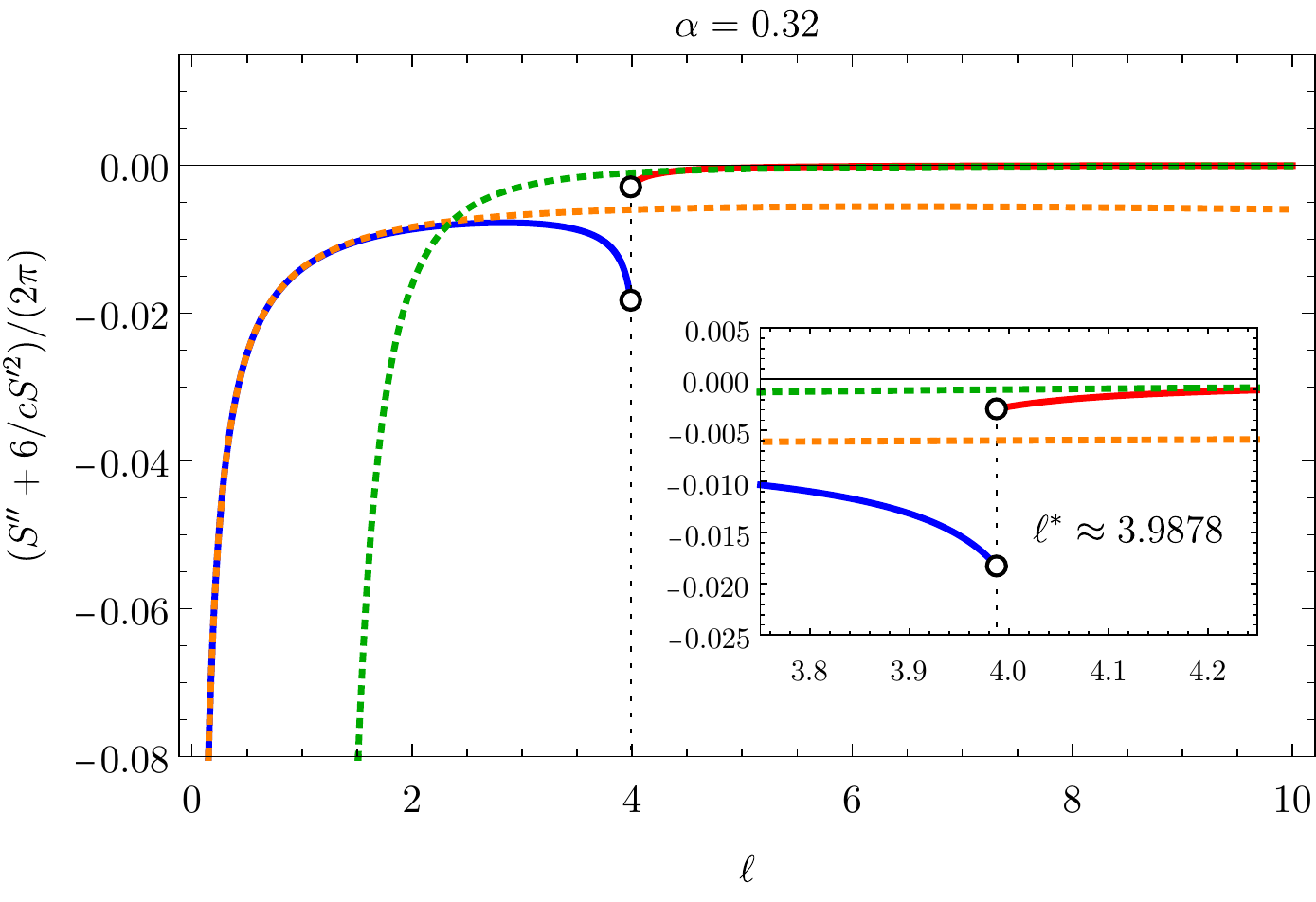}
		\caption{QNEC$_2$ for the domain wall solution with $\alpha=0.32$. Blue and red solid lines:  numerical results for QNEC$_2$ in the small and large $\ell$ branches left and right of the critical interval $\ell^*$.  Orange and green dashed lines: perturbative results for small $\ell$ and large $\ell$ given in \eqref{eq:QNECsmall} and \eqref{eq:QNEClargeI}, respectively.
		}
	\label{fig:DomainWallB4m0p04}
	\end{center}
\end{figure}

As a first example we chose $\alpha=0.32$ which lies comfortably in the interesting regime $\alpha>\alpha^\ast\approx0.27736$ where EE can be kinked and QNEC$_2$ discontinuous.
In Figure~\ref{fig:DomainWallB4m0p04} we show QNEC$_2$ as function of $\ell$ for the same setup and color coding as in Figure~\ref{fig:DomainWallSurfaces}. Blue and red curves again correspond to two different saddle points in the area functional.
Dotted orange and green curves are the perturbative small- and large-$\ell$ formulas.

Perturbation theory works remarkably well, except in a finite region close to the critical separation $\ell^\ast$ where the discontinuity is located.
As shown in section \ref{sec:1.5} QNEC$_2$ does not only jump but also has a divergence in form of a negative delta function at $\ell^\ast$ which we indicate by the black dashed line.
While the precise value of $\ell^\ast$ and size of the jump need to be determined numerically, we emphasize that the existence of the delta-divergence can only be deduced from analytic considerations. For $\ell<\ell^\ast$ QNEC$_2$ is not monotonic, but has a local maximum that leads to a finite gap.

\begin{figure}[htb]
	\begin{center}
	\includegraphics[height=0.24\textheight]{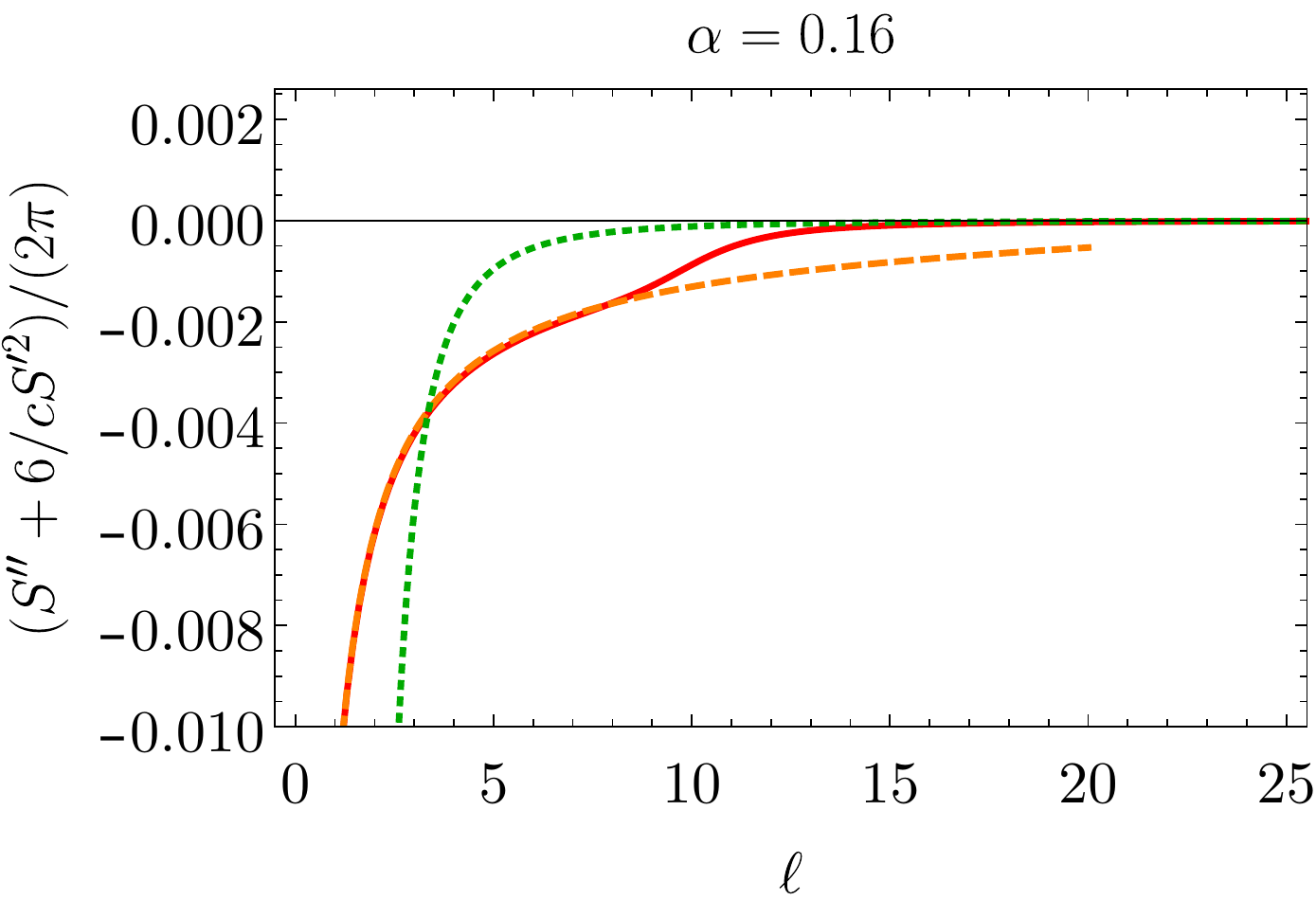}\quad	\includegraphics[height=0.24\textheight]{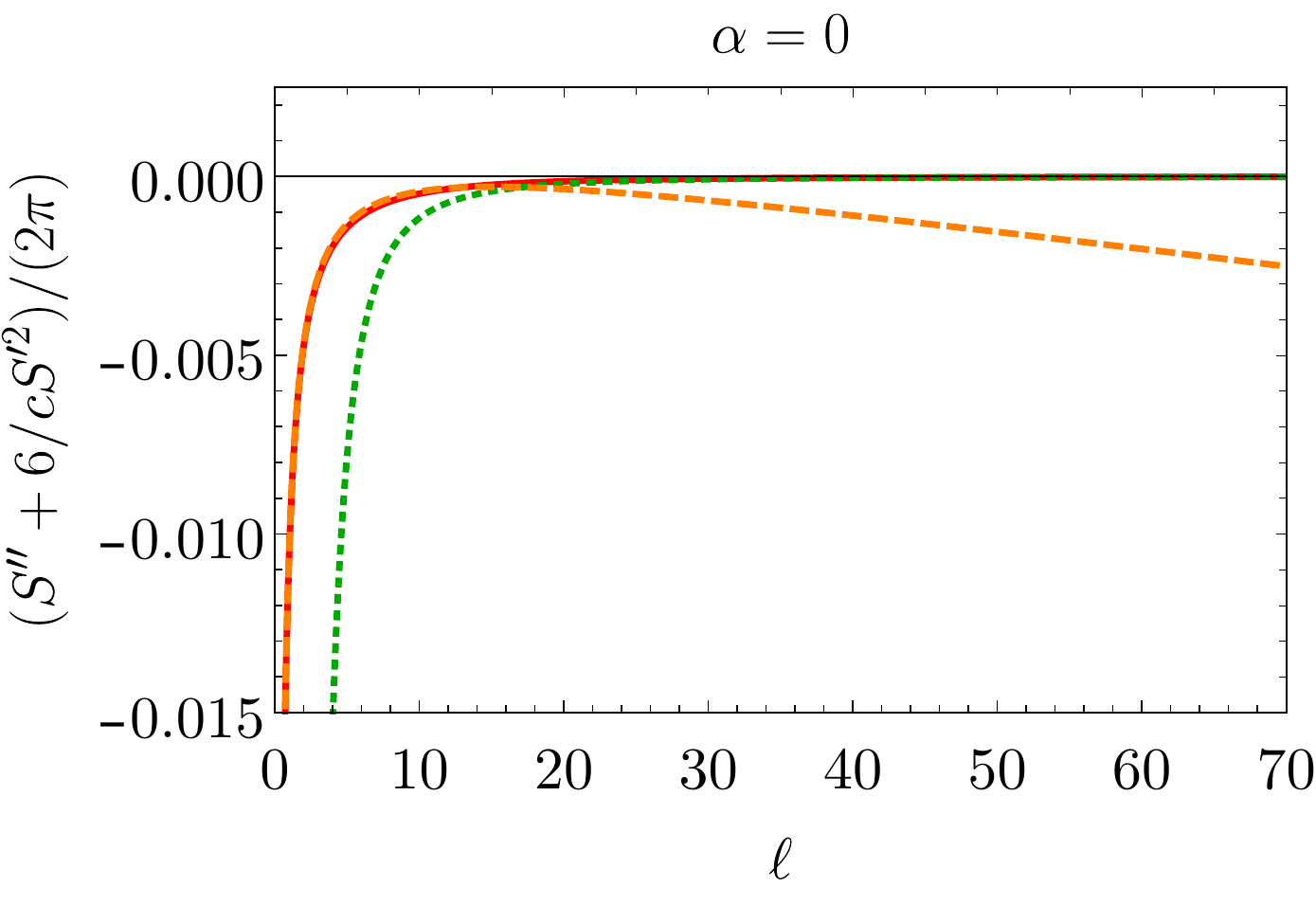}
		\caption{QNEC$_2$ for the domain wall solution with $\alpha=0.16$ (left) and $\alpha=0$ (right). Red solid lines: numeric result for QNEC$_2$.  Orange and green dashed lines: perturbative results for small $\ell$ and large $\ell$ given in \eqref{eq:QNECsmall} and \eqref{eq:QNEClargeI}, respectively. }
	\label{fig:DomainWallB4m0p02_QNEC}
	\end{center}
\end{figure}

The second example is $\alpha=0.16$ (see left Figure~\ref{fig:DomainWallB4m0p02_QNEC}). In this case the extremal surfaces are unique and EE is a smooth function of $\ell$. This results in QNEC$_2$ being smooth as well, even though is develops a shoulder around $\ell\approx 10$. The transition from the UV- to the IR-scaling regime clearly happens in this region.

Finally, the right plot in Figure~\ref{fig:DomainWallB4m0p02_QNEC} shows the case $\alpha=0$. Compared to the other examples QNEC$_2$ does not have any distinguished features in this case. The extremal surfaces are unique, EE and QNEC$_2$ are smooth and it is not possible to localize the transition between UV- and IR scaling from the numerical results. 

\subsection{Predicting phase transitions from ground state \texorpdfstring{QNEC$_2$}{QNEC2}}\label{sec:predict}
\begin{figure}[htb]
	\begin{center}
	\includegraphics[height=0.37\textheight]{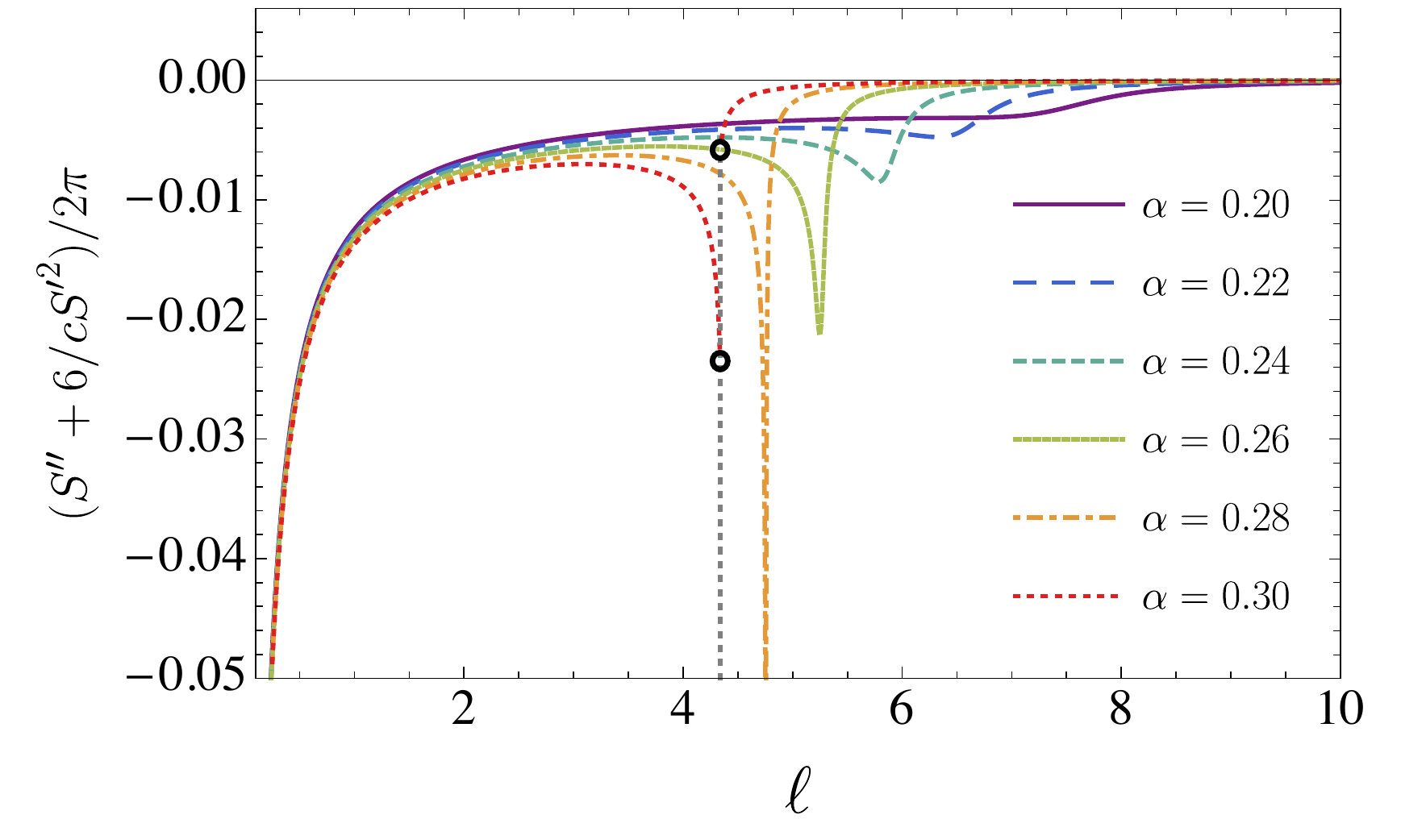}
		\caption{Ground state QNEC$_2$ as function of $\ell$ for $0.2\leq\alpha\leq0.3$.
		}
	\label{fig:QNECasTool}
	\end{center}
\end{figure}

A striking consequence of our numerical analysis for intermediate $\ell$ is the curious fact that QNEC$_2$ in ground states has features that allow to characterize the thermodynamic phase structure of the theory.
If ground state QNEC$_2$ is non-monotonic in $\ell$, which happens in our one-parameter family of theories for $\alpha>0.2$, then the theory has a first order phase transition at finite $T$. Increasing $\alpha$ makes the local minimum that is responsible for the non-monotonicity of QNEC$_2$ more pronounced until it eventually turns into a discontinuity for $\alpha\approx 0.277$. For $\alpha>0.277$ the phenomenon of multiple RT-surfaces and the related jump (plus delta function) in QNEC$_2$ occur. Also in this case the theory always has a first order phase transition. This is illustrated in Figure~\ref{fig:QNECasTool}, showing ground state QNEC$_2$ as function of $\ell$ for different values of $\alpha$.
Thus, spotting non-monotonicity or a jump in QNEC$_2$ for the ground state allows to predict a first order phase transition at some finite temperature.

Similarly, QNEC$_2$ for the ground state is monotonic and has a smooth, but clearly visible and localized transition region between the UV- and IR-scaling regime in case the theory has a second order phase transition at finite $T$. We show an example for this in the left plot of Figure~\ref{fig:DomainWallB4m0p02_QNEC}. If the theory at finite temperature has a smooth crossover then ground state QNEC$_2$ is a smooth, monotonic and featureless function of the interval $\ell$ without a clearly localized transition region between the UV- and IR-scaling regime. An example for this is shown in the right plot of Figure~\ref{fig:DomainWallB4m0p02_QNEC}.

This means ground state QNEC$_2$ can be used as diagnostic tool to identify and locate phase transitions. While the identification of ground state QNEC$_2$ and thermodynamic phase structure seems remarkable, we find it not to be precisely one-to-one. There are for example cases in which the theory has a first order phase transition but QNEC$_2$ is monotonic in $\ell$. In our model this is the case for $0.16<\alpha\leq 0.2$. But the converse statement ``if ground state QNEC$_2$ is non-monotonic then the theory has a first order phase transition'' we always find to be true.
We have checked for the potentials listed in Table \ref{tab:0} (two-dimensional cases) that similar features arise rather generically regardless of the value of the conformal weight, which suggests they are not artefacts of our particular choice of model.

%%%%%%%%%%%%%%%%%%%%%%%%%%%%%%%%%%%%%%%%%%
\subsection{Thermal states}\label{sec:QNECfiniteT}
%%%%%%%%%%%%%%%%%%%%%%%%%%%%%%%%%%%%%%%%%%

We consider now QNEC$_2$ for thermal states that are dual to black brane geometries discussed in section \ref{sec:3.3}, using metrics of the form 
\eq{
\extd s^2 = \frac{1}{z^2}\,\Big(-A(z)\,\extd t^2+\frac{\extd z^2}{A(z)} + S(z)\,\extd x^2\Big)\,.
}{eq:t0}
Again we start with a perturbative analysis, where we are mostly interested in the large $\ell$ limit. With no loss of generality we scale the temperature such that $2\pi T=1$ so that our only rermaining scale is the interval length $\ell$.

The exact expression for the scalar field
\eq{
\phi(z)=\epsilon\,\frac{4\sqrt{\pi}}{\Gamma(\tfrac14)^2}\,\sqrt{z}\,K\Big(\frac{1-z}{2}\Big) + {\cal O}(\eps^2)
}{eq:t1}
contains the backreaction parameter $\epsilon\ll 1$ and the elliptic integral $K$. The metric functions with no loss of generality are given by
\eq{
A(z)=1-z^2+\epsilon^2\,A_2(z)+{\cal O}(\epsilon^3)\qquad\qquad S(z)=1+\epsilon^2S_2(z)+{\cal O}(\epsilon^3)
}{eq:t2}
where for convenience we fixed the horizon to be located at $z_h=1$. The near boundary expansions of these three functions are given by
\begin{align}
    \phi(z) &= \epsilon\,\sqrt{z}\,\Big(1-\frac{4\pi^2}{\Gamma(\tfrac14)^4}\,z+\frac18\,z^2 + {\cal O}(z^3)\Big) \\
    A_2(z) &= -\frac14\,z\,\ln z + a_2 z^2 + {\cal O}(z^3) \\
    S_2(z) &= -\frac14\,z\,\ln z + \frac{3\pi^2}{\Gamma(\tfrac14)^4}\, z^2 + {\cal O}(z^3)
\end{align}
where we set to zero some of the leading and subleading coefficients, again with no loss of generality. 

The flux components of the boundary stress tensor 
\eq{
2\pi\langle T_{kk}\rangle = \frac{1}{4\GN}+\frac{\epsilon^2}{4\GN}\,\bigg(\frac{3\pi^2}{\Gamma(\tfrac 14)^4}-a_2\bigg) +{\cal O}(\epsilon^3)
}{eq:t3}
can be expressed in terms of the near horizon quantity $S_2(1)$ due to the existence of the radially conserved quantity
\eq{
Q = -\frac{S(z)^{3/2}}{2z}\frac{\extd}{\extd z}\frac{A(z)}{S(z)}\qquad\qquad \frac{\extd Q}{\extd z} = 0
}{eq:t5}
which yields $Q=1+\epsilon^2\,[3\pi^2/\Gamma(\tfrac14)^4-a_2]$ in the near boundary expansion and $Q=1+\tfrac12\,\epsilon^2S_2(1)$ at the horizon. Therefore, the left hand side of the QNEC$_2$ inequality \eqref{eq:QNEC2} reads
\eq{
2\pi\langle T_{kk}\rangle = \frac{1}{4\GN}+\frac{\epsilon^2}{8\GN}\,S_2(1) + {\cal O}(\epsilon^3)\,.
}{eq:t4}

To get the right hand side of the QNEC$_2$ inequality \eqref{eq:QNEC2} we follow the procedure in section \ref{sec:1.4}. See appendix \ref{app:int} for details on the integrals. The $\lambda$-dependent result for the renormalized area is given by
\eq{
{\cal A}^{\textrm{\tiny ren}}(\ell\gg 1) = (\ell + \lambda)\,\Big(1 + \frac{\epsilon^2}{2}\,S_2(1)\Big) + \frac{\la^2\epsilon^2}{16}\,\big(S_2'(1)-4S_2(1)\big)  
+ {\cal O}(e^{-\ell}) + {\cal O}(\lambda^3) + {\cal O}(\epsilon^3)
}{eq:area}
where prime means derivative with respect to $z$.

From the result \eqref{eq:area} for the renormalized area we obtain the expected volume law \eqref{eq:i6} plus corrections that are suppressed exponentially like $e^{-\ell}$ or at least cubic in the backreaction parameter $\eps$. Moreover, we obtain expressions for the quantities denoted by ${\cal A}_1$ and ${\cal A}_2$ in \eqref{eq:i15}.
\eq{
{\cal A}_1(\ell) = 1+ \frac{\eps^2}{2}\,S_2(1) +\dots \qquad\qquad {\cal A}_2(\ell) = \frac{\eps^2}{8}\,\big(S_2'(1)-4S_2(1)\big)+\dots
}{eq:areas}
The ellipses denote terms suppressed by $e^{-\ell}$ or by $\eps^3$. Therefore, the right hand side of the QNEC$_2$ inequality \eqref{eq:QNEC2} reads
\eq{
\partial_\lambda^2 S + \frac{6}{c}\,\big(\partial_\lambda S\big)^2 = \frac{c}{6} + \frac{c\eps^2}{12}\,S_2(1) + \frac{c\eps^2}{48}\,S_2'(1) + {\cal O}(e^{-\ell}) + {\cal O}(\eps^3)
}{eq:t7}
where for clarity we wrote explicitly the $\lambda$-derivatives and kept the notation that prime denotes $z$-derivative.

Using $4\GN=6/c$ we see that left \eqref{eq:t4} and right \eqref{eq:t7} hand sides of QNEC$_2$ almost cancel and the QNEC$_2$ inequality turns into a convexity condition for the function $S_2$ at the horizon.
\eq{
2\pi\langle T_{kk}\rangle - \partial_\lambda^2 S - \frac{6}{c}\,\big(\partial_\lambda S\big)^2 = -\frac{c\eps^2}{48}\,S_2'(1) \geq 0
}{eq:t8}
We have checked both numerically and analytically that this condition is satisfied. It is also possible to argue on general grounds from the second law of black hole mechanics that the inequality \eqref{eq:t8} has to be satisfied.\footnote{This goes back to the genesis of the QNEC$_2$ inequality, which was associated with the quantum focusing conjecture and the generalized second law \cite{Bousso:2015mna}. Since the QNEC$_2$ inequality must be respected the fact that \eqref{eq:t8} holds is unsurprising.} Namely, one can view our backreaction calculation as a consequence of a dynamical process where initially $\eps$ was zero and then a small amount of scalar matter is thrown into the black hole. The quantity $S_2(1)$ is proportional to the additional entropy density generated in this process and thus has to be positive. From a field theory perspective it also has to be monotonically increasing with energy~$\sim 1/z$. This means that $S_2$ expressed as function of $z$ must be monotonically decreasing, which is precisely the inequality \eqref{eq:t8}. 

We turn now to the evaluation of QNEC$_2$ at arbitrary values of the interval length $\ell$.
In boost invariant ground states it is possible to write QNEC$_2$ entirely in terms of $\ell$-derivatives of EE. 
For this a simple shooting algorithm is sufficient to determine EE as function of $\ell$ and subsequently QNEC$_2$.
The situation in black brane geometries is more complicated and one has to evaluate genuine lightlike derivatives of EE.
We do this using the relaxation approach of \cite{Ecker:2019ocp} with either a known geodesic in AdS$_3$ or previously obtained numerical solutions as initial guess.

These ansatz geodesics are then relaxed to discrete one-parameter families of geodesics with one endpoint shifted by lightlike vectors $k^\mu_{i,\pm}=i\delta(1,±1)$. Since we only consider spatially homogeneous and time-reversal invariant states, QNEC$_2$ does not depend on the orientation of the null vector and also not on which endpoint we vary. We typically produce a family of seven geodesics with $i={−3,−2,...,3}$ and set the size of the increment to $\delta= 10^{-5}$. From the length of these geodesics we compute the corresponding EEs and generate a third order polynomial fit $S≈c_0+c_1\delta+c_2\delta^2+c_3\delta^3$ from which we extract first and second derivative at $\delta= 0$. More details on the numerical implementation can be found in \cite{Ecker:2018jgh}.

\begin{figure}[htb]
	\begin{center}
     \includegraphics[height=0.17\textheight]{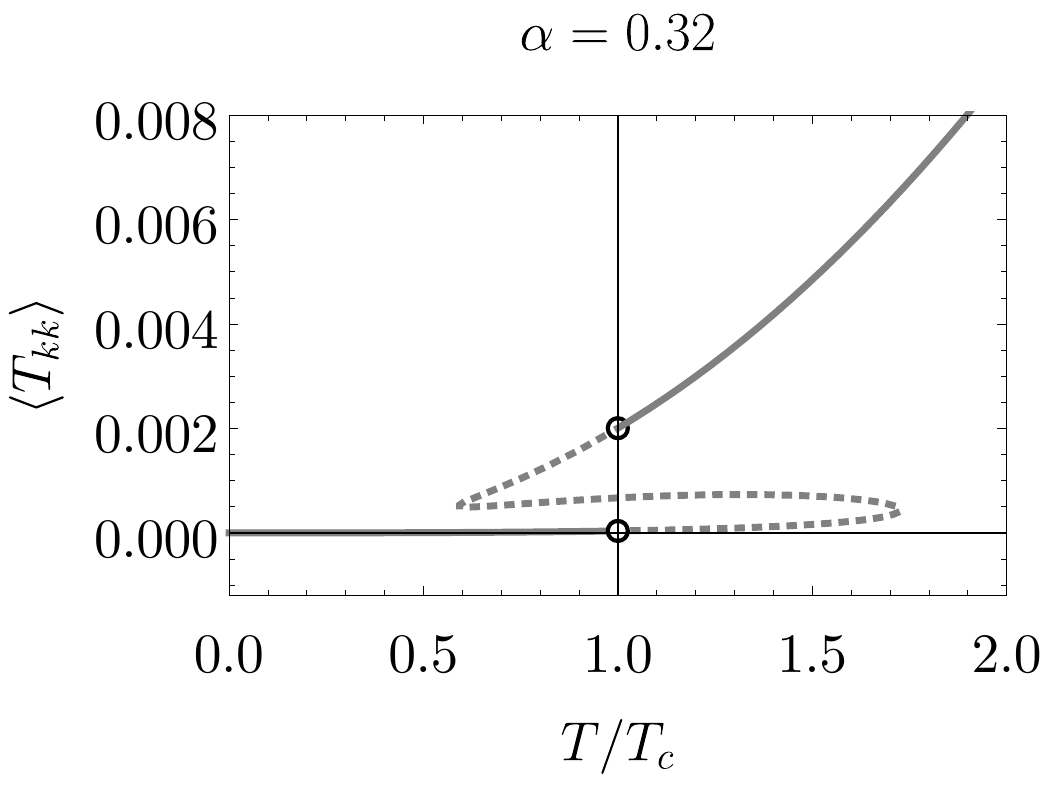}	\includegraphics[height=0.17\textheight]{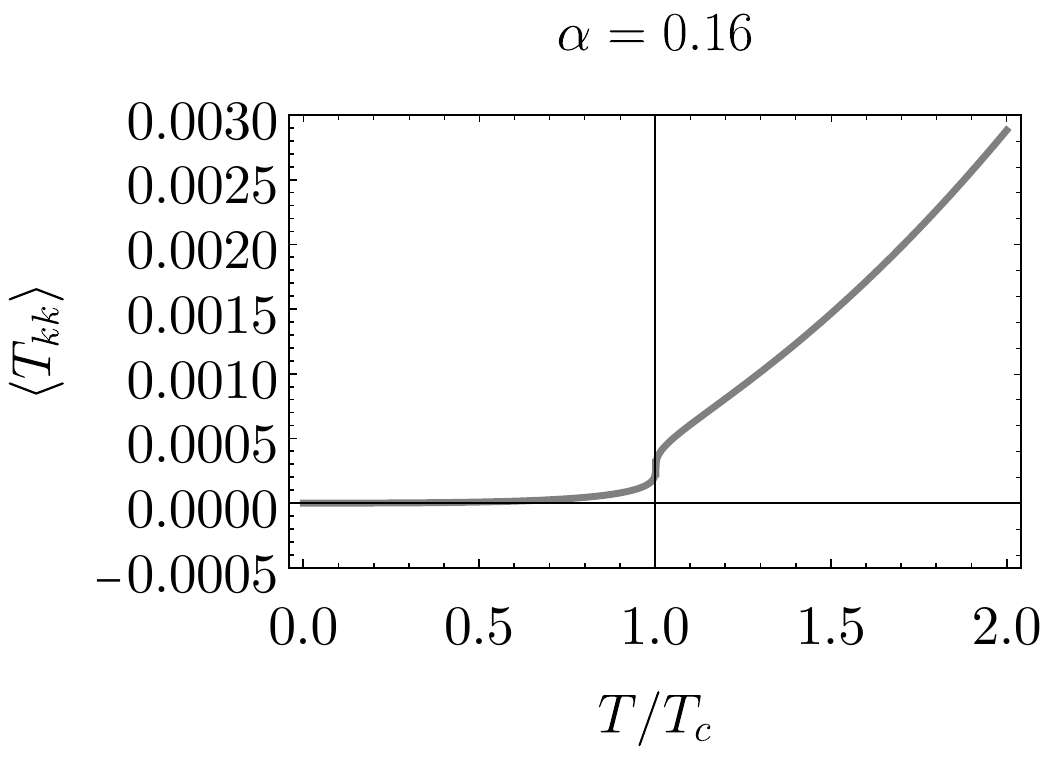} 
     \includegraphics[height=0.17\textheight]{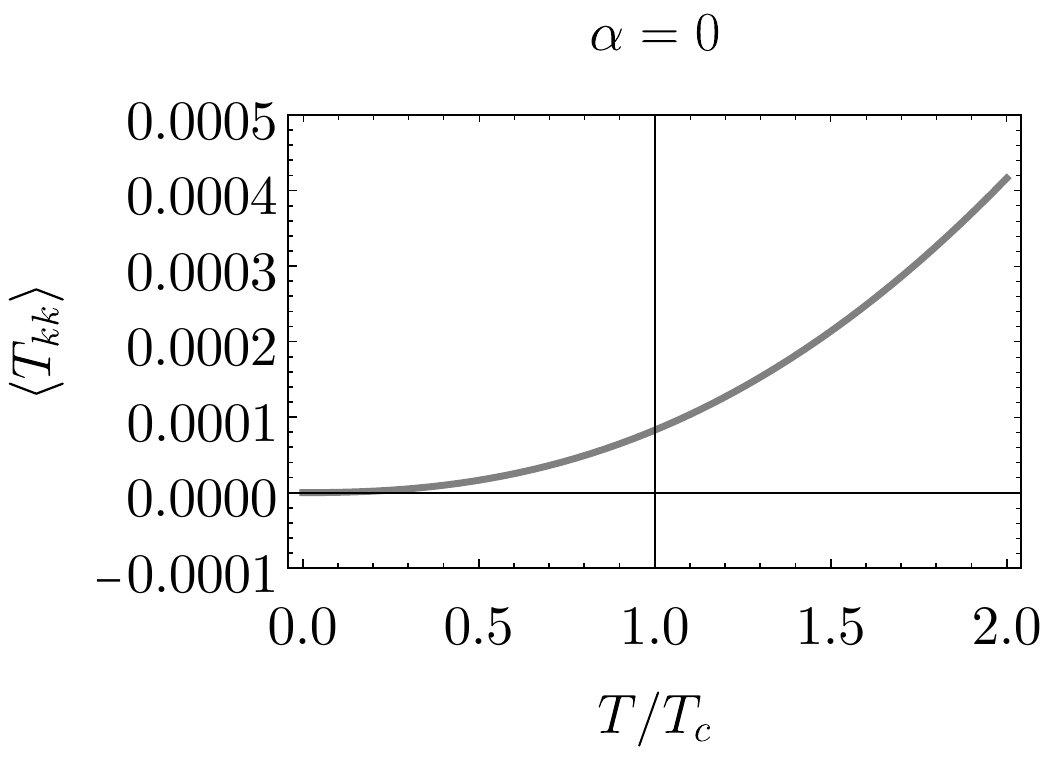}
	 \caption{$\langle T_{kk}\rangle$ as function of temperature for different values of $\alpha$.}
	\label{fig:Tkk}
	\end{center}
\end{figure}

We recall that $\alpha$ determines the thermodynamic phase structure of the model and can be tuned to realize either a first or second order phase transition or a smooth crossover. Like for EE  we study now QNEC$_2$ in specific examples for each of these cases.
It is useful to first look at the left hand side of QNEC$_2$ \eqref{eq:QNEC2} before comparing it with the right hand side.
In Figure~\ref{fig:Tkk} we show $\langle T_{kk}\rangle$ as function of $T$ for our three different values of $\alpha$.
As expected $\langle T_{kk}\rangle$ displays features that are characteristic for the specific kind of phase transitions in the respective model.
Since $\langle T_{kk}\rangle$ is a positive, but not necessarily single-valued function of $T$, all examples we consider satisfy in addition to QNEC$_2$ also the classical null energy condition.

For $\alpha=0.32$ (left plot in Figure~\ref{fig:Tkk}) $\langle T_{kk}\rangle$ is multi-valued around the critical temperature $T_c$, where it has a finite jump. Thermodynamically unpreferred branches are dashed and preferred ones solid. For $T<T_c$ the energy momentum tensor is close to the vacuum value $\langle T_{kk}\rangle=0$. The `critical' state at $T=T_c$ requires special attention. As shown in \cite{Janik:2017ykj,Attems:2019yqn,Bea:2020ees} there exists an infinite number of degenerate phases at $T=T_c$, which on the gravity side have inhomogenous horizons and can be seen as a mixture of small and large black branes. We expect this to also be the case in our two-dimensional model. Because we limit our study to homogeneous states, we cannot make precise statements about EE and QNEC$_2$ at $T=T_c$, only that they will inherit the inhomogeneous structure of the bulk geometry.

In the middle of Figure~\ref{fig:Tkk} we plot $\langle T_{kk}\rangle$ for $\alpha=0.16$ where the system has a second order phase transition. In this case $\langle T_{kk}\rangle$ is a single-valued, continuous and monotonic function of $T$. For $T<T_c$ $\langle T_{kk}\rangle$ remains close to zero and the transition between UV- and IR-scaling regime is tightly localized around $T=T_c$. For $T>T_c$, i.e., in the large black hole branch $\langle T_{kk}\rangle$ depends strongly on $T$ and approaches at large $T$ the universal form given by \eqref{eq:t4}.

Finally in the right plot of Figure~\ref{fig:Tkk} we show $\langle T_{kk}\rangle$ for $\alpha=0$. We recall that we defined $T_c\approx 0.01$ as the temperature where the speed of sound has a local minimum. Here the curve does not have any distinct features at $T=T_c$ and the transition between UV- and IR-regime is not localized.

\begin{figure}[htb]
	\begin{center}
	\includegraphics[width=0.75\textwidth]{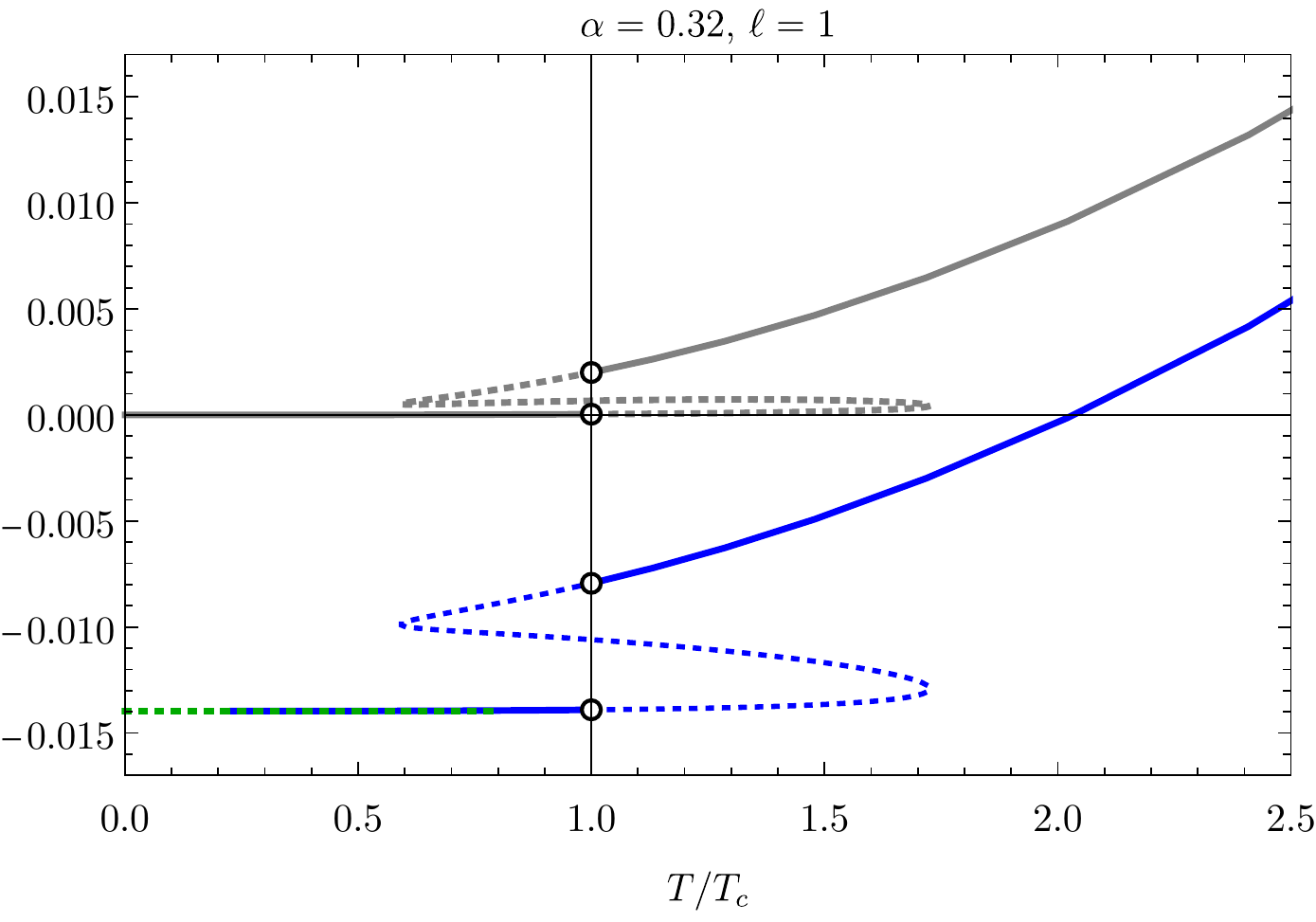}
		\caption{QNEC$_2$ as function of $T$ for black brane solution with $\alpha=0.32$ and $\ell=1$. Blue lines show the QNEC$_2$ expression $\frac{1}{2\pi}\left(S''+\tfrac{6}{c}(S')^2\right)$ and gray lines show $\langle T_{kk}\rangle$, where solid (dashed) parts of the curves correspond to thermodynamically stable (unstable) phases. 
		In addition, the dashed green curve shows the value of QNEC$_2$ of the corresponding $T=0$ domain wall solution.}
	\label{fig:QNECblackBrane11}
	\end{center}
\end{figure}

Let us now discuss the results for QNEC$_2$. We begin with $\alpha=0.32$ for which the theory has a first order phase transition at $T=T_c$ and QNEC$_2$ as function of $T$ has a jump. If we choose $\ell$ in the range $3.870<\ell<4.058$ there is, like for domain walls, an additional discontinuity due to the existence of multiple RT-surfaces. In Figure~\ref{fig:QNECblackBrane11} we show the results for $\ell=1$, which is well below this range. Blue and gray lines are left and right hand side of QNEC$_2$ and solid (dotted) lines indicate thermodynamically preferred (unpreferred) solutions.
For $T<T_c$, i.e., in the small black brane branch, QNEC$_2$ grows only very slowly with $T$ and remains very close to the $T=0$ limit given by the domain wall result shown as dashed green line.

\begin{figure}[htb]
	\begin{center}
	\includegraphics[width=0.75\textwidth]{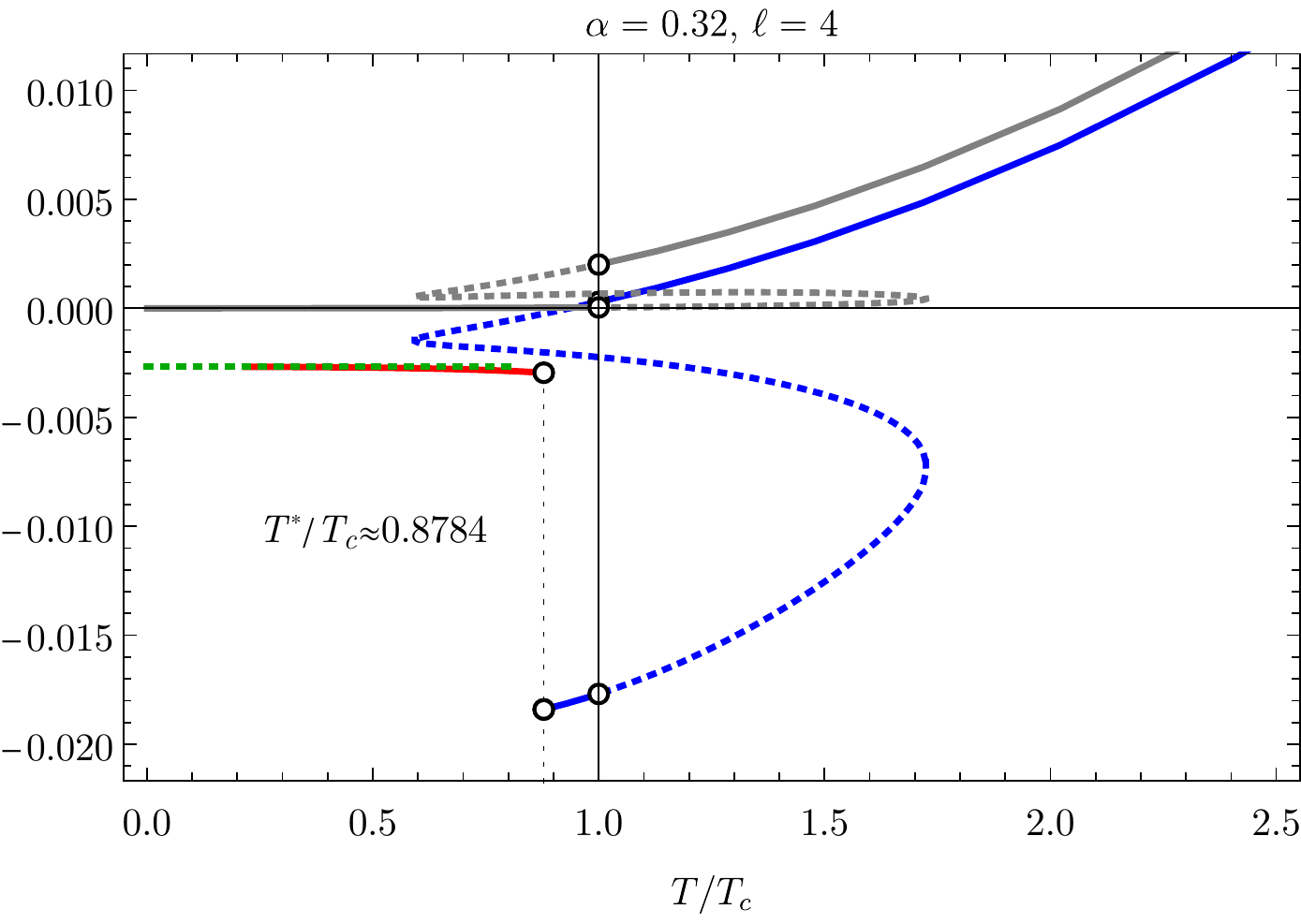}
		\caption{QNEC$_2$ as function of $T$ for black brane solution with $\alpha=0.32$ and $\ell=4$.
		 Blue and red lines show the QNEC$_2$ expression $\frac{1}{2\pi}\left(S''+\tfrac{6}{c}(S')^2\right)$ and gray lines show $\langle T_{kk}\rangle$, where solid (dashed) parts of the curves correspond to thermodynamically stable (unstable) phases. Red and blue segments correspond to different branches of RT surfaces that dominate at small and large separations, respectively.
		In addition, the dashed green curve shows the value of QNEC$_2$ of the corresponding $T=0$ domain wall solution.}
	\label{fig:QNECblackBranel4}
	\end{center}
\end{figure}

In Figure~\ref{fig:QNECblackBranel4} we show the case $\ell=4$ in which QNEC$_2$ has an especially rich structure.
This example has in addition to the phase transition at $T=T_c$ another transition at $T=T^\ast\approx 0.8784\, T_c$ due to the existence of multiple RT-surfaces. Here the two saddle points for the RT-surfaces displayed in red and blue exchange dominance which leads to a jump in QNEC$_2$. From the analysis in section \ref{sec:1.5} we know there is actually also a negative delta function which we indicate by the dashed black line. Curiously, QNEC$_2$ is not a monotonic function of $T$ in this case. Starting from small $T$ where it nicely fits the corresponding domain wall result shown as dashed green curve, QNEC$_2$ decreases monotonically until it falls at $T=T^\ast$ to the smaller value of the blue branch. This short blue branch is then monotonically increasing until the jump at $T=T_c$. Following the blue curve at $T>T_c$ QNEC$_2$ grows monotonically and approaches \eqref{eq:t7} in the limit of large $T$. We emphasize that the two critical temperatures $T^\ast$ and $T_c$ are different in nature. The former is related to the existence of multiple RT surfaces, while the latter is related to the thermal phase transition. Additionally we find $T^\ast<T_c$, because multiple RT surfaces only exist in the small black brane branch with $T<T_c$.

\begin{figure}[htb]
	\begin{center}
	\includegraphics[width=0.75\textwidth]{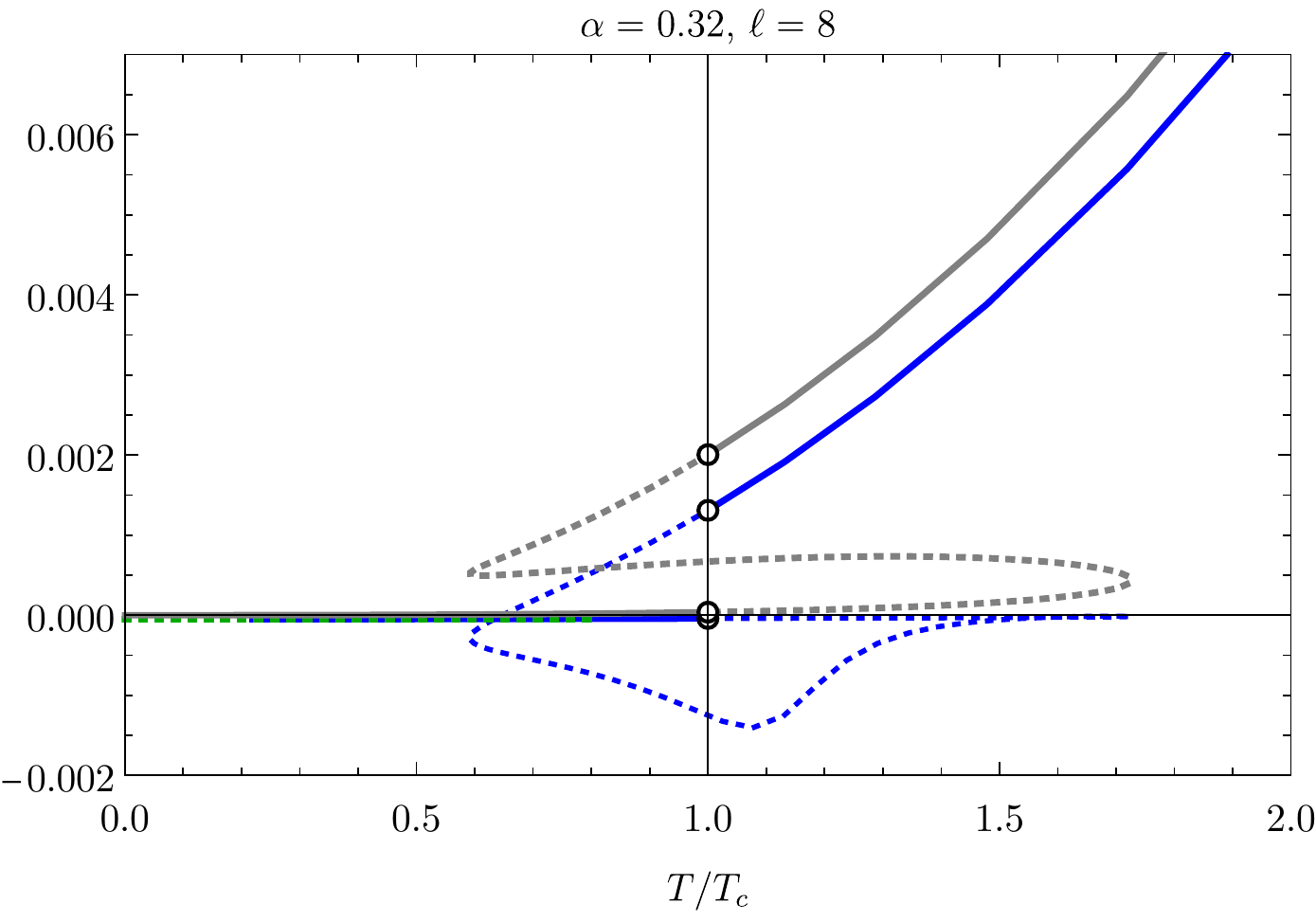}
		\caption{QNEC$_2$ as function of $T$ for black brane solution with $\alpha=0.32$ and $\ell=8$. 
		Blue lines show the QNEC$_2$ expression $\frac{1}{2\pi}\left(S''+\tfrac{6}{c}(S')^2\right)$ and gray lines show $\langle T_{kk}\rangle$, where solid (dashed) parts of the curves correspond to thermodynamically stable (unstable) phases. 
		In addition, the dashed green curve shows the value of QNEC$_2$ of the corresponding $T=0$ domain wall solution.
		}
	\label{fig:QNECblackBrane18}
	\end{center}
\end{figure}

The third case we analyze is $\ell=8$ shown in Figure~\ref{fig:QNECblackBrane18}.
This case is again outside (this time above) the regime where multiple RT-surfaces exist. Hence there is only the discontinuity at the thermal phase transition. Curiously in the thermodynamically disfavoured branch (dashed line) QNEC$_2$ has two self-intersections, one at $T<T_c$ and another (barely visible) close to the turning point at $T>T_c$. At these intersections QNEC$_2$ takes the same value in two different phases which makes QNEC$_2$ not a good probe to distinguish different phases. As one can deduce from Figure~\ref{EE-1st-PT} (right) for $\ell=8$ we are for $T<T_c$ always in the non-gapped plateau like IR-scaling regime where QNEC$_2$ is almost saturated. This is different to the aforementioned case $\ell=4$ in which QNEC$_2$ for $T<T_c$ is in the gapped UV-scaling regime and not close to saturation.
For the cases $\alpha=0.16$ and $\alpha=0$ QNEC$_2$ is a monotonic and continuous function of both $T$ and $\ell$.

\begin{figure}
    \centering
    \includegraphics[height=0.195\textheight]{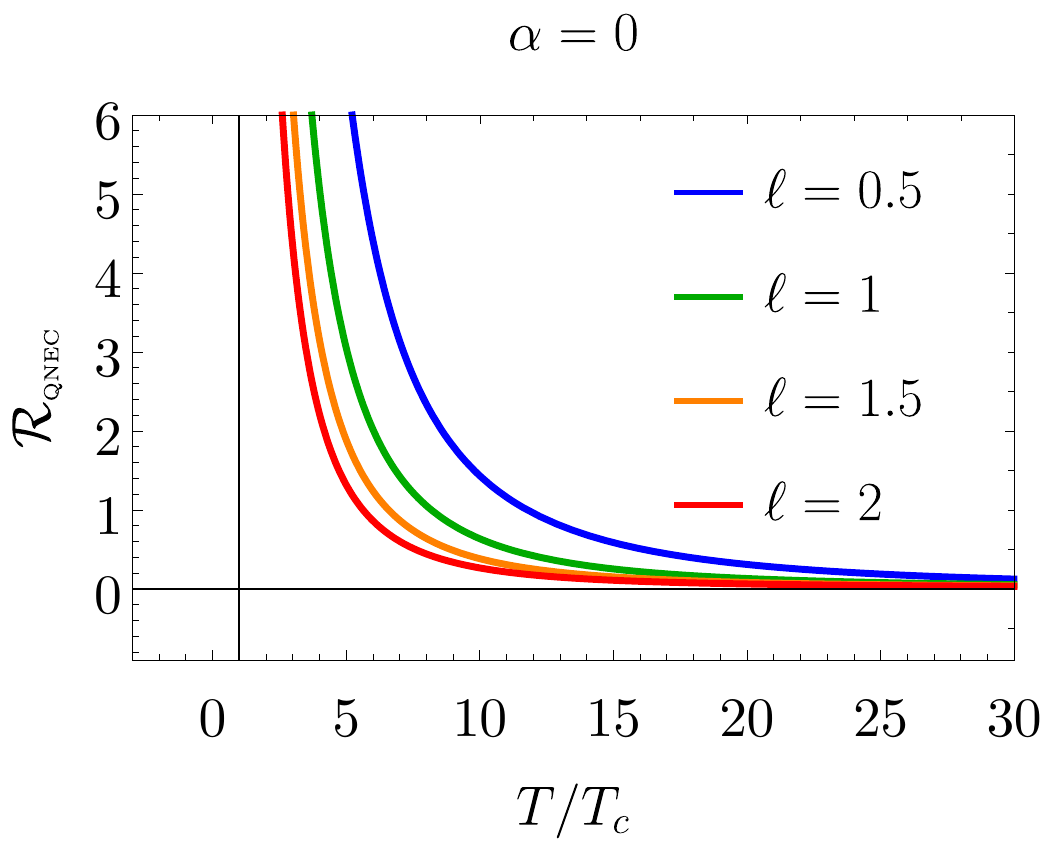}\,
    \includegraphics[height=0.195\textheight]{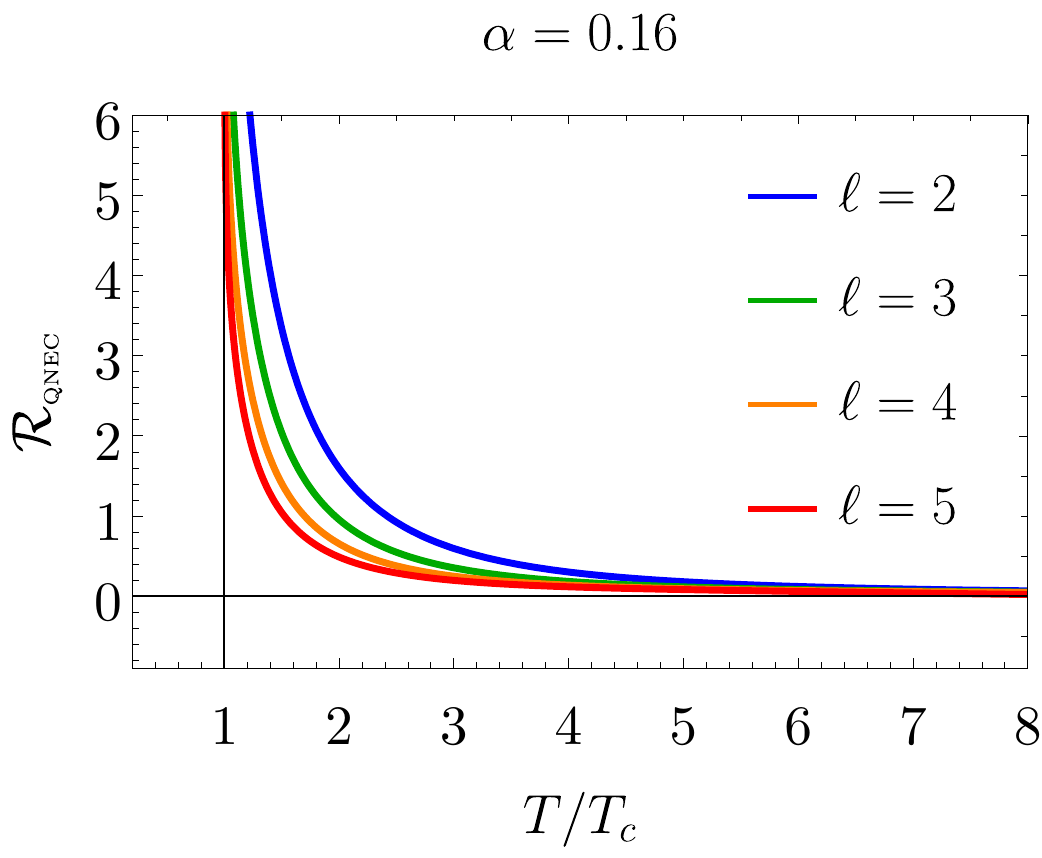}\,
    \includegraphics[height=0.195\textheight]{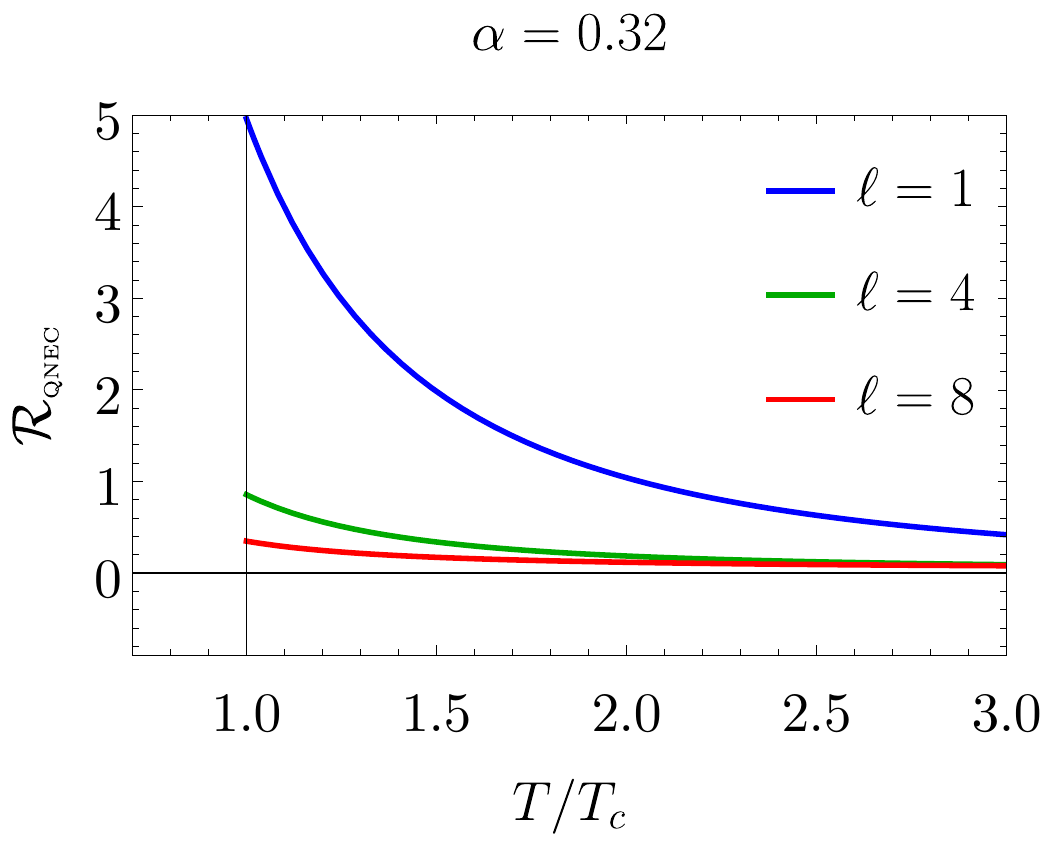}
    \caption{QNEC non-saturation ratio ${\cal R}_{\textrm{\tiny QNEC}}$ as function of temperature $T>T_c$ for various values of interaction strength $\alpha$ and interval length $\ell$}
    \label{fig:R}
\end{figure}

Finally, we quantify the QNEC non-saturation by plotting in Figure~\ref{fig:R} the dimensionless ratio
\eq{
{\cal R}_{\textrm{\tiny QNEC}} := 1 - \frac{S''+\frac6c\,(S^\prime)^2}{2\pi\,\langle T_{kk}\rangle}
}{eq:RQNEC}
for various values of $\ell$ and the three values of $\alpha$ discussed above, for temperatures greater than the critical. When QNEC$_2$ saturates ${\cal R}_{\textrm{\tiny QNEC}}$ vanishes and otherwise is positive, so this quantity is a good measure for how far we are from saturation. As evident from the plots, at large temperature the ratio ${\cal R}_{\textrm{\tiny QNEC}}$ tends to zero, which means that QNEC is approximately saturated at large temperature. We see this analytically by comparing with \eqref{eq:t8} which also yields a vanishing ratio in the limit of small $\epsilon$.

%For completeness we show the cases $\alpha=0.16$ and $\alpha=0$ in Figure~\ref{fig:QNECblackBrane}. In both cases QNEC$_2$ is a monotonic and continuous function of both $T$ and $\ell$.
%\begin{figure}[htb]
%	\begin{center}
%	\includegraphics[height=0.25\textheight]{QNEC2ndOrderPT.pdf}\quad	\includegraphics[height=0.25\textheight]{QNEC2Xover.pdf}
%		\caption{QNEC$_2$ for the black brane solution with $\alpha=0.16$ (left) and for $\alpha=0$ (right) for different values of $\ell$.}
%	\label{fig:QNECblackBrane}
%	\end{center}
%\end{figure}

%%%%%%%%%%%%%%%%%%%%%%%%
\section{Conclusion}\label{sec:Summary}
%%%%%%%%%%%%%%%%%%%%%%%%

After summarizing in section \ref{sec:1} general aspects of QNEC$_2$ in deformed holographic CFTs we focused on a specific example from section \ref{sec:model} onwards, namely a deformed CFT$_2$ dual to Einstein gravity with a massive scalar field \eqref{eq:action} and asymptotically AdS$_3$ boundary conditions. For sake of specificity we fixed the conformal weight of the operator dual to the scalar field to $\Delta=\frac32$ and considered a one-parameter family of potentials \eqref{eq:angelinajolie}, which led already to a rich phase structure, with crossovers, second- and first-order phase transitions that we analyzed in detail in sections \ref{sec:thermo}-\ref{sec:QNEC} using QNEC$_2$ (and associated quantities like EE and the Casini--Huerta $c$-function) as diagnostic tool. For small and large intervals we were able to provide closed-form expressions, while for intermediate intervals we relied on numerics.

Unexpectedly our numerical studies reveal that QNEC$_2$ in ground states displays certain features that are characteristic for the thermodynamic phase structure of the theory. For example, if ground state QNEC$_2$ is non-monotonic in $\ell$, then the theory has a first order phase transition at finite temperature.
Furthermore, our examples show that QNEC$_2$ as function of $T$ can have a very rich and non-trivial structure that includes multiple discontinuities caused by thermal phase transitions and multiple RT surfaces. Using QNEC$_2$ as tool for detecting phase transitions is similar in spirit to using entanglement as a probe of confinement \cite{Klebanov:2007ws}, but the type of phase transition discussed in our work differs from the confinement-deconfinement phase transition

We address now some generalizations of our results. Even without changing the model one could consider more general states than the ones considered in the present work, by dropping the assumption of stationarity and/or translation invariance. In full generality this can be done only numerically, but the numerical routines to determine QNEC$_2$ are typically not hard to implement and computationally not very demanding. 

The most straightforward model modification is to change the mass of the scalar field and the potential. In this way one could scan the model space and study QNEC$_2$ as function of the conformal weight $\Delta$ and additional parameters in the potential, which may reveal novel features. A particularly interesting set of examples would be critical states at $T=T_c$ with first order phase transition \cite{Janik:2017ykj}, which we did not analyze in this paper. Another extension would be to consider more general potentials related to exotic holographic RG flows classified in \cite{Kiritsis:2016kog}. 

Adding a Maxwell field leads to a wider class of models with even richer phase structure and phenomenology, like holographic superconductors \cite{Gubser:2008px,Hartnoll:2008vx} or other holographic condensed matter models \cite{Charmousis:2010zz}, that can again be analyzed along the lines of the present work, using QNEC$_2$ as diagnostic tool.  Investigating the effects of hyperscaling violation \cite{Gouteraux:2011ce,Huijse:2011ef} and breaking of translation invariance \cite{Andrade:2013gsa} from a QNEC$_2$ perspective could be rewarding as well.

It could also be of interest to venture beyond the supergravity approximation and consider $1/c$ corrections to bulk and boundary theories, see section 5 in \cite{Ecker:2019ocp} and refs.~therein for such corrections to QNEC$_2$.

Finally, it will be interesting to consider holographic correspondences beyond asymptotically AdS$_3$/deformed CFT$_2$. There are two different types of generalizations (which can also be combined): either one keeps the dimension, but relaxes the asymptotic AdS$_3$ behavior, e.g.~by considering flat space holography and the associated generalization of QNEC$_2$ \cite{Grumiller:2019xna}, or one goes to higher dimensions and uses QNEC \eqref{eq:QNEC} instead of QNEC$_2$ as diagnostic tool for phase transitions and other phenomenological aspects.

\section*{Note added}

While finishing our manuscript reference \cite{Baggioli:2020cld} appeared on the arXiv, which uses a similar logic as in section \ref{sec:predict} to identify quantum phase transitions. Instead of QNEC they use a generalized version of the Casini--Huerta $c$-function \cite{Chu:2019uoh} as diagnostic tool.

%%%%%%%%%%%%%%%%%%%%%%%%
\section*{Acknowledgement}
%%%%%%%%%%%%%%%%%%%%%%%%

This work was supported by the Austrian Science Fund (FWF), projects P~28751-N27, P~30822-N27, P~32581-N27 and W~1252-N27. 
CE was in addition supported by the Delta-Institute for Theoretical Physics (D-ITP) that is funded by the Dutch Ministry of Education, Culture and Science (OCW). HS was partly supported by Guangdong Major Project of Basic and Applied Basic Research No. 2020B0301030008, the National Natural Science Foundation of China with Grant No.12035007, Science and Technology Program of Guangzhou No. 2019050001.

\begin{appendix}
%%%%%%%%%%%%%%%%%%%%%%%%
\section{Near-boundary analysis and UV/IR relation}\label{section:near-bdry}
%%%%%%%%%%%%%%%%%%%%%%%%

Like in the main text we fix the conformal weight of the operator dueal to the scalar field to $\Delta=\frac32$. The metric functions in Fefferman--Graham gauge
\begin{equation}
\extd s^2=\frac{\extd z^2-F(z)\extd t^2+G(z)\extd x^2}{z^2}
\end{equation}
are expanded near the boundary as (generalized) Taylor--Maclaurin series of the radial coordinate $z$.
\begin{equation}
F(z)=\sum_{i=0}^\infty f_{i}\, z^i\qquad\qquad G(z)=\sum_{i=0}^\infty g_{i}\, z^i\qquad\qquad \phi(z)=z^{1/2}\sum_{i=0}^\infty \phi_{i}\, z^i
\end{equation}
The leading order coefficients are given by $f_{0}=g_{0}=1$ while $\phi_0=\source$ is the source of the scalar field. The one-point functions are related to the normalizable modes $f_2$ and $\phi_1$. Using on-shell conditions the first few terms are given by
\bea
&&F(z )=1-\frac{\phi_0^2}{4}z+f_2 z ^2 +\mathcal{O}(z^3)\\
&&G(z )=1-\frac{\phi_0^2}{4}z-\Big(\frac{\phi_0^4}{16}+\frac{3\phi_0\phi_1}{4}+f_2\Big)\,z^2 +\mathcal{O}(z^3)\\
&&\phi (z )=z^{1/2}\left(\phi_0+\phi_1z +\mathcal{O}(z^{2})\right)\,.
\eea

By contrast, the metric functions in Gubser gauge $\phi(r)=j^{2-\Delta}r$ where we set $\source = 1$ from now on
\bea
\extd s^2=e^{2A}\left(-H \extd t^2+\extd x^2\right)+e^{2B}\,\frac{\extd r^2}{H}
\eea
have asymptotic expansions involving also logarithmic terms.
\eq{
A(r)=-2\,\log r+\sum_{i=0}^\infty a_{2i}\, r^{2i}  \qquad B(r)=\log\frac{2}{ r} +\sum_{i=0}^\infty b_{2i}\, r^{2i} \qquad H(r)=\sum_{i=0}^\infty h_{2i}\, r^{2i}
}{eq:nolabel}
The non-normalizable modes are fixed as $a_{0}= b_{0}=0$, $h_{0}=1$. The expectation value of the boundary stress tensor and the operator dual to the scalar field are related to the normalizable modes $a_2$ and $h_4$. Using on-shell conditions the first few terms are given by
\bea
&&A(r)=-2\,{\log r}+a_{2} r^2
+\mathcal{O}\left(r^4\right)\\
&&B(r)=\log\frac{2}{r} -\Big(a_2+\frac{1}{8}\Big) r^2
+\mathcal{O}\left(r^4\right)\\
&&H(r)=1+h_4r^4\Big(1-\frac{1+24a_2}{12}\,r^2+\mathcal{O}\left(r^4\right) \Big)\,.
\eea

In Gubser gauge the UV and IR data are related as follows. Knowing the functions $A$ and $B$ one can integrate \eqref{eq:EOM3} to find
\be
H(r)=H(r_0)+C \int\limits_{r_0}^{r}e^{-2A(r')+B(r')}\extd r'
\label{eq:H}
\ee  
where $H(r_0)$ and $C$ are constants of integration. By choosing $r_0=r_h$ and noting that $H$ is the blackening function the expression \eqref{eq:H} reduces to 
\be
H(r)=C \int_{r_h}^{r}e^{-2A(r')+B(r')}\extd r'
\ee  
Differentiating with respect to $r$ and evaluating at the horizon yields
\be\label{eq:Hp}
H'(r_h)=C \, e^{-2A(\phi_h)+B(\phi_h)}\,.
\ee
Comparing equations \eqref{eq:Hp} and \eqref{eq:EOM1} at the horizon determines the integration constant. 
\be
C=e^{2A(\phi_h)+B(\phi_h)}V'(r_h)=-16 \pi \GN\, s\, T
\label{eq:C}
\ee
Finally, expanding both sides in equation \eqref{eq:Hp} near the boundary yields $C=2h_4$, which
combined with \eqref{eq:C} establishes a relation between boundary and horizon data.
\be
h_4=\frac{1}{2}e^{2A(\phi_h)+B(\phi_h)}V'(r_h)
\ee

%%%%%%%%%%%%%%%%%%%%%%%%%%%%%%%%%%%%%%%%
\section{Holographic renormalization}\label{appendix-HRG}
%%%%%%%%%%%%%%%%%%%%%%%%%%%%%%%%%%%%%%%%

In general holographic renormalization is based on the details of the theory, namely the conformal weight of the scalar field and the scalar potential.  As in the rest of our work we fix $\Delta=\frac32$. It turns out that the counter-terms are
\bea
I_{\textrm{ct}}&=&\frac{1}{16\pi \GN^2}\int_{\partial M}\extd^2x\sqrt{\gamma}\,\Big[\ell_{\textrm{\tiny{AdS}}} \log(\rho)\, R+\frac{W}{\ell_{\textrm{\tiny{AdS}}}} \Big]  \\
&=&\frac{1}{16\pi \GN^2}\int_{\partial M}\extd^2x \sqrt{\gamma}\,\Big[{\ell_{\textrm{\tiny{AdS}}}}\log(\rho)\,R -\frac{2}{\ell_{\textrm{\tiny{AdS}}}}-\frac{1}{4\ell_{\textrm{\tiny{AdS}}}}\,\phi^2-\frac{\alpha}{8\ell_{\textrm{\tiny{AdS}}}}\, \phi^4\Big]\nn
\eea
where $\gamma$ is the determinant of the boundary metric and in the second line we use the superpotential \eqref{eq:superpotential}. Using the superpotential as counter-term fixes the ambiguous coefficient of the finite $\phi^4$ term to the unique value that gives zero free energy for the ground state dual to the domain wall geometry. The near boundary solution for static and stationary solutions with fixed AdS is given in appendix \ref{section:near-bdry}.

The one point functions are given as functional derivatives of the generating functional
\be
\langle {\mathcal{O}_{\phi}} \rangle = \lim_{z\rightarrow0} \frac{z^{-\Delta }}{\sqrt{-\gamma}}\frac{\delta \Gamma}{\delta\phi}\qquad\qquad
\langle T_{ij} \rangle =2 \lim_{z\rightarrow0} \frac{z^{-1}}{\sqrt{-\gamma}}\frac{\delta\Gamma}{\delta\gamma^{ij}}\label{EMT}
\ee
with the holographically renormalized action
\be
\Gamma=I_{\textrm{\tiny{bulk}}}+I_{\textrm{\tiny{GHY}}}+I_{\textrm{\tiny{ct}}}
\ee
that consists of the bulk action $I_{\textrm{\tiny{bulk}}}$, the Gibbons--Hawking--York boundary term $I_{\textrm{\tiny{GHY}}}$ and the holographic counter-term $I_{\textrm{\tiny{ct}}}$.

The near boundary expansion from appendix \ref{section:near-bdry} yields the one point functions 
\begin{align}
\langle {\mathcal{O}_{\phi}} \rangle &=\frac{1}{8\pi\GN}\,\Big(\frac{1}{2}\phi_{1}- \frac{\alpha}{4} {\phi_0^3} \Big)\label{vevPhi-1}\\
\langle T_{tt} \rangle &=\frac{1}{8\pi\GN}\,\Big(-2 f_{2}+\frac{1+2\alpha}{16}\phi_0^4- \phi_0 \phi_{1}\Big)\label{EMT-tt}\\
\langle T_{xx} \rangle &=\frac{1}{8\pi\GN}\,\Big(-2f_{2}+\frac{1-2\alpha}{16}\phi_0^4-\frac{1}{2}\phi_0 \phi_{1}\Big)\label{EMT-xx}
\end{align}
These one-point functions respect the anticipated Ward identity.
\be
\langle T^{i}_{~i} \rangle=\phi_0\, \langle \mathcal{O}_{\phi} \rangle\label{ward}
\ee
The fact that the trace of the energy-momentum tensor is non-zero in general is due to the breaking of conformal symmetry in the presence of a dimensionful source.

%%%%%%%%%%%%%%%%%%%%%%%%%%%%%%%%%%%%%%%%%%
\section{Evaluation of typical \texorpdfstring{QNEC$_2$}{QNEC2} integrals for large intervals} \label{app:int}
%%%%%%%%%%%%%%%%%%%%%%%%%%%%%%%%%%%%%%%%%%

We follow here the discussion in section \ref{sec:1.4} and evaluate the integrals in the limit of large intervals, $\ell\gg 1$, for various background geometries. 

We start with domain wall solutions, where we can set the null deformation parameter to zero, $\lambda=0$, since the dual QFT state is boost invariant. As stated in the main text, for Case 0 we parametrize the turning point as $\rho_\ast=\ln\delta$ with small positive $\delta$, where $\rho$ refers to the domain wall radial coordinate \eqref{eq:metricDomainWall}. The spatial integral \eqref{eq:i12} simplifies to
\eq{
\ell\,e^{A_\ast} = 2\int\limits_{\ln\de}^\infty\extd\rho\,\Big(\frac{1}{\sqrt{1-y}}-\sqrt{1-y}\Big)
}{eq:app2}
where $y$ is defined in \eqref{eq:y}. The renormalized area integral \eqref{eq:i15} yields
\eq{
{\cal A}_{\textrm{\tiny{ren}}}(\ell) = - 2\rho_\ast + 2\int\limits_{\ln\de}^\infty\extd\rho\,\Big(\frac{1}{\sqrt{1-y}}-1\Big)\,.
}{eq:app3}
The two integrals above can be converted into compact integrals
\eq{
2\int\limits_{\rho_\ast}^\infty\extd\rho\,f(y)=\int\limits_0^1\extd y\,\frac{f(y)}{y\,\frac{\extd A(\rho(y))}{\extd\rho}}
}{eq:app4}
which require the knowledge of $\rho$ as function of $y$. For Case 0 this relation reads
\eq{
\rho=\rho_\ast-\frac18\,e^{-\rho_\ast}-\frac12\,\ln y+W\big(\tfrac18\,\sqrt{y}\,\exp(-\rho_\ast+\tfrac18\,e^{-\rho_\ast})\big)
}{ae:app5}
with the Lambert-$W$ function. Its expansion $W(\exp(x))=x-\ln x+\frac{\ln x}{x}+{\cal O}(\tfrac{\ln^2x}{x^2})$ permits to evaluate the two integrals above perturbatively in $\de$.
\begin{subequations}
\label{eq:app7}
\begin{align}
&2\int\limits_{\ln\de}^\infty\extd\rho\,\Big(\frac{1}{\sqrt{1-y}}-\sqrt{1-y}\Big) = 16\de -128\ln(2)\,\de^2 +  {\cal O}(\de^3) \\
&2\int\limits_{\ln\de}^\infty\extd\rho\,\Big(\frac{1}{\sqrt{1-y}}-1\Big) = 16\ln(2)\,\de + \frac{16}{3}\,\big(\pi^2-24\ln 2-12\ln^2 2\big)\,\de^2 +  {\cal O}(\de^3)
\end{align}
\end{subequations}
The first one yields a result for the small parameter $\de$ in terms of the interval $\ell$,
\eq{
\de = \frac{1}{8\ln\frac{\ell}{16}} + \dots
}{eq:app6}
where the ellipsis denotes subleading terms. The second one together with the term $-2\rho_\ast$ in \eqref{eq:app3}, when multiplied by $\frac c6$, yields the result for renormalized EE displayed in the main text \eqref{eq:S0large}.

Case I is qualitatively different from Case 0 since there is a finite lower bound on the radial coordinate, $\ln\alpha$. For large intervals the turning point is close to this lower bound, so as explained in the main text we parametrize it as $\rho_\ast=(1-\delta)\,\ln\alpha$, again with small and positive $\delta$. Both integrals \eqref{eq:app7} are now order ${\cal O}(\de^2)$, so that the leading contribution to area comes from the term $-2\rho_\ast$. Plugging $A_\ast=1/(16 \alpha \ln(\alpha)\,\de)$ into equation \eqref{eq:app2} allows to express $\de$ in terms of $\ell$.
\eq{
\delta = -\frac{1}{16\alpha\ln(\alpha)\,\ln\ell} + \dots
}{eq:app9}
The renormalized area
\eq{
{\cal A}_{\textrm{\tiny{ren}}} = -2\rho_\ast + \dots = -2\ln\alpha - \frac{1}{8\alpha\,\ln\ell} + \dots
}{eq:app10}
multiplied by $\frac c6$ then leads to the result \eqref{eq:SIlarge} for EE, where we have displayed one additional subleading term.

Case II can be treated most easily in the limit of large $|\alpha|$. Like in Case 0 the radial coordinate is unbounded, so we parametrize again $\rho_\ast=\ln\delta$ with small positive $\delta$. In the large $|\alpha|$ limit the function $A(\rho)$ is linear in the radial coordinate,
\eq{
A(\rho) = \Big(1-\frac{1}{16\alpha}\Big)\,\rho + \frac{1+\ln(-\alpha)}{16\alpha} + {\cal O}(\alpha^{-3})
}{eq:app11}
which means that the measure factor in the right hand side of \eqref{eq:app4} is rather trivial and yet significant, since it provides the ratio between IR and UV values of the central charge. The simplicity of the measure factor allows to perform all integrals in closed form and yields the result \eqref{eq:SIIlarge} in the main text. It turns out that the same result is true at small $|\alpha|$.

Finally, we consider the three relevant integrals for the black brane in the large $\ell$ limit, where the backreaction calculation discussed in section \ref{sec:QNECfiniteT} applies.

The deformed entangling region integral [$A_\ast=A(z_\ast)$ and $S_\ast=S(z_\ast)$]
\eq{
\frac{\ell+\lambda}{2}=\int\limits_0^{(\ell+\lambda)/2}\extd x  = \int\limits_{z_\ast}^0\frac{\extd z}{\sqrt{\Lambda^2 S^2(z)-S(z)A(z)+\frac{S^2(z)A(z)}{z^2}\,z^2_\ast\,\Big(\frac{1}{S_\ast}-\frac{\Lambda^2}{A_\ast}\Big)}}
}{eq:bh8}
the time-shift integral
\eq{
\frac\lambda2=\int\limits_0^{\lambda/2}\extd t  = \Lambda\,\int\limits_{z_\ast}^0\frac{\extd z}{A(z)\,\sqrt{\Lambda^2-\frac{A(z)}{S(z)}+\frac{A(z)}{z^2}\,z^2_\ast\,\Big(\frac{1}{S_\ast}-\frac{\Lambda^2}{A_\ast}\Big)}} \,.
}{eq:bh9}
and the renormalized area integral ($z_{\textrm{\tiny{cut}}}$ tends to $+0$ when the cutoff is removed)
\eq{
{\cal A}_{\textrm{\tiny{ren}}} = 2z_\ast\sqrt{\frac{1}{S_\ast}-\frac{\Lambda^2}{A_\ast}}\, \int\limits_{z_\ast}^{z_{\textrm{\tiny{cut}}}}\frac{\extd z}{z^2\,\sqrt{\Lambda^2-\frac{A(z)}{S(z)}+\frac{A(z)}{z^2}\,z^2_\ast\,\Big(\frac{1}{S_\ast}-\frac{\Lambda^2}{A_\ast}\Big)}} -2\ln z_{\textrm{\tiny{cut}}}
}{eq:bh11}
for large $\ell$ all lead to essentially two types of integral kernels.
\begin{subequations}
\label{eq:int1}
\begin{align}
I_1[h(y);\,\De] &= \int\limits_0^1\extd y\,\frac{h(y)}{\sqrt{1-y}(1-y+\De\,y)^{3/2}} \\
I_2[h(y);\,\De] &= \int\limits_0^1\extd y\,\frac{h(y)}{\sqrt{1-y}(1-y+\De\,y)^{5/2}}
\end{align}
\end{subequations}
Here $h(y)$ is a function that is continuous in the interval $[0,1]$ and Taylor expandable around $y=1$, while $\De=2\de-\de^2$ is a small parameter. The integration variable $y$ is related to our original radial coordinate $z$ by the simple coordinate transformation $y=z^2/(1-\de)^2$.

We evaluate such integrals perturbatively in $\De$, displaying one more order than we need. 
\begin{align}
 I_1[h(y);\,\De] &= \frac{2h(1)}{\De} + h'(1)\,\ln\frac{\De}{4} + 2h'(1) + {\cal I}_1[h(y)] + {\cal O}(\De\,\ln\De)\\
 I_2[h(y);\,\De] &= \frac{4h(1)}{3\De^2} + \frac{2h(1)-2h'(1)}{3\De} -\frac12h''(1)\ln\frac{\De}{4} -\frac43h''(1) + {\cal I}_2[h(y)] + {\cal O}(\De\ln\De)
\end{align}
The ${\cal O}(1)$ integrals are defined as ($h^{(m)}$ denotes the $m^{\textrm{th}}$ derivative of $h$)
\eq{
{\cal I}_n[h(y)] = \int\limits_0^1\extd y\,\frac{h(y)-\sum_{m=0}^n \frac{(y-1)^m}{m!}\,h^{(m)}(1)}{(1-y)^{n+1}}\,.
}{eq:int2}
In the rest of the appendix we apply these formulas to the three integrals mentioned above.

The time-shift integral \eqref{eq:bh9} yields
\eq{
\la = \frac{\La}{\de}\,\Big(1 + \frac{\eps^2}{4}\,\big(3A_2'(1)+2S_2(1)+S_2'(1)\big) \Big)\,\big(1+{\cal O}(\de\,\ln\de)+{\cal O}(\eps^3)\big)
}{eq:int3}
where we assumed $A_2(1)=0$ since we keep the horizon fixed at $z=1$. In this appendix prime denotes derivative with respect to $y$ (and not with respect to $z$, as it does in the main text). Moreover, we set $A_2'(1)=0$ since we also keep the temperature fixed. Solving \eqref{eq:int3} for $\La$ shows that it scales linearly in $\lambda$ and, to leading order, linearly in $\de$. The entangling region interval \eqref{eq:bh8} allows to solve $\de$ in terms of $\ell$, $\lambda$ and $\eps$. 
\eq{
\delta = 2e^{-\ell}\,\Big(1+\eps^2 \ell\,f - \lambda\,\big(1+\eps^2\,(\ell-1)\,f\big) + \frac34\,\lambda^2\,\big(1+\eps^2(\ell-2)\,f\big)\Big)+\dots
}{eq:int5}
with
\eq{
f := \frac{S_2'(1)}{4} - \frac{S_2(1)}{2}\,.
}{eq:int4}
We deduce from \eqref{eq:int5} that each power of $\de$ is suppressed at large $\ell$ by an instanton-like factor. Note that the suppression by $e^{-\ell}$ is compatible with general expectations and precisely agrees with the exponent derived for holographic EE in \cite{Fischler:2012ca}. Finally, the area integral \eqref{eq:bh11} yields the result \eqref{eq:area} stated in the main text.

\end{appendix}

\bibliographystyle{fullsort}
\bibliography{references}

\providecommand{\href}[2]{#2}\begingroup\raggedright\begin{thebibliography}{10}

\bibitem{Ryu:2006bv}
S.~Ryu and T.~Takayanagi, ``{Holographic derivation of entanglement entropy
  from AdS/CFT},'' {\em Phys. Rev. Lett.} {\bf 96} (2006) 181602,
\href{http://www.arXiv.org/abs/hep-th/0603001}{{\tt hep-th/0603001}}.
%%CITATION = HEP-TH/0603001;%%.

\bibitem{Harlow:2014yka}
D.~Harlow, ``{Jerusalem Lectures on Black Holes and Quantum Information},''
  {\em Rev. Mod. Phys.} {\bf 88} (2016) 015002,
  \href{http://www.arXiv.org/abs/1409.1231}{{\tt 1409.1231}}.

\bibitem{Hayden:2016cfa}
P.~Hayden, S.~Nezami, X.-L. Qi, N.~Thomas, M.~Walter, and Z.~Yang,
  ``{Holographic duality from random tensor networks},'' {\em JHEP} {\bf 11}
  (2016) 009, \href{http://www.arXiv.org/abs/1601.01694}{{\tt 1601.01694}}.

\bibitem{VanRaamsdonk:2016exw}
M.~Van~Raamsdonk, ``{Lectures on Gravity and Entanglement},'' in {\em
  {Theoretical Advanced Study Institute in Elementary Particle Physics}: {New
  Frontiers in Fields and Strings}}, pp.~297--351.
\newblock 2017.
\newblock \href{http://www.arXiv.org/abs/1609.00026}{{\tt 1609.00026}}.

\bibitem{Almheiri:2020cfm}
A.~Almheiri, T.~Hartman, J.~Maldacena, E.~Shaghoulian, and A.~Tajdini, ``{The
  entropy of Hawking radiation},''
  \href{http://www.arXiv.org/abs/2006.06872}{{\tt 2006.06872}}.

\bibitem{Bousso:2015mna}
R.~Bousso, Z.~Fisher, S.~Leichenauer, and A.~C. Wall, ``{Quantum focusing
  conjecture},'' {\em Phys. Rev.} {\bf D93} (2016), no.~6, 064044,
\href{http://www.arXiv.org/abs/1506.02669}{{\tt 1506.02669}}.
%%CITATION = ARXIV:1506.02669;%%.

\bibitem{Bousso:2015wca}
R.~Bousso, Z.~Fisher, J.~Koeller, S.~Leichenauer, and A.~C. Wall, ``{Proof of
  the Quantum Null Energy Condition},'' {\em Phys. Rev.} {\bf D93} (2016),
  no.~2, 024017,
\href{http://www.arXiv.org/abs/1509.02542}{{\tt 1509.02542}}.
%%CITATION = ARXIV:1509.02542;%%.

\bibitem{Malik:2019dpg}
T.~A. Malik and R.~Lopez-Mobilia, ``{Proof of the quantum null energy condition
  for free fermionic field theories},'' {\em Phys. Rev. D} {\bf 101} (2020),
  no.~6, 066028, \href{http://www.arXiv.org/abs/1910.07594}{{\tt 1910.07594}}.

\bibitem{Koeller:2015qmn}
J.~Koeller and S.~Leichenauer, ``{Holographic Proof of the Quantum Null Energy
  Condition},'' {\em Phys. Rev.} {\bf D94} (2016), no.~2, 024026,
\href{http://www.arXiv.org/abs/1512.06109}{{\tt 1512.06109}}.
%%CITATION = ARXIV:1512.06109;%%.

\bibitem{Balakrishnan:2017bjg}
S.~Balakrishnan, T.~Faulkner, Z.~U. Khandker, and H.~Wang, ``{A General Proof
  of the Quantum Null Energy Condition},'' {\em JHEP} {\bf 09} (2019) 020,
  \href{http://www.arXiv.org/abs/1706.09432}{{\tt 1706.09432}}.

\bibitem{Ceyhan:2018zfg}
F.~Ceyhan and T.~Faulkner, ``{Recovering the QNEC from the ANEC},'' {\em
  Commun. Math. Phys.} {\bf 377} (2020), no.~2, 999--1045,
  \href{http://www.arXiv.org/abs/1812.04683}{{\tt 1812.04683}}.

\bibitem{PhysRevLett.123.121602}
D.~Grumiller, P.~Parekh, and M.~Riegler, ``Local quantum energy conditions in
  non-lorentz-invariant quantum field theories,'' {\em Phys. Rev. Lett.} {\bf
  123} (Sep, 2019) 121602.

\bibitem{Wall:2011kb}
A.~C. Wall, ``{Testing the Generalized Second Law in 1+1 dimensional Conformal
  Vacua: An Argument for the Causal Horizon},'' {\em Phys. Rev.} {\bf D85}
  (2012) 024015,
\href{http://www.arXiv.org/abs/1105.3520}{{\tt 1105.3520}}.
%%CITATION = ARXIV:1105.3520;%%.

\bibitem{Gubser:2008ny}
S.~S. Gubser and A.~Nellore, ``{Mimicking the QCD equation of state with a dual
  black hole},'' {\em Phys. Rev.} {\bf D78} (2008) 086007,
\href{http://www.arXiv.org/abs/0804.0434}{{\tt 0804.0434}}.
%%CITATION = ARXIV:0804.0434;%%.

\bibitem{Janik:2015iry}
R.~A. Janik, J.~Jankowski, and H.~Soltanpanahi, ``{Nonequilibrium Dynamics and
  Phase Transitions in Holographic Models},'' {\em Phys. Rev. Lett.} {\bf 117}
  (2016), no.~9, 091603,
\href{http://www.arXiv.org/abs/1512.06871}{{\tt 1512.06871}}.
%%CITATION = ARXIV:1512.06871;%%.

\bibitem{Attems:2016ugt}
M.~Attems, J.~Casalderrey-Solana, D.~Mateos, I.~Papadimitriou,
  D.~Santos-Oliván, C.~F. Sopuerta, M.~Triana, and M.~Zilhão,
  ``{Thermodynamics, transport and relaxation in non-conformal theories},''
  {\em JHEP} {\bf 10} (2016) 155,
\href{http://www.arXiv.org/abs/1603.01254}{{\tt 1603.01254}}.
%%CITATION = ARXIV:1603.01254;%%.

\bibitem{Janik:2016btb}
R.~A. Janik, J.~Jankowski, and H.~Soltanpanahi, ``{Quasinormal modes and the
  phase structure of strongly coupled matter},'' {\em JHEP} {\bf 06} (2016)
  047,
\href{http://www.arXiv.org/abs/1603.05950}{{\tt 1603.05950}}.
%%CITATION = ARXIV:1603.05950;%%.

\bibitem{Attems:2017ezz}
M.~Attems, Y.~Bea, J.~Casalderrey-Solana, D.~Mateos, M.~Triana, and M.~Zilhao,
  ``{Phase Transitions, Inhomogeneous Horizons and Second-Order
  Hydrodynamics},'' {\em JHEP} {\bf 06} (2017) 129,
\href{http://www.arXiv.org/abs/1703.02948}{{\tt 1703.02948}}.
%%CITATION = ARXIV:1703.02948;%%.

\bibitem{Janik:2017ykj}
R.~A. Janik, J.~Jankowski, and H.~Soltanpanahi, ``{Real-Time dynamics and phase
  separation in a holographic first order phase transition},'' {\em Phys. Rev.
  Lett.} {\bf 119} (2017), no.~26, 261601,
  \href{http://www.arXiv.org/abs/1704.05387}{{\tt 1704.05387}}.

\bibitem{Brown:1986nw}
J.~Brown and M.~Henneaux, ``{Central Charges in the Canonical Realization of
  Asymptotic Symmetries: An Example from Three-Dimensional Gravity},'' {\em
  Commun. Math. Phys.} {\bf 104} (1986) 207--226.

\bibitem{Banados:1998gg}
M.~Banados, ``{Three-dimensional quantum geometry and black holes},'' {\em AIP
  Conf. Proc.} {\bf 484} (1999), no.~1, 147--169,
  \href{http://www.arXiv.org/abs/hep-th/9901148}{{\tt hep-th/9901148}}.

\bibitem{Roberts:2012aq}
M.~M. Roberts, ``{Time evolution of entanglement entropy from a pulse},'' {\em
  JHEP} {\bf 12} (2012) 027, \href{http://www.arXiv.org/abs/1204.1982}{{\tt
  1204.1982}}.

\bibitem{Sheikh-Jabbari:2016znt}
M.~Sheikh-Jabbari and H.~Yavartanoo, ``{Excitation entanglement entropy in two
  dimensional conformal field theories},'' {\em Phys. Rev. D} {\bf 94} (2016),
  no.~12, 126006, \href{http://www.arXiv.org/abs/1605.00341}{{\tt 1605.00341}}.

\bibitem{Holzhey:1994we}
C.~Holzhey, F.~Larsen, and F.~Wilczek, ``{Geometric and renormalized entropy in
  conformal field theory},'' {\em Nucl. Phys. B} {\bf 424} (1994) 443--467,
  \href{http://www.arXiv.org/abs/hep-th/9403108}{{\tt hep-th/9403108}}.

\bibitem{Calabrese:2004eu}
P.~Calabrese and J.~L. Cardy, ``{Entanglement entropy and quantum field
  theory},'' {\em J. Stat. Mech.} {\bf 0406} (2004) P06002,
  \href{http://www.arXiv.org/abs/hep-th/0405152}{{\tt hep-th/0405152}}.

\bibitem{Ecker:2019ocp}
C.~Ecker, D.~Grumiller, W.~van~der Schee, M.~M. Sheikh-Jabbari, and P.~Stanzer,
  ``{Quantum Null Energy Condition and its (non)saturation in 2d CFTs},'' {\em
  SciPost Phys.} {\bf 6} (2019), no.~3, 036,
\href{http://www.arXiv.org/abs/1901.04499}{{\tt 1901.04499}}.
%%CITATION = ARXIV:1901.04499;%%.

\bibitem{Khandker:2018xls}
Z.~U. Khandker, S.~Kundu, and D.~Li, ``{Bulk Matter and the Boundary Quantum
  Null Energy Condition},'' {\em JHEP} {\bf 08} (2018) 162,
\href{http://www.arXiv.org/abs/1803.03997}{{\tt 1803.03997}}.
%%CITATION = ARXIV:1803.03997;%%.

\bibitem{Mezei:2019sla}
M.~Mezei and J.~Virrueta, ``{The Quantum Null Energy Condition and Entanglement
  Entropy in Quenches},''
\href{http://www.arXiv.org/abs/1909.00919}{{\tt 1909.00919}}.
%%CITATION = ARXIV:1909.00919;%%.

\bibitem{Ecker:2017jdw}
C.~Ecker, D.~Grumiller, W.~van~der Schee, and P.~Stanzer, ``{Saturation of the
  Quantum Null Energy Condition in Far-From-Equilibrium Systems},'' {\em Phys.
  Rev. D} {\bf 97} (2018), no.~12, 126016,
  \href{http://www.arXiv.org/abs/1710.09837}{{\tt 1710.09837}}.

\bibitem{Myers:2012ed}
R.~C. Myers and A.~Singh, ``{Comments on Holographic Entanglement Entropy and
  RG Flows},'' {\em JHEP} {\bf 04} (2012) 122,
\href{http://www.arXiv.org/abs/1202.2068}{{\tt 1202.2068}}.
%%CITATION = ARXIV:1202.2068;%%.

\bibitem{Liu:2012eea}
H.~Liu and M.~Mezei, ``{A Refinement of entanglement entropy and the number of
  degrees of freedom},'' {\em JHEP} {\bf 04} (2013) 162,
  \href{http://www.arXiv.org/abs/1202.2070}{{\tt 1202.2070}}.

\bibitem{Casini:2004bw}
H.~Casini and M.~Huerta, ``{A Finite entanglement entropy and the c-theorem},''
  {\em Phys. Lett.} {\bf B600} (2004) 142--150,
\href{http://www.arXiv.org/abs/hep-th/0405111}{{\tt hep-th/0405111}}.
%%CITATION = HEP-TH/0405111;%%.

\bibitem{Casini:2006es}
H.~Casini and M.~Huerta, ``{A c-theorem for the entanglement entropy},'' {\em
  J. Phys.} {\bf A40} (2007) 7031--7036,
\href{http://www.arXiv.org/abs/cond-mat/0610375}{{\tt cond-mat/0610375}}.
%%CITATION = COND-MAT/0610375;%%.

\bibitem{Breitenlohner:1982bm}
P.~Breitenlohner and D.~Z. Freedman, ``{Positive Energy in anti-De Sitter
  Backgrounds and Gauged Extended Supergravity},'' {\em Phys. Lett. B} {\bf
  115} (1982) 197--201.

\bibitem{Breitenlohner:1982jf}
P.~Breitenlohner and D.~Z. Freedman, ``{Stability in Gauged Extended
  Supergravity},'' {\em Annals Phys.} {\bf 144} (1982) 249.

\bibitem{Chomaz:2003dz}
P.~Chomaz, M.~Colonna, and J.~Randrup, ``{Nuclear spinodal fragmentation},''
  {\em Phys. Rept.} {\bf 389} (2004) 263--440.

\bibitem{Bea:2020ees}
Y.~Bea, O.~J. Dias, T.~Giannakopoulos, D.~Mateos, M.~Sanchez-Garitaonandia,
  J.~E. Santos, and M.~Zilhao, ``{Crossing a large-$N$ phase transition at
  finite volume},'' \href{http://www.arXiv.org/abs/2007.06467}{{\tt
  2007.06467}}.

\bibitem{Balasubramanian:2014sra}
V.~Balasubramanian, B.~D. Chowdhury, B.~Czech, and J.~de~Boer, ``{Entwinement
  and the emergence of spacetime},'' {\em JHEP} {\bf 01} (2015) 048,
  \href{http://www.arXiv.org/abs/1406.5859}{{\tt 1406.5859}}.

\bibitem{Fischler:2012ca}
W.~Fischler and S.~Kundu, ``{Strongly Coupled Gauge Theories: High and Low
  Temperature Behavior of Non-local Observables},'' {\em JHEP} {\bf 05} (2013)
  098,
\href{http://www.arXiv.org/abs/1212.2643}{{\tt 1212.2643}}.
%%CITATION = ARXIV:1212.2643;%%.

\bibitem{Ecker:2018jgh}
C.~Ecker, {\em {Entanglement Entropy from Numerical Holography}}.
\newblock PhD thesis, Vienna, Tech. U., 9, 2018.
\newblock \href{http://www.arXiv.org/abs/1809.05529}{{\tt 1809.05529}}.

\bibitem{Attems:2019yqn}
M.~Attems, Y.~Bea, J.~Casalderrey-Solana, D.~Mateos, and M.~Zilh{\~a}o,
  ``{Dynamics of Phase Separation from Holography},'' {\em JHEP} {\bf 01}
  (2020) 106, \href{http://www.arXiv.org/abs/1905.12544}{{\tt 1905.12544}}.

\bibitem{Klebanov:2007ws}
I.~R. Klebanov, D.~Kutasov, and A.~Murugan, ``{Entanglement as a probe of
  confinement},'' {\em Nucl. Phys. B} {\bf 796} (2008) 274--293,
  \href{http://www.arXiv.org/abs/0709.2140}{{\tt 0709.2140}}.

\bibitem{Kiritsis:2016kog}
E.~Kiritsis, F.~Nitti, and L.~Silva~Pimenta, ``{Exotic RG Flows from
  Holography},'' {\em Fortsch. Phys.} {\bf 65} (2017), no.~2, 1600120,
  \href{http://www.arXiv.org/abs/1611.05493}{{\tt 1611.05493}}.

\bibitem{Gubser:2008px}
S.~S. Gubser, ``{Breaking an Abelian gauge symmetry near a black hole
  horizon},'' {\em Phys. Rev. D} {\bf 78} (2008) 065034,
  \href{http://www.arXiv.org/abs/0801.2977}{{\tt 0801.2977}}.

\bibitem{Hartnoll:2008vx}
S.~A. Hartnoll, C.~P. Herzog, and G.~T. Horowitz, ``{Building a Holographic
  Superconductor},'' {\em Phys. Rev. Lett.} {\bf 101} (2008) 031601,
  \href{http://www.arXiv.org/abs/0803.3295}{{\tt 0803.3295}}.

\bibitem{Charmousis:2010zz}
C.~Charmousis, B.~Gouteraux, B.~S. Kim, E.~Kiritsis, and R.~Meyer, ``{Effective
  Holographic Theories for low-temperature condensed matter systems},'' {\em
  JHEP} {\bf 11} (2010) 151, \href{http://www.arXiv.org/abs/1005.4690}{{\tt
  1005.4690}}.

\bibitem{Gouteraux:2011ce}
B.~Gouteraux and E.~Kiritsis, ``{Generalized Holographic Quantum Criticality at
  Finite Density},'' {\em JHEP} {\bf 12} (2011) 036,
  \href{http://www.arXiv.org/abs/1107.2116}{{\tt 1107.2116}}.

\bibitem{Huijse:2011ef}
L.~Huijse, S.~Sachdev, and B.~Swingle, ``{Hidden Fermi surfaces in compressible
  states of gauge-gravity duality},'' {\em Phys. Rev. B} {\bf 85} (2012)
  035121, \href{http://www.arXiv.org/abs/1112.0573}{{\tt 1112.0573}}.

\bibitem{Andrade:2013gsa}
T.~Andrade and B.~Withers, ``{A simple holographic model of momentum
  relaxation},'' {\em JHEP} {\bf 05} (2014) 101,
  \href{http://www.arXiv.org/abs/1311.5157}{{\tt 1311.5157}}.

\bibitem{Grumiller:2019xna}
D.~Grumiller, P.~Parekh, and M.~Riegler, ``{Local quantum energy conditions in
  non-Lorentz-invariant quantum field theories},'' {\em Phys. Rev. Lett.} {\bf
  123} (2019), no.~12, 121602, \href{http://www.arXiv.org/abs/1907.06650}{{\tt
  1907.06650}}.

\bibitem{Baggioli:2020cld}
M.~Baggioli and D.~Giataganas, ``{Detecting Topological Quantum Phase
  Transitions via the c-Function},''
  \href{http://www.arXiv.org/abs/2007.07273}{{\tt 2007.07273}}.

\bibitem{Chu:2019uoh}
C.-S. Chu and D.~Giataganas, ``{$c$-Theorem for Anisotropic RG Flows from
  Holographic Entanglement Entropy},'' {\em Phys. Rev. D} {\bf 101} (2020),
  no.~4, 046007, \href{http://www.arXiv.org/abs/1906.09620}{{\tt 1906.09620}}.

\end{thebibliography}\endgroup

\end{document}